\documentclass[useAMS,usenatbib]{mn2e}
\usepackage{graphicx}
\usepackage{hyperref}

\newcommand{\mnras}{MNRAS}
\newcommand{\apj}{ApJ}
\newcommand{\apjs}{ApJS}
\newcommand{\nat}{Nature}

\newcommand{\aap}{A\&A}
\newcommand{\araa}{ARA\&A}

\newcommand{\nLAE}{247}
\newcommand{\npairs}{195}
\newcommand{\Nneigmax}{15}

\newcommand{\UVBHM}{1.14\times10^{-20}}
\newcommand{\UVB}{0.44\times10^{-20}}

\newcommand{\SBcgs}{\ erg s$^{-1}$ cm$^{-2}$ arcsec$^{-2}$}

\title{Stacking the Cosmic Web in Fluorescent Ly$\alpha$ Emission with MUSE}
\author[Sofia G. Gallego et al.]{Sofia G. Gallego$^{1}$, Sebastiano Cantalupo$^{1}$, Simon Lilly$^{1}$, Raffaella Anna Marino$^{1}$, 
\newauthor Gabriele Pezzulli$^{1}$, Joop Schaye$^{2}$,  Lutz Wisotzki$^{3}$, Roland Bacon$^{4}$, Hanae Inami$^{4}$,
\newauthor Mohammad Akhlaghi$^{4}$, Sandro Tacchella$^{1}$, Johan Richard$^{4}$, Nicolas Bouche$^{5}$,
\newauthor Matthias Steinmetz$^{3}$ \& Marcella Carollo$^{1}$
\thanks{E-mail:
gallegos@phys.ethz.ch}
\\$^{1}$Institute for Astronomy, Department of Physics, ETH Z\"urich, CH-8093 Z\"urich, Switzerland
\\$^{2}$Leiden Observatory, Leiden University, PO Box 9513, NL-2300 RA Leiden, the Netherlands
\\$^{3}$Leibniz-Institut f\"ur Astrophysik Potsdam, AIP, An der Sternwarte 16, D-14482 Potsdam, Germany
\\$^{4}$Univ Lyon, Univ Lyon1, Ens de Lyon, CNRS, Centre de Recherche Astrophysique de Lyon UMR5574, F-69230, Saint-Genis-Laval, France
\\$^{5}$CNRS/IRAP, 9 Avenue Colonel Roche, F-31400 Toulouse, France
}
\begin{document}

\date{\today}

\pagerange{\pageref{firstpage}--\pageref{lastpage}} \pubyear{2017}

\maketitle

\label{firstpage}

\begin{abstract}
Cosmological simulations suggest that most of the matter in the Universe is distributed along filaments connecting galaxies. Illuminated by the cosmic UV background (UVB), these structures are expected to glow in fluorescent Ly$\alpha$ emission 
with a Surface Brightness (SB) that is well below current observational limits for individual detections.
Here, we perform a stacking analysis of the deepest MUSE/VLT data using three-dimensional regions (subcubes) with orientations determined by the position of neighbouring Ly$\alpha$ galaxies (LAEs) at $3<z<4$.
Our method should increase the probability of detecting filamentary Ly$\alpha$ emission, provided that these structures are Lyman Limit Systems (LLSs). By stacking 390 oriented subcubes we reach a $2\sigma$ sensitivity level of $\mathrm{SB}\approx0.44\times10^{-20}$\SBcgs\  in an aperture of $1\,\mathrm{arcsec^2}\times 6.25\,\mathrm{\AA}$, which is three times below the expected fluorescent Ly$\alpha$ signal from the Haardt-Madau 2012 (HM12) UVB at $z\sim3.5$.
No detectable emission is found on intergalactic scales,
implying that at least two thirds 
of our subcubes do not contain oriented LLSs for a HM12 UVB.
On the other hand, significant emission is detected in the circum-galactic medium (CGM) of galaxies 
in the direction of the neighbours.
The signal is stronger for galaxies with a larger number of neighbours and appears to be independent of any other galaxy properties such as luminosity, redshift and neighbour distance.
We estimate that preferentially oriented satellite galaxies cannot contribute significantly to this signal, suggesting instead that gas densities in the CGM are typically larger in the direction of neighbouring galaxies on cosmological scales. 

\end{abstract}


\section{Introduction}

Our standard cosmological paradigm predicts that structures in the Universe grew from initial Gaussian quantum fluctuations into a ``Cosmic Web'' of intergalactic filaments
\citep[e.g.,][]{Peebles1975,Bond1996} 
where galaxies form and evolve. 
However, most of the baryonic material in these filaments is expected to be too diffuse to form stars.

In the local universe, it has been empirically demonstrated that the large-scale distribution of galaxies and the velocity field are consistent with the predicted filamentary structure \citep[see e.g.,][]{Libeskind2015}. At high-redshift, the first evidence of an Intergalactic Medium (IGM) came from the analysis of absorption lines in the spectra of quasars (see \citeauthor{Rauch1998} \citeyear{Rauch1998} for a review).
Unfortunately, given the one-dimensional nature of these absorption probes we have still little direct information on the spatial distribution and small-scale properties of intergalactic gas. 
Direct imaging of the Cosmic Web is in principle possible through fluorescent Ly$\alpha$ emission \citep{Hogan1987,Gould1996,Haiman2001,Cantalupo2005}. 
In particular, it is expected that gaseous filaments illuminated by ionizing radiation from the cosmic UV background (UVB) or local sources should emit Ly$\alpha$ radiation following hydrogen recombinations. 
For self-shielded gas clouds, about 60\% of the incident ionizing radiation should be converted to fluorescent Ly$\alpha$ emission (\citeauthor{Gould1996} \citeyear{Gould1996}, but see \citeauthor{Cantalupo2005} \citeyear{Cantalupo2005}). Such clouds therefore act as a kind of mirror of the UVB if they are away from bright UV sources such as quasars.
In the spectra of a quasar, self-shielded clouds correspond to Lyman-limit systems (LLSs; with column densities of neutral hydrogen $N_\mathrm{HI}>10^{17.2}$ cm$^{-2}$) and Damped Ly$\alpha$ systems (DLA; with $N_\mathrm{HI}>10^{20.3}$ cm$^{-2}$) although the latter are typically much rarer \citep{Peroux2001,Prochaska2010,Noterdaeme2014}.
Ly$\alpha$ imaging of the typical LLSs should then provide direct constraints on the value of the cosmic UVB.

Previous attempts to detect fluorescent Ly$\alpha$ emission induced by the cosmic UVB have been unsuccessful. The deepest spectroscopic observation (a 92hr exposure with the VLT/FORS2 instrument)
conducted so far reached a 1$\sigma$ surface brightness (SB) limit of $8\times10^{-20}$\SBcgs\  
per arcsec$^{2}$ aperture 
at $z\approx3$ \citep{Rauch2008}. This observation used a long slit probing a total area of $2''\times453''$ ($\approx$0.25 arcmin$^2$) and a redshift range of $2.67<z<3.75$. Given the large redshift range probed and the observed incident rate of 
LLSs of about 1.5 per unit redshift at $z\approx3$, \citep[e.g.,][]{Prochaska2010}, a large number of fluorescently emitting LLSs could have been detected in this study. This null result implied an upper limit on the UVB ionisation rate of $\Gamma_{\mathrm{HI}}<2.7\times10^{-12}$\,s$^{-1}$ at 1$\sigma$ at $z\approx3$. 

What are other observational and theoretical constraints on the cosmic UVB? 
Using the so called ``proximity effect", i.e. the decrease in the number density of Ly$\alpha$ forest lines in proximity of quasars due to the increased ionizing radiation, 
we can put limits on the average intensity of the UVB at the Lyman-limit, e.g.  $\mathrm{J = (9 \pm 4)\times 10^{-22}\, erg\, cm^{-2}\, s^{-1}\,Hz^{-1}\,sr^{-1}}$ 
(\citeauthor{DallAglio2008} \citeyear{DallAglio2008}; see also, e.g., \citeauthor{Carswell1987} \citeyear{Carswell1987}; \citeauthor{Bajtlik1988} \citeyear{Bajtlik1988}; \citeauthor{Scott2000} \citeyear{Scott2000}; \citeauthor{Calverley2011} \citeyear{Calverley2011}). 
These measurements, however, may be affected by clustering in the proximity of quasars or errors in the estimates of the quasars' ionizing luminosities and systemic redshifts. 
An alternative method uses the mean flux in the Ly$\alpha$ forest in combination with numerical simulations where the UVB is adjusted until the 
mean flux in artificial Ly$\alpha$ forest spectra matches the real data \citep[see e.g.,][]{Rauch1997,Bolton2005,Faucher2008,Becker2013}.
This method typically gives systematically lower values (by about a factor of 2 to 3, depending on the UVB spectral energy distribution) for the amplitude of the UVB compared 
to the proximity-effect measurements, although different  works in the literature have discrepancies of a factor up to 2 due to different IGM temperatures in the simulations 
(see e.g., \citeauthor{Becker2013} \citeyear{Becker2013} for a discussion). 

Overall, these studies suggest that the UVB hydrogen ionisation rate
should be around $0.8\times10^{-12}$ s$^{-1}$ at $z\approx3.5$ with very little evolution
in the redshift range $2.5<z<4.5$ \citep[e.g.,][]{Becker2013}. 
Predictions made with synthesis UVB models, e.g., \citet{Haardt1996}, \citet{Faucher2009}, \citeauthor{Haardt2012} (\citeyear{Haardt2012}, hereafter HM12), produce similar values of $\Gamma_{\mathrm{HI}}$ 
but suggest a more pronounced redshift evolution, mostly due to the assumed fraction of ionizing photons from galaxies and from the
extrapolation of the observed quasar luminosity functions to the faint-end.
In particular, the models from HM12 predict $\Gamma_{\mathrm{HI}} \approx 0.95\times10^{-12}\,\mathrm{s}^{-1}$ at $z=2.5$ and lower by a factor of 1.7 at $z=4$.

Given these low values of $\Gamma_{\mathrm{HI}}$, it is clear that the expected fluorescent emission from the UVB is out of reach for current facilities. 
Indeed, the expected UVB fluorescence Surface Brightness (SB) for  $\Gamma_{\mathrm{HI}}=0.7\times10^{-12}$ s$^{-1}$ is
$\UVBHM$\SBcgs\
at redshift $z=3.5$ \citep[see e.g.,][]{Cantalupo2005}.
One way to overcome this limitation is to look in the vicinity of bright quasars whose radiation can enhance the incident ionizing radiation by several orders of magnitude \citep{Cantalupo2005,Kollmeier2010}. In recent years, quasar-induced fluorescent emission has been 
detected by means of specifically designed narrow-band (NB) filters and with the new MUSE integral-field spectrograph 
(see \citeauthor{Cantalupo2012} \citeyear{Cantalupo2012}; \citeauthor{Cantalupo2014} \citeyear{Cantalupo2014}; \citeauthor{Hennawi2015} \citeyear{Hennawi2015}; \citeauthor{Borisova2016} \citeyear{Borisova2016} and \citeauthor{Cantalupo2016} \citeyear{Cantalupo2016} for a review).
In addition to providing a new observational window on the Circumgalactic Medium (CGM) of galaxies hosting quasars, these observations can constrain the quasar emission properties.
However, they do not give us any constraints on the cosmic UVB.

Without the boosting effect of quasars there are no alternatives for the detection of fluorescent emission from the UVB with current facilities other
 than stacking a series of deep observations. 
Typical Ly$\alpha$ stacking methods used so far in the literature assume a circularly symmetric distribution of emission. Cosmological simulations suggest instead that the gas distribution between galaxies should be filamentary
and that the filaments should be oriented preferentially towards neighbouring galaxies \citep[e.g.,][]{Bond1996,Gheller2015}. 

In this study, we develop and apply the idea of an ``oriented stacking" approach\begin{footnote}{We notice that a similar idea was also proposed in \citet{vandeVoort2013}, however a quantitative analysis was not presented in that study.}\end{footnote}
 using Ly$\alpha$ emitting galaxies detected in deep MUSE cubes as reference points for the
three-dimensional orientation of each stacking element. If neighbouring galaxies are indeed connected by (straight) filaments and if these filaments contain LLSs, then our oriented-stacking
method should boost the signal-to-noise ratio of UVB-induced fluorescence in IGM filaments by about the square root of the number of stacking elements. 
As we show in this paper, by using the deepest MUSE datacubes currently available and by staking more than 300 individual, ``re-oriented" subcubes around galaxies we are able
to achieve a nominal $3\sigma$ detection limit of $\mathrm{SB}\approx0.78\times10^{-20}$\SBcgs\  in an aperture of $0.4\,\mathrm{arcsec^2}$ for a pseudo NB of width $6.25\,\mathrm{\AA}$, well below the expected fluorescent signal from the values of the cosmic UVB reported above.
In case of a positive detection, this method could also provide direct information on the size and distribution of LLSs and intergalactic filaments in emission away from quasars and therefore in a more typical environment,
thus giving us constraints on the size and morphological properties of these systems.

The paper is organised as follows. 
In Section 2 we describe the data and the selected galaxy catalogue. Section 3 describes the stacking procedure, sample selection and coordinate transformations. Results and discussion are shown in Sections 4 and 5 respectively.
Throughout the paper we assume a flat $\Lambda$CDM cosmology with $\mathrm{H_0}=69.6\,\mathrm{km\,s^{-1}\,Mpc^{-1}}$, $\mathrm{\Omega_\mathrm{m}}=0.286$ and $\mathrm{\Omega_\Lambda}=0.714$ \citep{Bennett2014}.

\section[data]{The Data}
MUSE is a second generation instrument mounted on the Very Large Telescope (VLT) at the Paranal Observatory, Chile, and part of the European Southern Observatory (ESO). It is a panoramic integral-field spectrograph with a field of view of $1'\times1'$ and a wavelength range of $470\,\mathrm{nm}< \lambda< 940\,\mathrm{nm}$ \citep{Bacon2010} with a spatial and wavelength sampling of $0.2''\times0.2''$ and $1.25\,\mathrm{\AA}$, respectively.
To date, two very deep integrations (total exposure time per field of about 27 to 31 hours) have been obtained during commissioning and as a part of the MUSE Guaranteed Time of Observations (GTO):
 the Hubble Deep Field South (HDFS) \citep{Bacon2010} and the MUSE Ultra Deep Field (UDF) (Bacon et al. 2017 submitted).
The HDFS observation was obtained during the last commissioning run of MUSE with a 27 hour exposure time in a field of $1\,\mathrm{arcmin}^2$. The UDF observations consist of a mosaic of nine 10-hour exposure fields obtained during GTO of the MUSE Consortium, plus an overlapping 31-hour exposure in a $1.15\,\mathrm{arcmin}^2$ field. 

For the HDFS we use an improved data reduction obtained with the CubExtractor package (Cantalupo in prep., see also \citeauthor{Borisova2016} \citeyear{Borisova2016} for a short description) that will be presented in a separate paper. 
The full data reduction of the UDF field is described in Bacon et al. 2017 submitted \citep[see also][]{Conseil2016}.
In this paper we use both the HDFS field and the deepest part of the UDF observation, called UDF-10 (hereafter UDF), which have similar depths.

The catalogue of LAEs in the HDFS was extracted from \citet{Bacon2015} and contains 89 LAEs including 26 LAEs not detected in the HST WFPC2 deep broad-band images. 
For the UDF, we use a preliminary LAE catalogue (Inami et al. submitted), that, combined with the HDFS catalogue, gives us a total of  $\nLAE$ LAEs .

During the stacking procedure, we center on the 3-d peak of the Ly$\alpha$ emission that we have re-estimated for each individual galaxy
with respect to the original catalogues. 
Finally, we discard LAEs with low confidence levels (e.g., objects with low signal-to-noise or possible interlopers) and sources closer than $2''$ to the
edge. When pairs of LAE (i.e. objects within $3''$ from each other) are present in the catalogue, we discard the faintest of the two. 
Applying these criteria we discarded 36 LAEs from the initial catalogues.

\section{Stacking Procedure}

In this section we explain how we obtained a set of oriented subcubes around galaxies 
in the direction of their neighbours for the stacking procedure. 
As discussed in Section 1, if galaxies are connected by filaments
with column densities equal to or higher than those of LLSs, our stacking analysis should significantly enhance the
expected fluorescent Ly$\alpha$ signal. 

\subsection{Sample Selection}

\begin{figure}
\begin{center}
\includegraphics[trim={0cm 0cm 1.8cm 0cm},clip, width=3in]{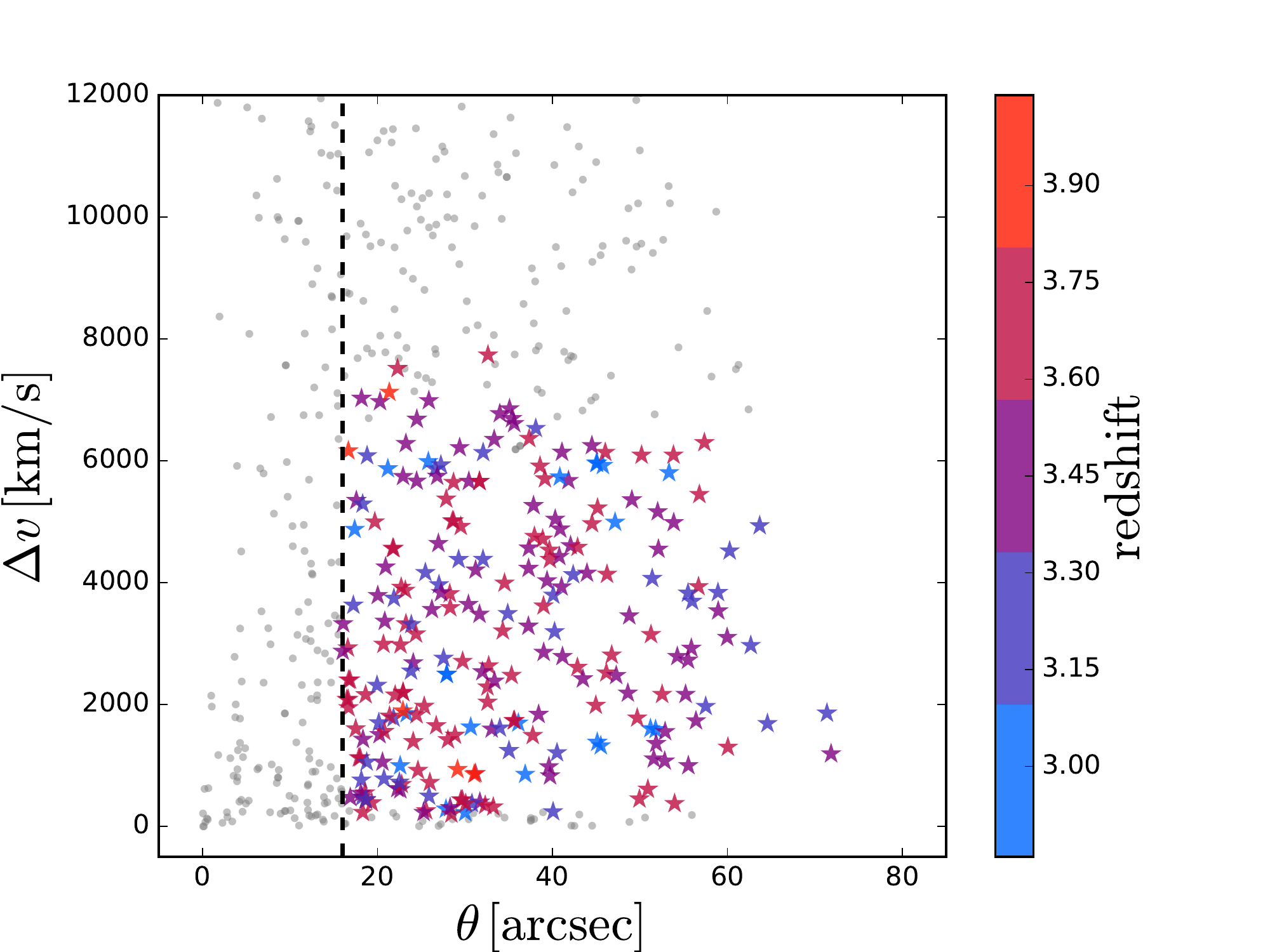}
\includegraphics[trim={0cm 0cm 1.8cm 0cm},clip, width=3in]{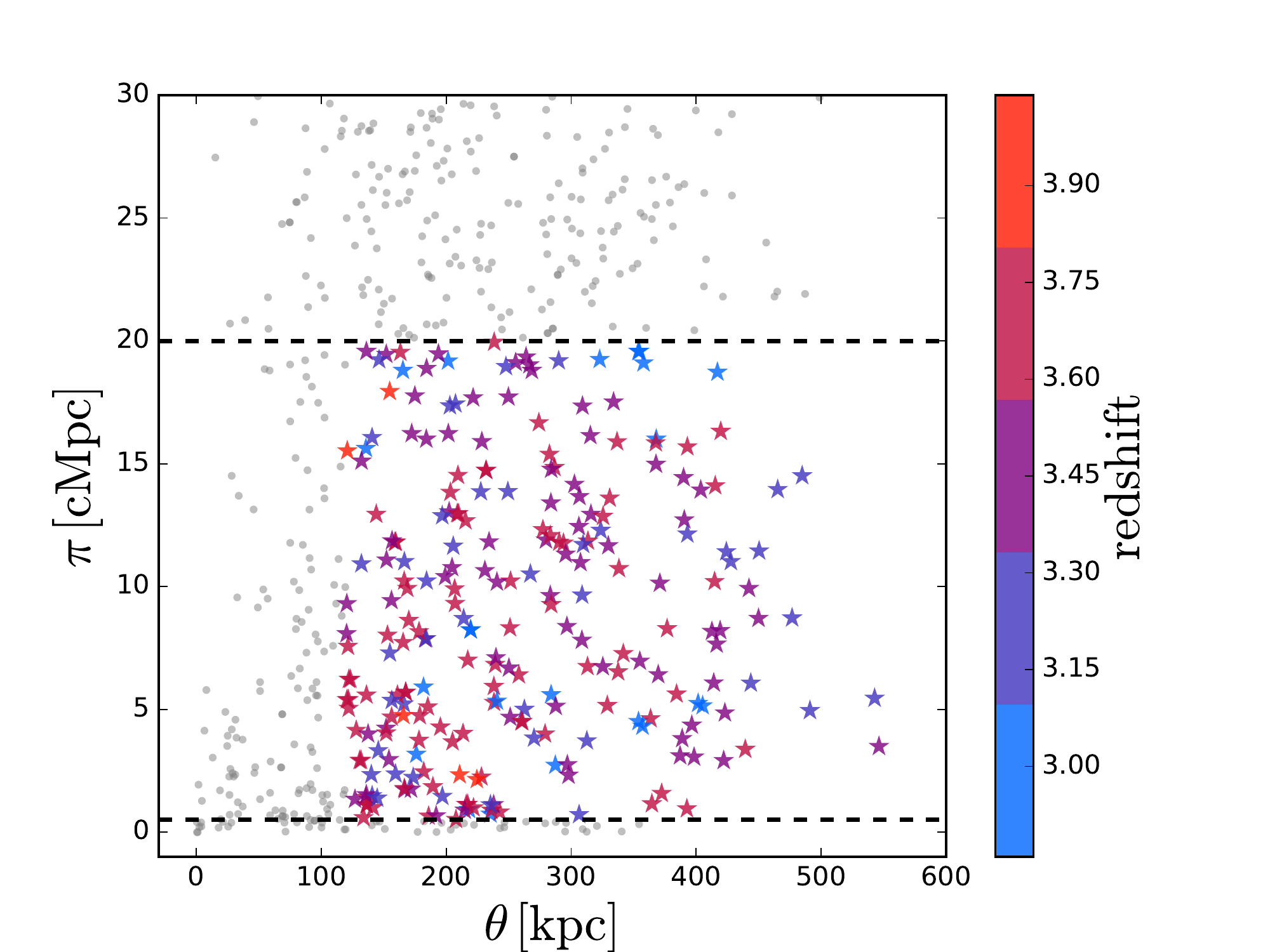}
   \caption{Projected ($\theta$) versus line of sight ($\Delta v$ and $\pi$) distribution of LAE neighbours, color coded by their average redshift. The top panel presents the observationally-derived quantities 
   while the bottom panel shows the inferred values given our chosen cosmological parameters. The dashed lines represent our selection criteria in both projected and line of sight distances.
   Notice that our projected distance range is much smaller than the line of sight separations because of the limited MUSE field of view ($1'\times1'$). Therefore, any neighbours at distances 
   larger than 600 kpc in projected space will not be present in our catalogue. Gray points represent neighbours outside our selection criteria.}    
   \label{pitheta}
\end{center}
\end{figure} 

As a first criterion for our stacking procedure we select a set of galaxy neighbours within line of sight comoving distances ($\pi$) between 0.5 and 20 Mpc (cMpc).
Our choice of the distance upper limit is driven by the need of a large sample of galaxies to reach the required fluorescent emission levels (discussed in Section 1).
However, we limit this distance to 20 cMpc because we expect that the probability 
that two galaxies are connected by a filament 
should rapidly decrease with galaxy distance \citep[see e.g.,][Fig. 8]{Gheller2015}. We find that 20 cMpc is the best compromise between these
two factors. Because we are mostly interested in intergalactic scales and because of uncertainties due to peculiar velocities, 
we limit the smaller distances to 0.5 cMpc. 

 Figure \ref{pitheta} shows the positions of the neighbour-combinations on the projected/comoving distance space, color coded by their average redshift. 
We select 
 the redshift range of $2.9<z<4$ where $z=2.9$ is the minimum Ly$\alpha$ redshift covered by MUSE and we restrict the range to $z<4$ to minimize cosmological SB redshift-dimming effect.
Moreover, we discard neighbours closer than $16''$ to avoid confusion between the Ly$\alpha$ emission coming from the galaxies or their CGM and the potential filamentary structure,
 We did not use a larger projected distance limit to avoid reducing too much the number of subcubes available for the stacking analysis. 
 Within this particular distance range a single LAE can have up to $\Nneigmax$ neighbours. 

The final sample consists of a set of 96 LAEs and $\npairs$ LAE neighbours, or equivalently $2 \times \npairs$ orientations. This corresponds to a cumulative exposure time of $\sim10'000$ hours.

\subsection{Subcubes transformation and stacking}

For each individual LAE in our sample we select a region -- centred on the LAE -- with spatial size of $32''\times32''$ and wavelength width of $12.5\,\mathrm{\AA}$ that has been re-oriented with respect to the
original datacube applying the coordinate transformation described below.

Before coordinate transformation, we performed continuum subtraction using a median filter approach as in \citet{Borisova2016} and masking continuum sources to avoid any continuum flux contamination.
Moreover, we masked a small fraction of wavelength layers in correspondence of bright sky-lines to avoid being contaminated by skyline residuals.

\begin{figure}
\begin{center}
\includegraphics[trim={0cm 0cm 0cm 0cm},clip, width=3in]{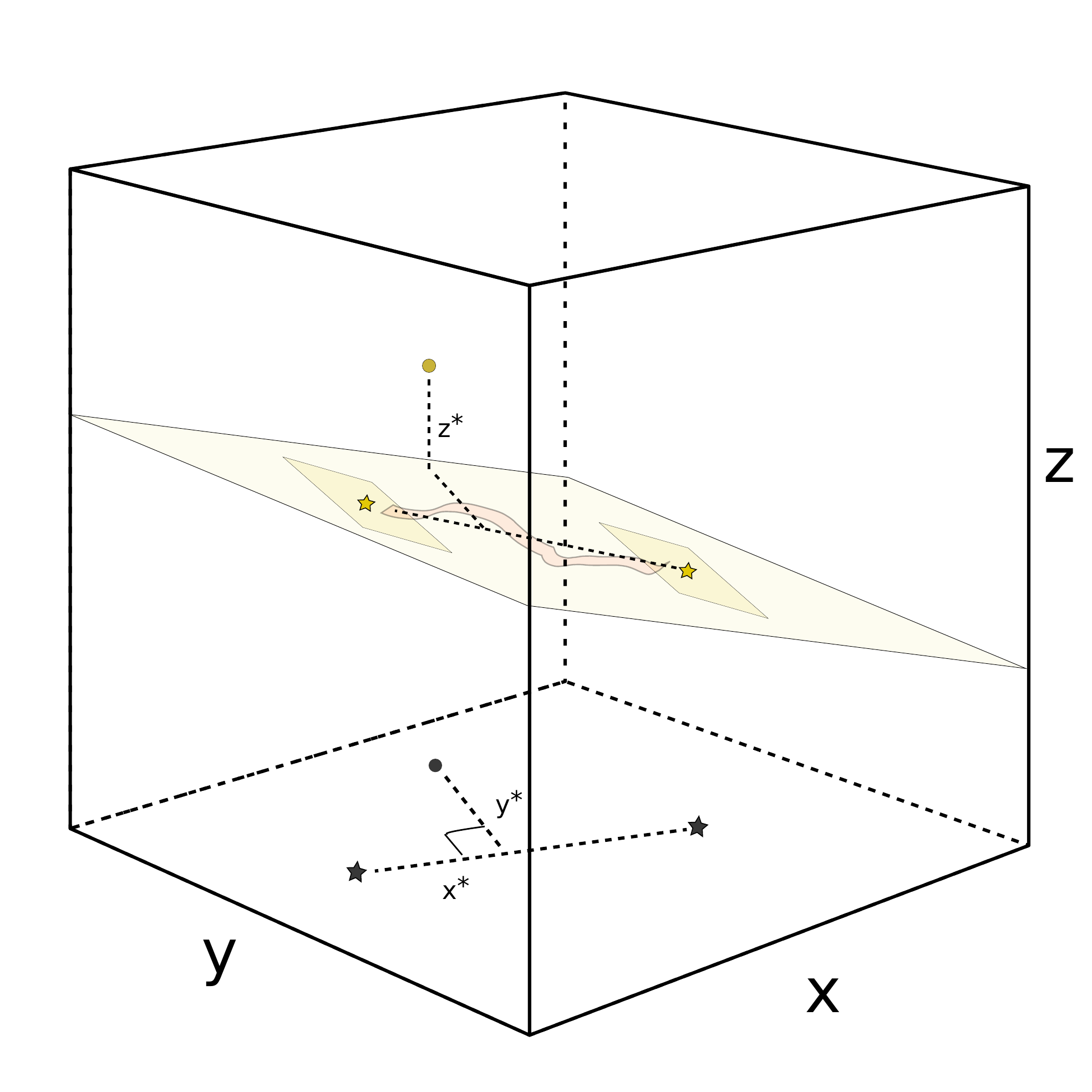}
   \caption{Cartoon representation of the sub-cube coordinate transformation (see text for details). The yellow stars represent the positions of the LAEs in the cube and the pink region depicts a possible filament. 
   The yellow point indicates the position of a particular voxel in the cube. The same objects in projected coordinates are represented in black.
   The light-yellow layer containing the 2 LAE positions represents the plane for which the transformed $z$-coordinate ($z^*$) is equal to zero, whereas the yellow regions around each LAE represent the central layer of the extracted and transformed subcubes with a spatial size of $32''\times32''$. }
   \label{cube}
\end{center}
\end{figure} 

Then, we apply a 2-d transformation of the spatial coordinates in such a way that the resulting angle between the LAE and its neighbour is always zero with respect to the x-axis of the transformed coordinates. 
This means that for each voxel in the cube with coordinates $\textbf{c}=(x,y)$ (independent of z)
there will be a new set of coordinates $\textbf{c}^*=(x^*,y^*)$ defined by:

\begin{center}
\begin{equation}
x^*=\frac{\textbf{u}\,.\,\textbf{d}}{|\textbf{d}|},\,\,\,y^*=\frac{|\textbf{u}\times \textbf{d}|}{|\textbf{d}|}
\end{equation}
\end{center}

Where $\textbf{d}=\textbf{c}_n-\textbf{c}_l$ is the projected distance between the LAEs, $\textbf{c}_l$ and $\textbf{c}_n$ are the spatial coordinates of the galaxy and its neighbour respectively,
and $\textbf{u}=\textbf{c}-\textbf{c}_l$. 

The third coordinate is derived by \emph{shearing} the $z$ coordinate ($\lambda$) with respect to the LAEs as:

\begin{center}
\begin{equation}
z^*=z-z_l-\frac{(z_n-z_l)\,x^*}{|\textbf{d}|} \ .
\end{equation}
\end{center}

The use of the shear is driven by the necessity of accounting for the maximum possible emission of the filament along the direction of the neighbour, by assuming that the wavelength is equivalent
to a distance (therefore omitting any effect of peculiar velocities)
and that the filaments are in a straight line between the galaxies. By using this method we also preserve the spectral shape of the Ly$\alpha$ emission coming from the LAEs and their surroundings. Among our sample, the shear is normally distributed around zero with a standard deviation of $10\,\mathrm{pixels}$. In these coordinates $(0,0,0)$ and $(|\textbf{d}|,0,0)$ are the new positions of the LAE and its neighbour respectively.
Figure \ref{cube} shows a cartoon representation of the coordinate transformation procedure in the cube for a given combination of LAEs.

\begin{figure*}
\begin{center}
\includegraphics[trim={.2cm 3.8cm 1.97cm 0cm},clip, width=1.832in]{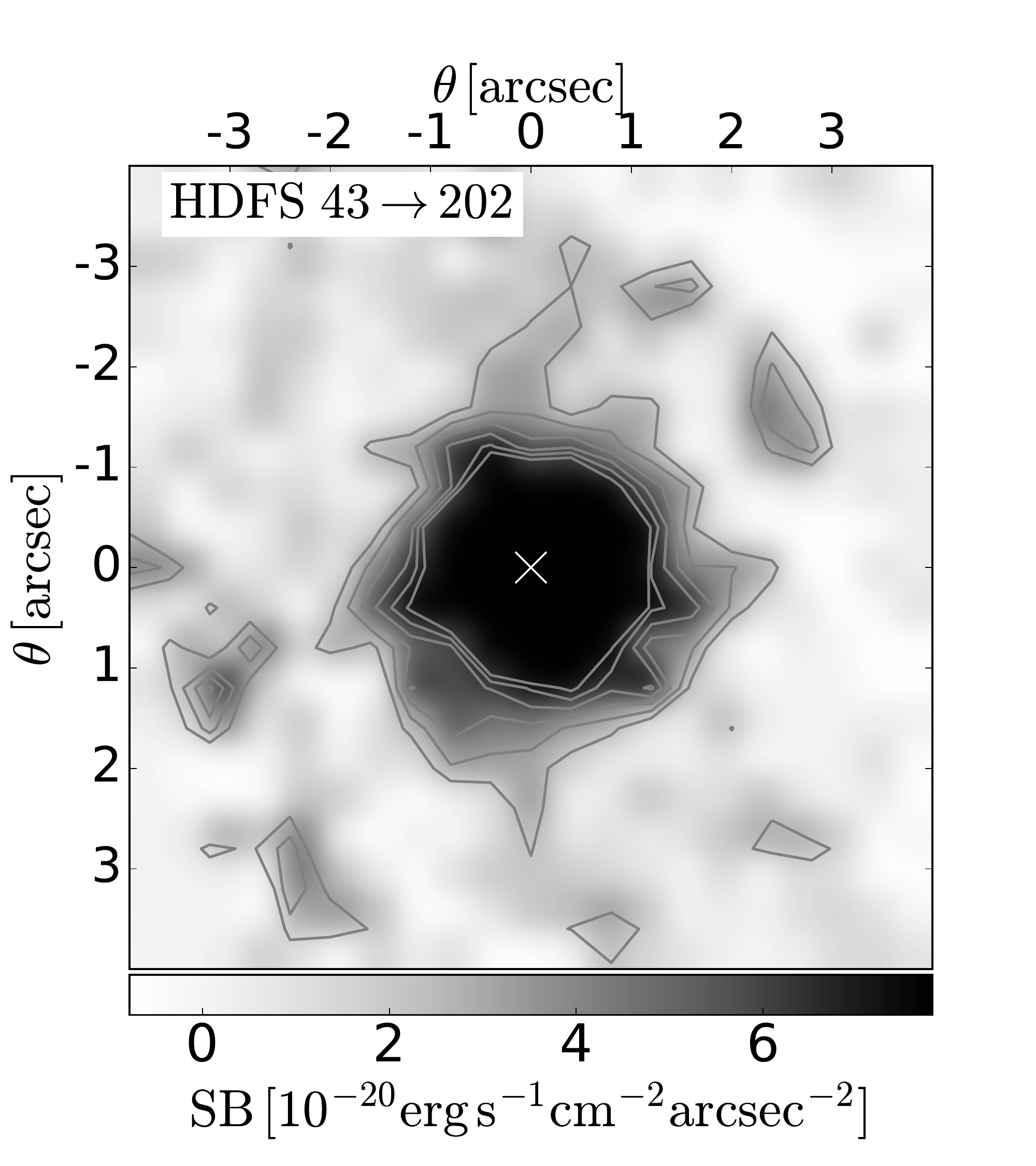}
\includegraphics[trim={2.4cm 3.8cm 1.98cm 0cm},clip, width=1.61in]{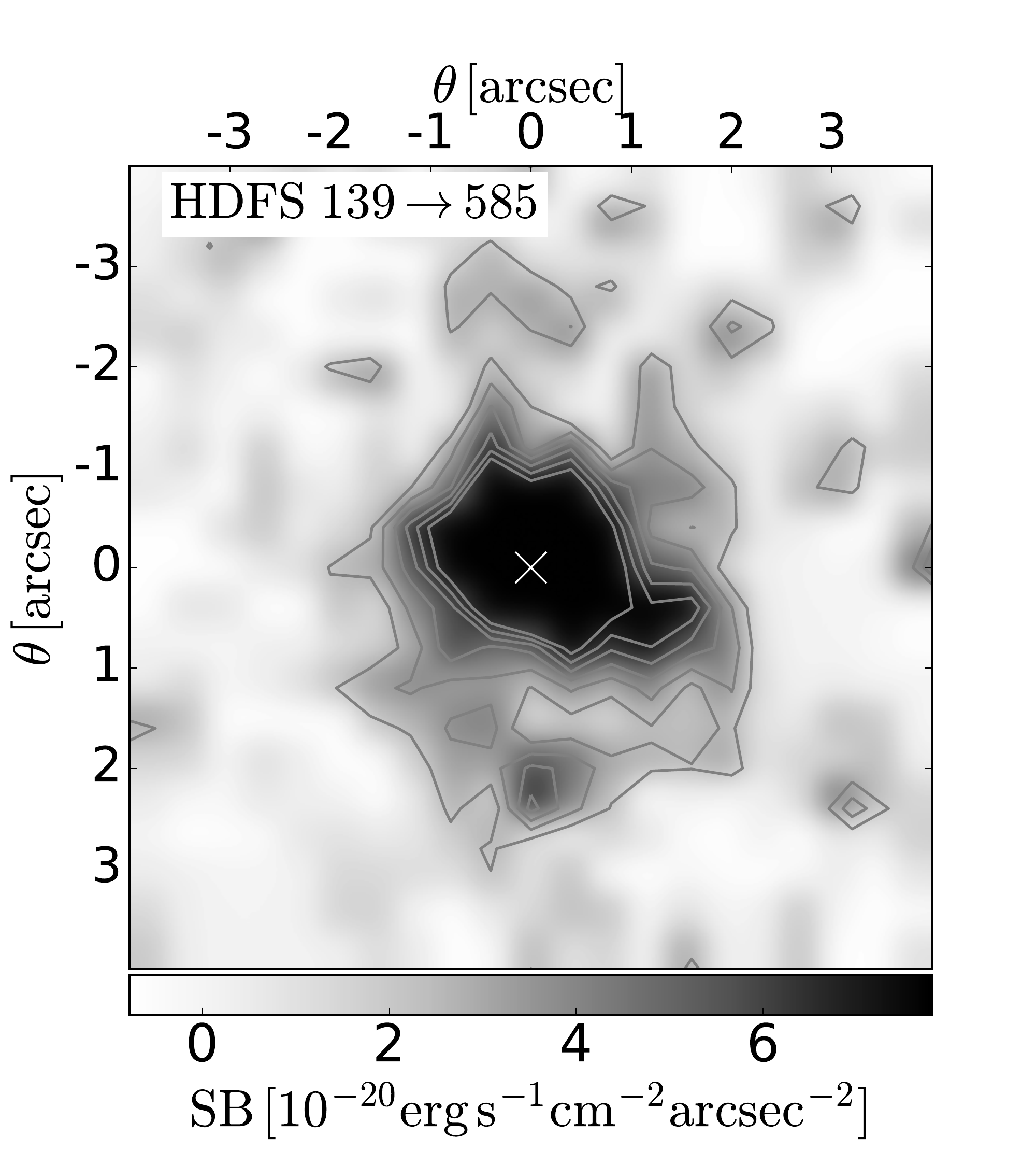}
\includegraphics[trim={2.4cm 3.8cm 1.98cm 0cm},clip, width=1.61in]{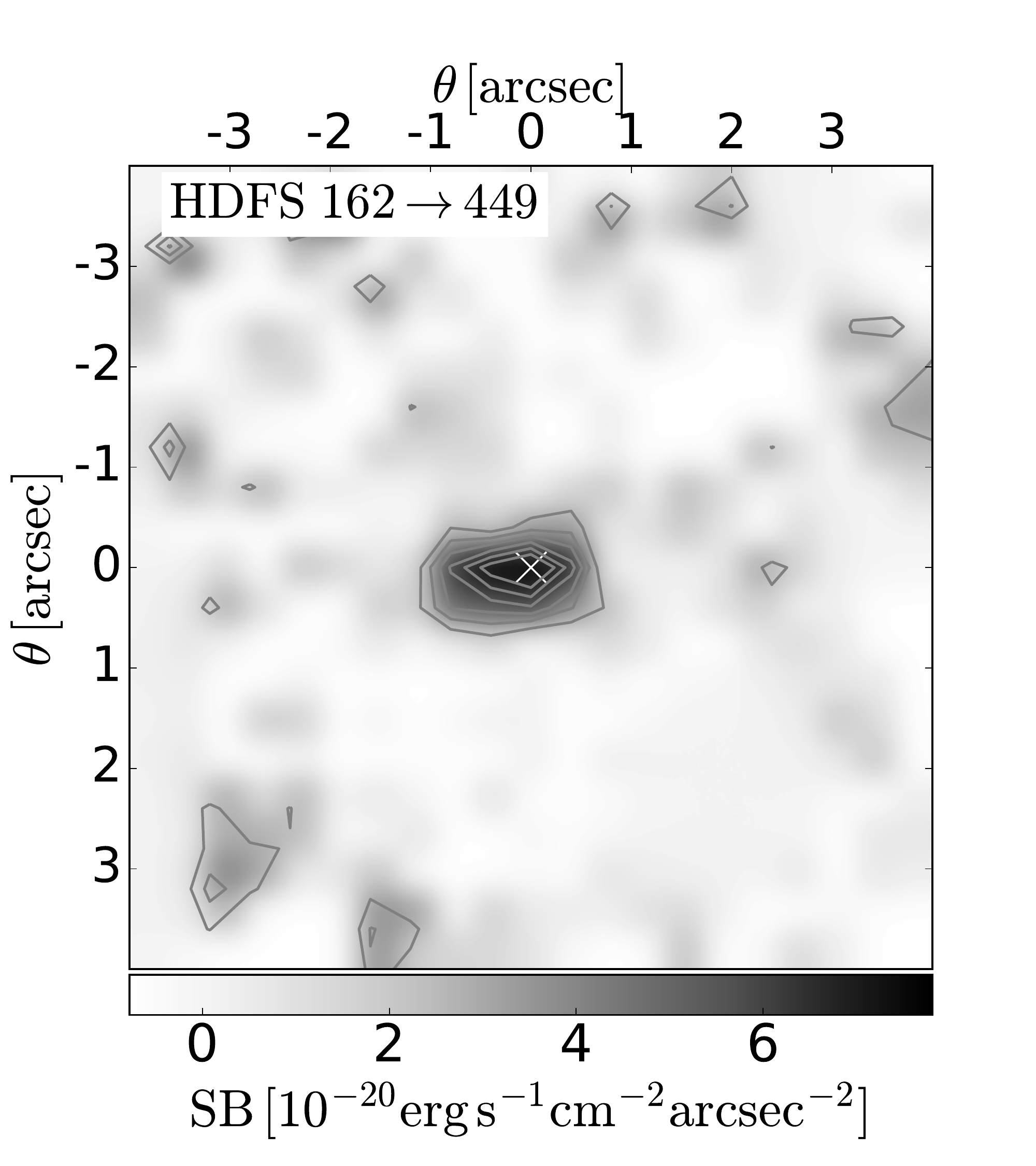}
\includegraphics[trim={2.4cm 3.8cm 1.98cm 0cm},clip, width=1.61in]{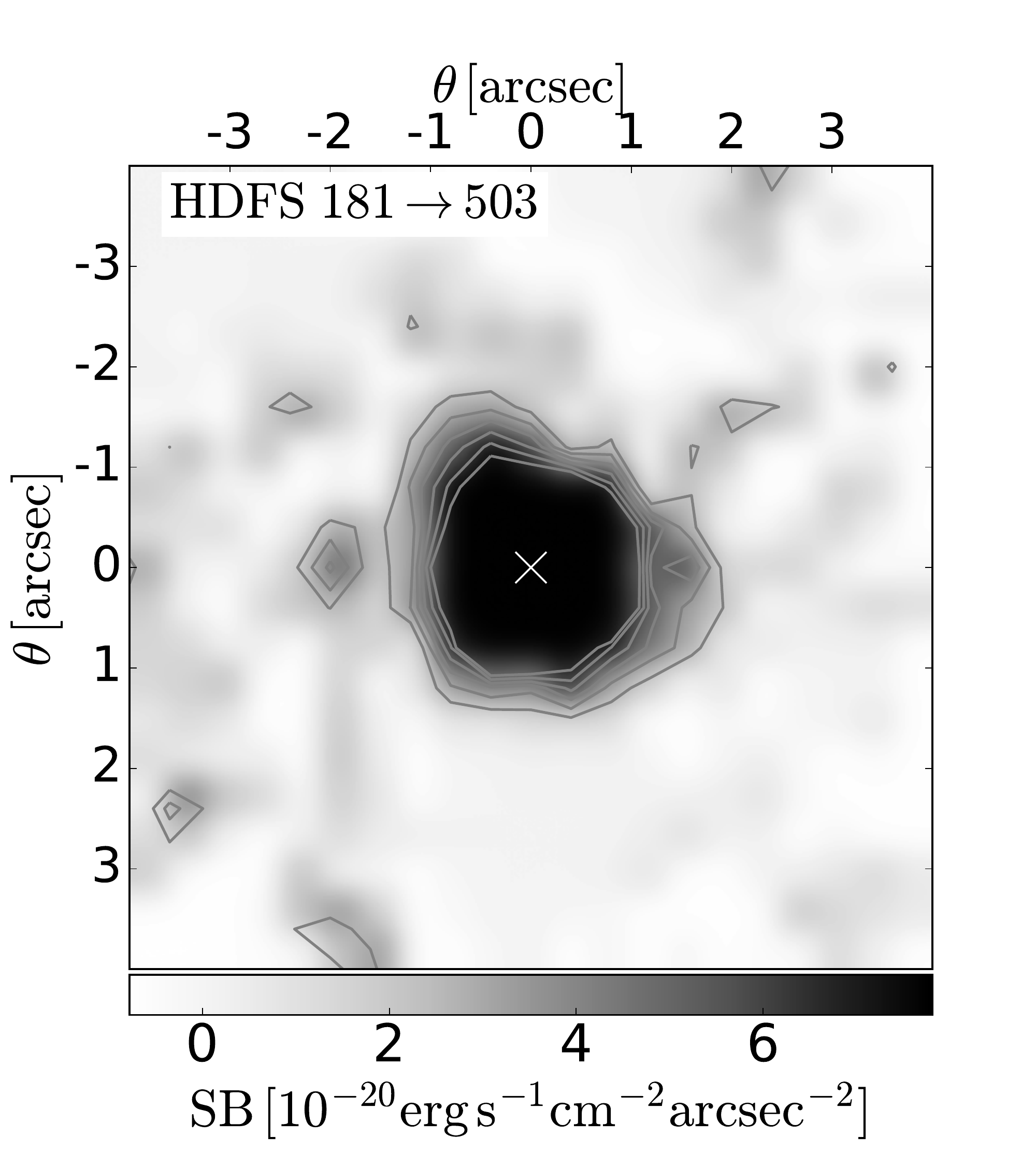}

\includegraphics[trim={.2cm 3.8cm 1.97cm 3cm},clip, width=1.832in]{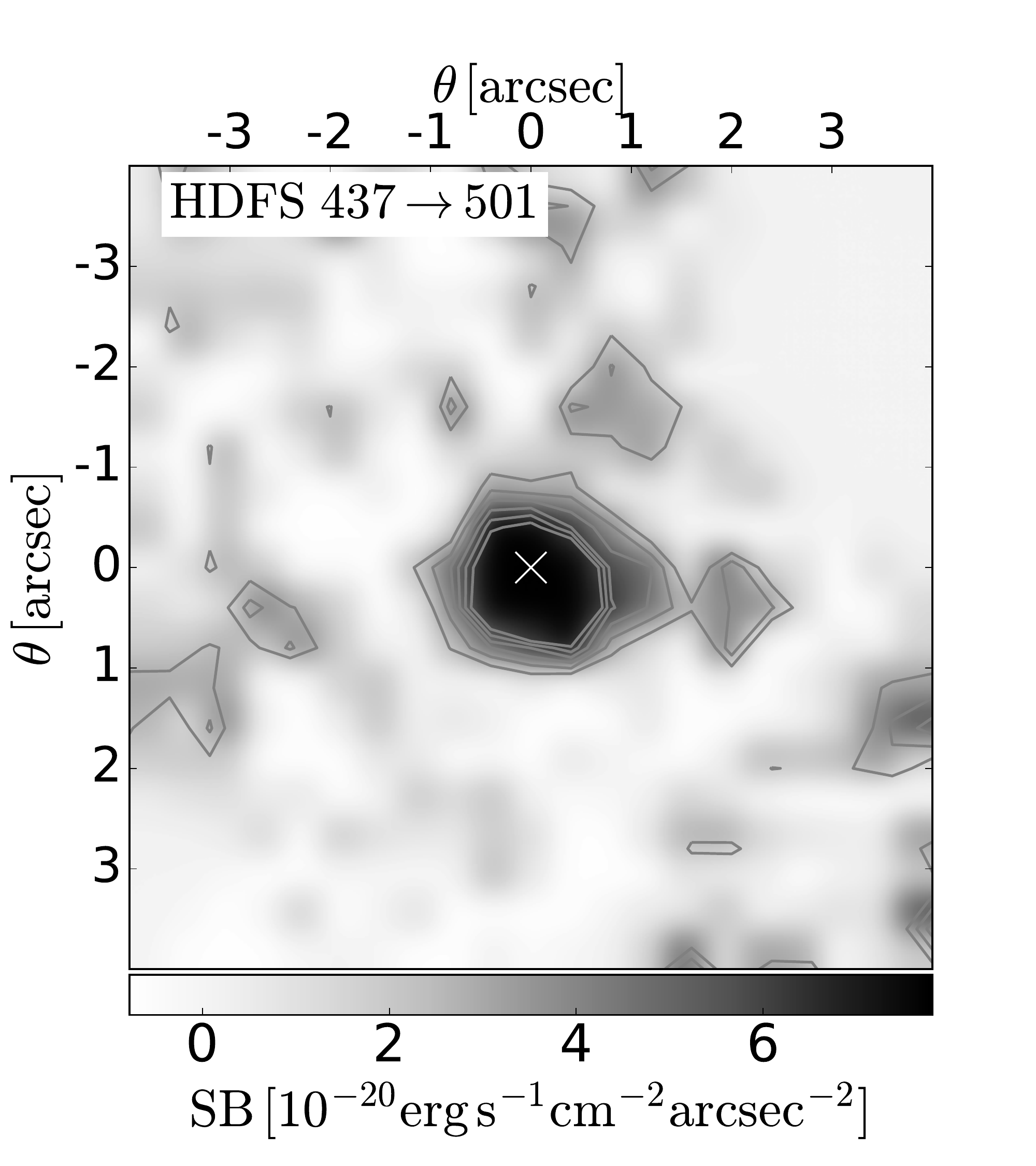}
\includegraphics[trim={2.4cm 3.8cm 1.98cm 3cm},clip, width=1.61in]{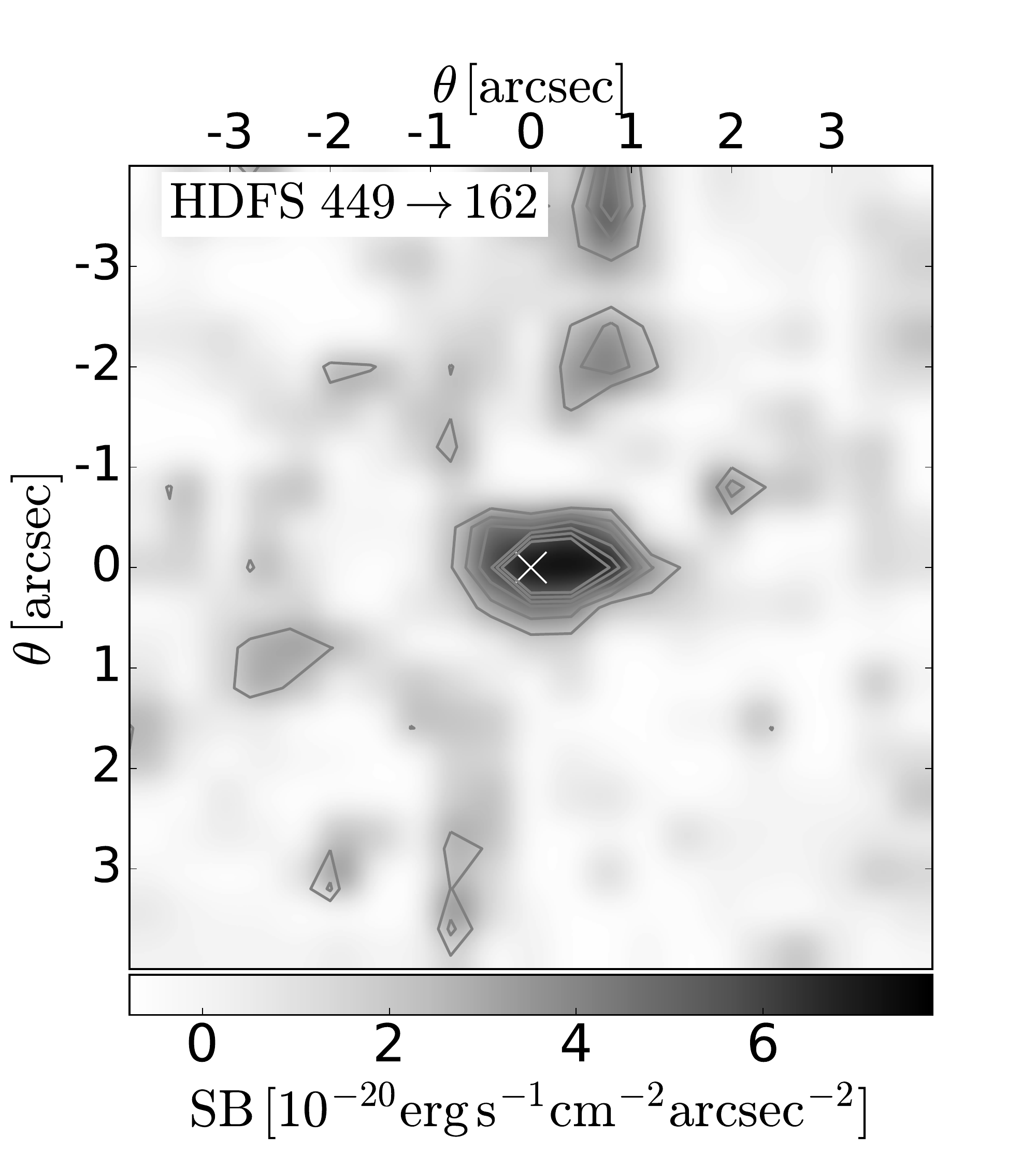}
\includegraphics[trim={2.4cm 3.8cm 1.98cm 3cm},clip, width=1.61in]{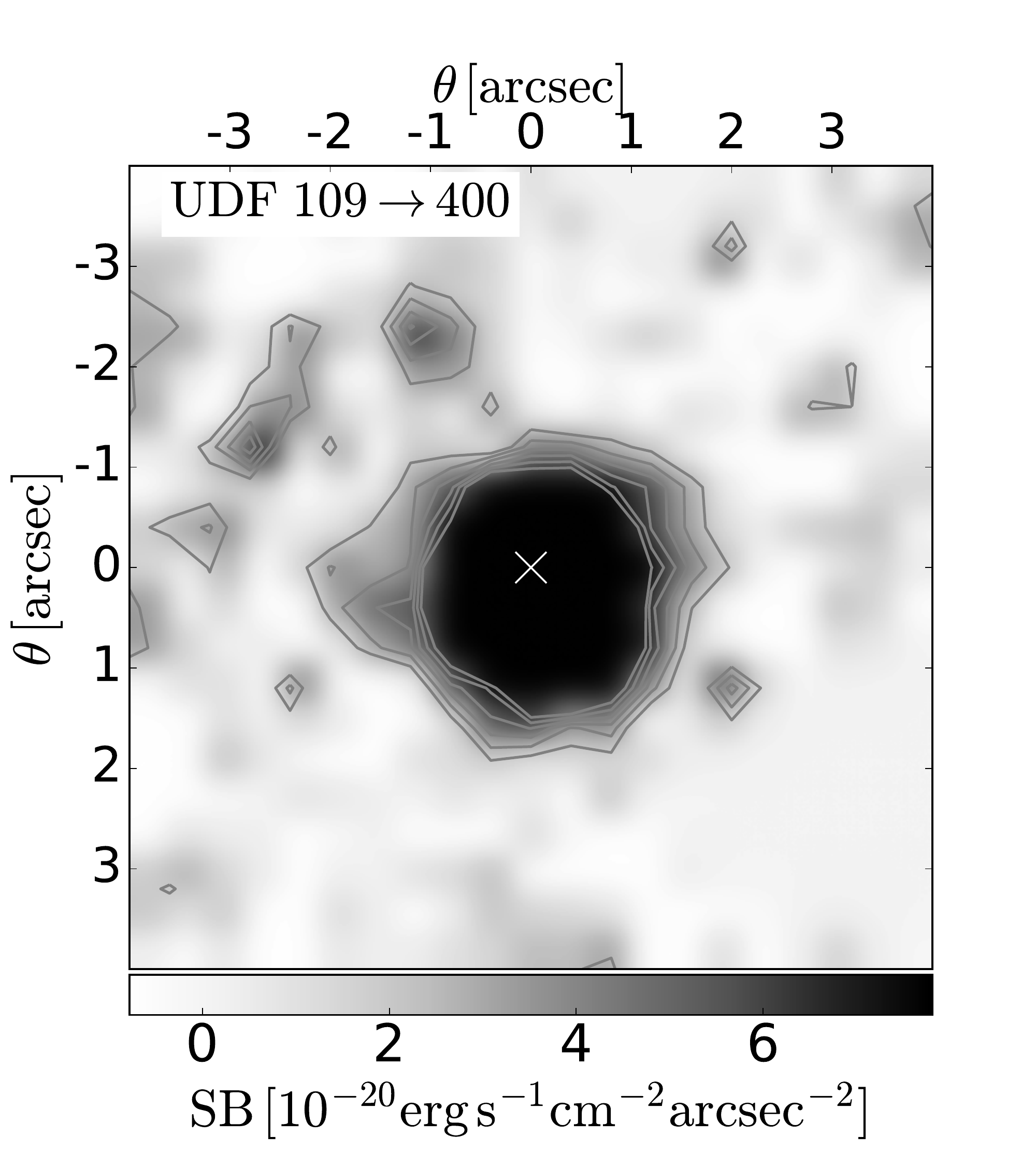}
\includegraphics[trim={2.4cm 3.8cm 1.98cm 3cm},clip, width=1.61in]{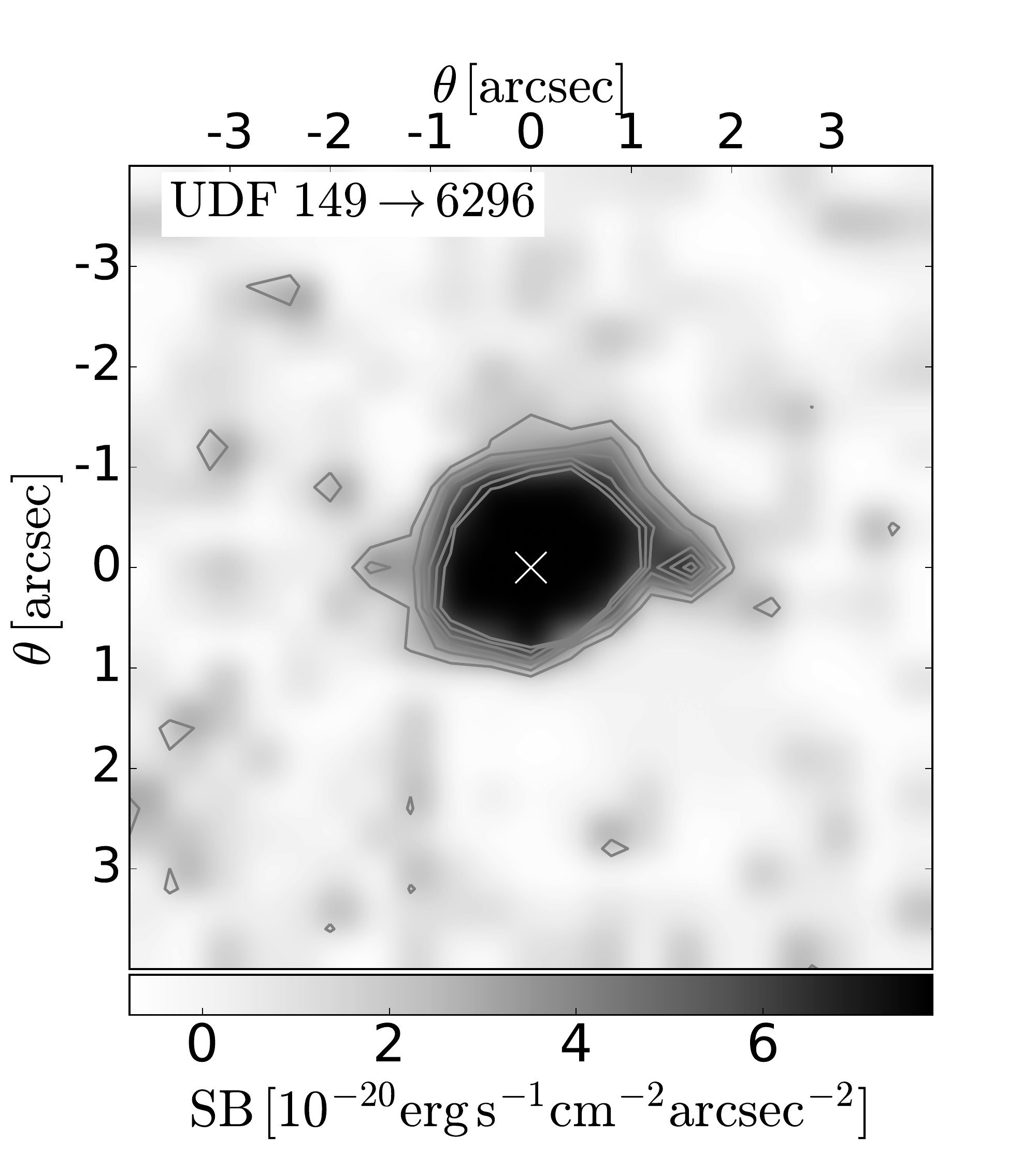}

\includegraphics[trim={.2cm 3.8cm 1.97cm 3cm},clip, width=1.832in]{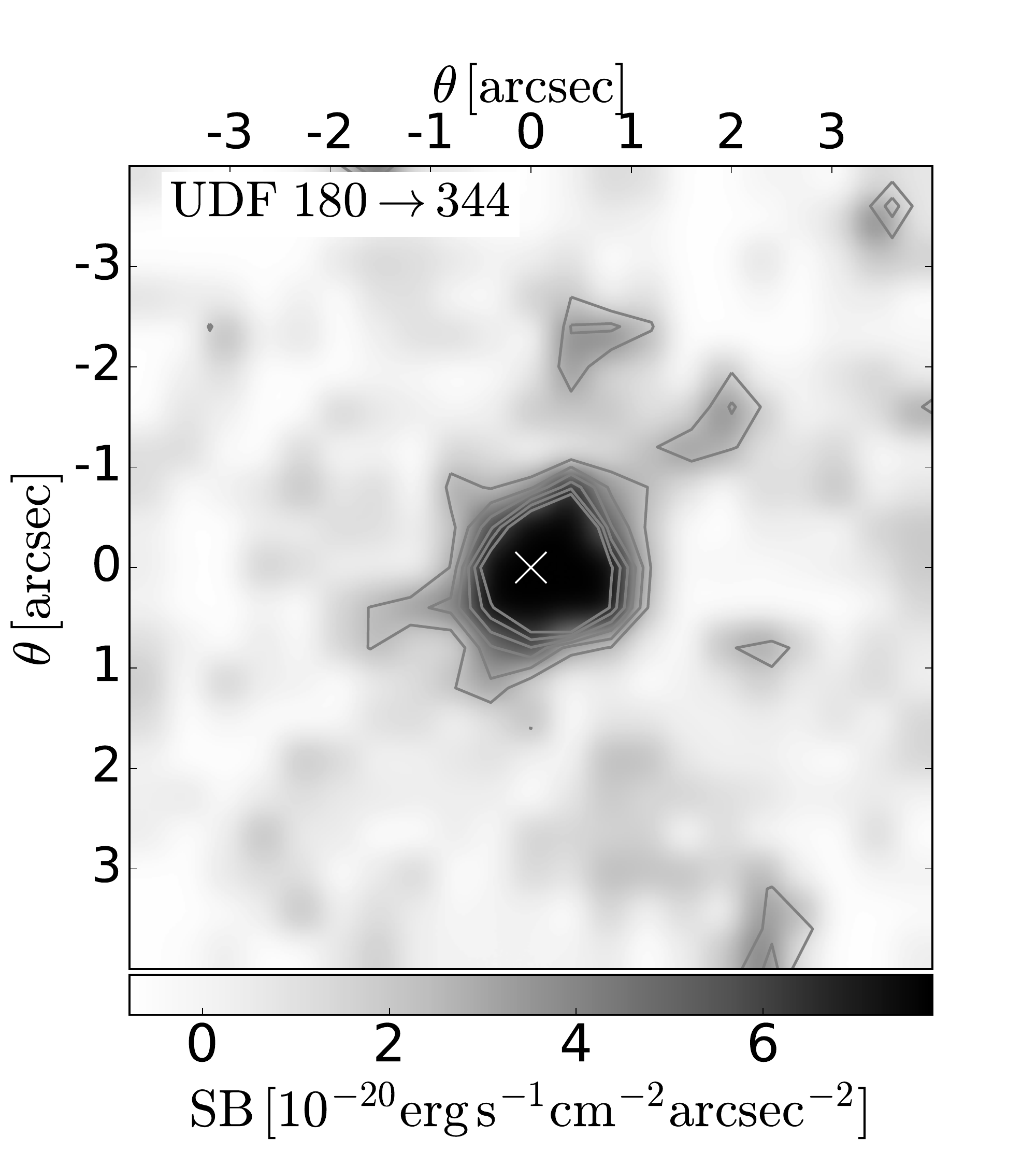}
\includegraphics[trim={2.4cm 3.8cm 1.98cm 3cm},clip, width=1.61in]{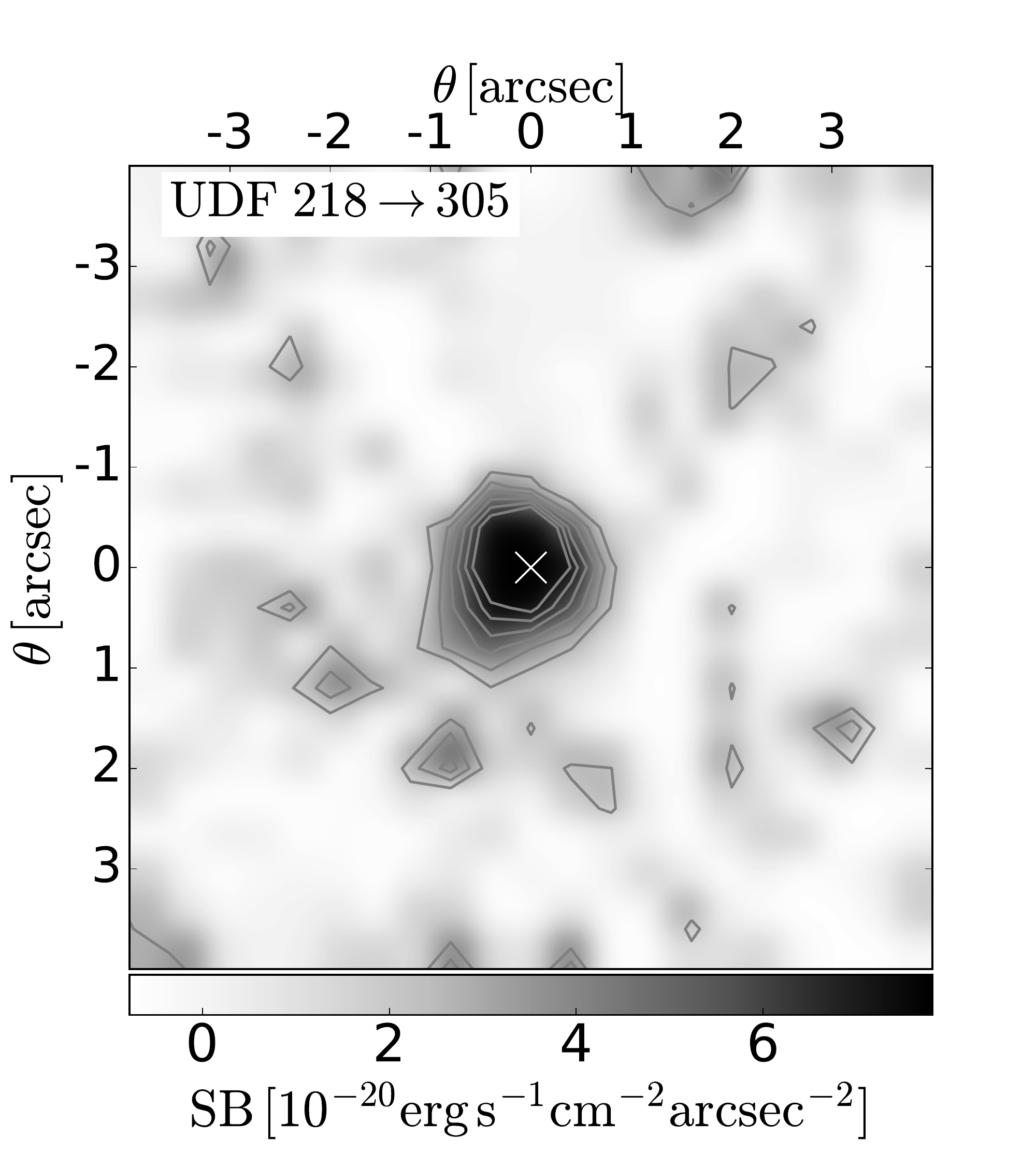}
\includegraphics[trim={2.4cm 3.8cm 1.98cm 3cm},clip, width=1.61in]{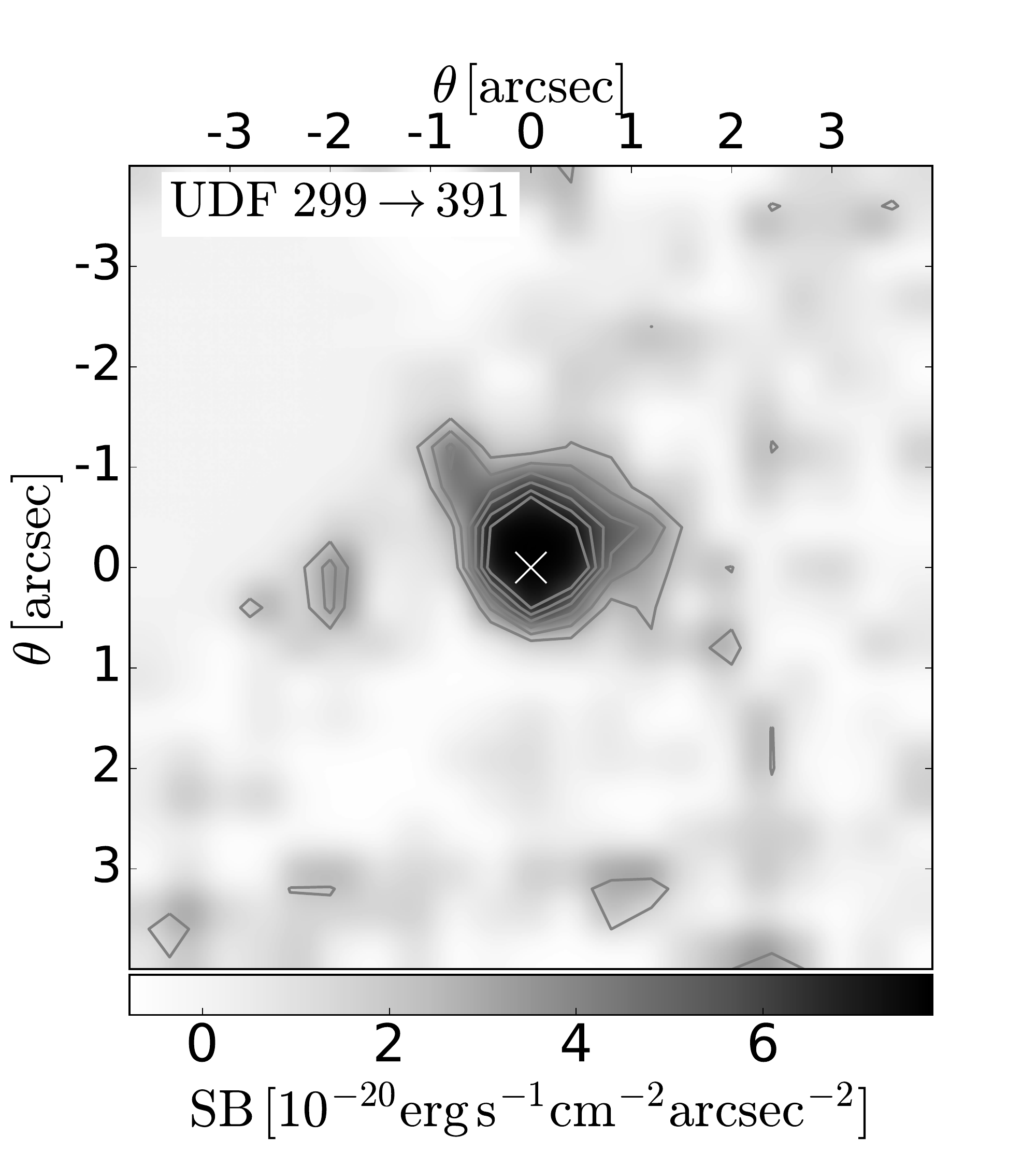}
\includegraphics[trim={2.4cm 3.8cm 1.98cm 3cm},clip, width=1.61in]{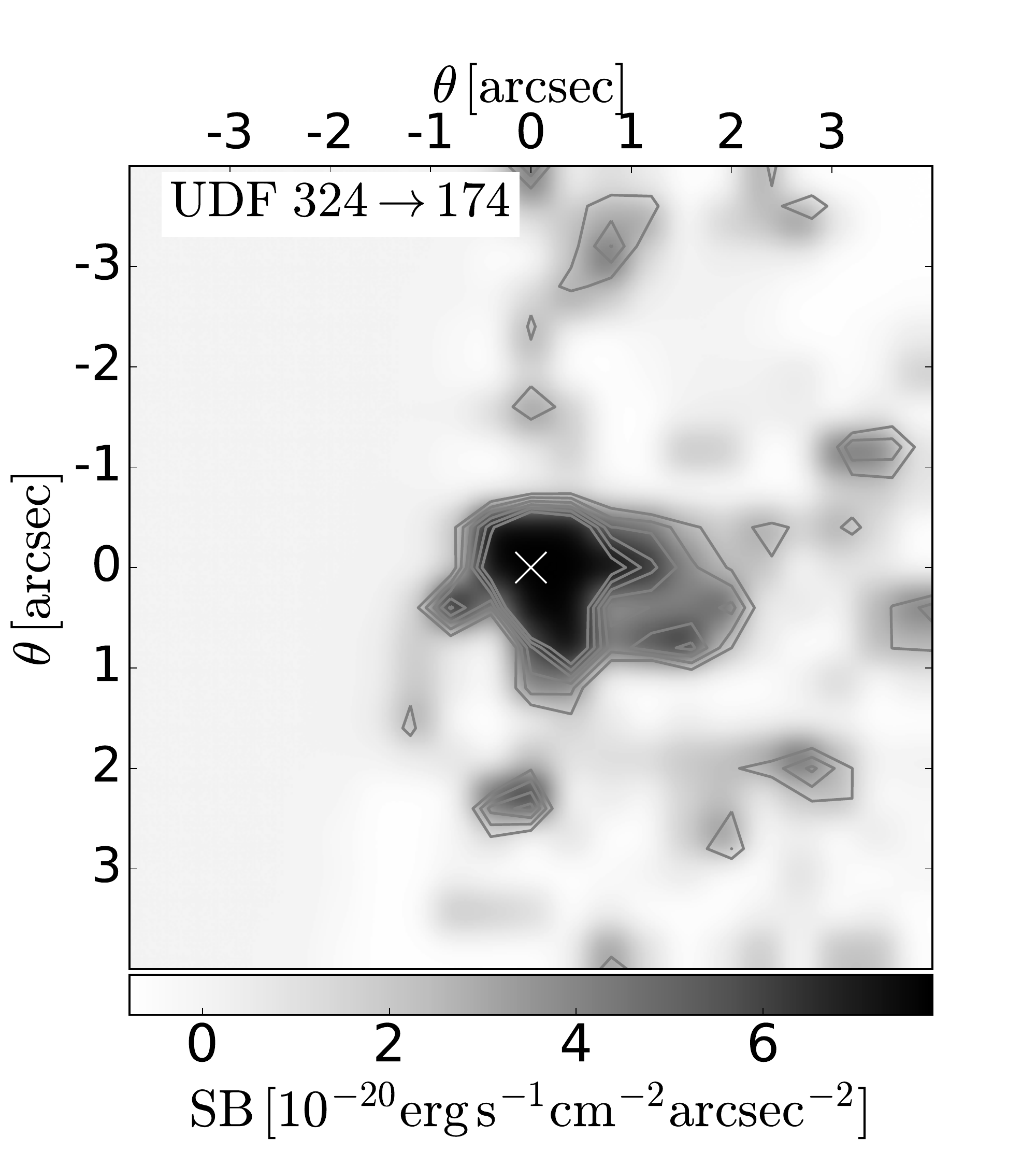}

\includegraphics[trim={.2cm .4cm 1.97cm 3cm},clip, width=1.832in]{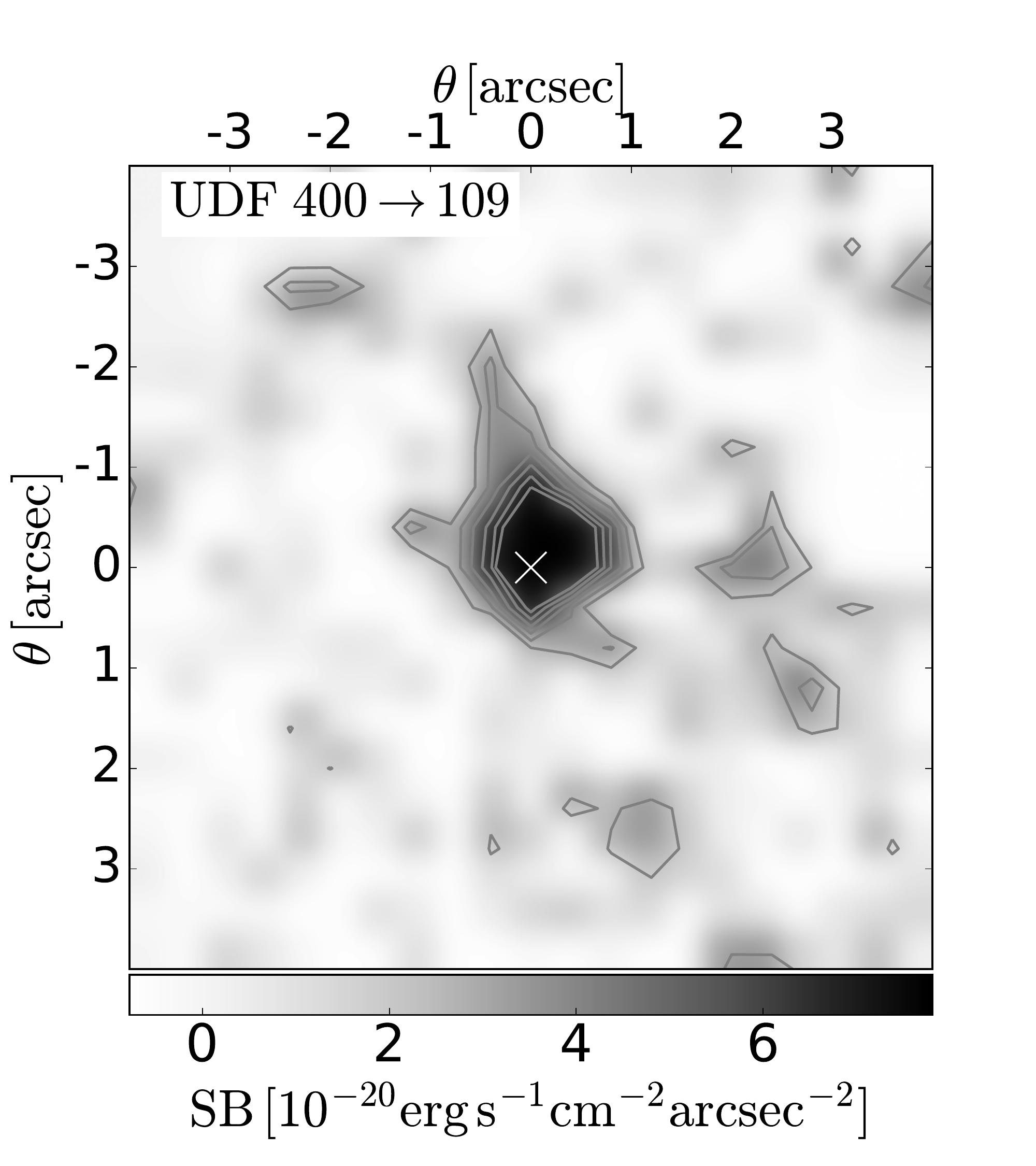}
\includegraphics[trim={2.4cm .4cm 1.98cm 3cm},clip, width=1.61in]{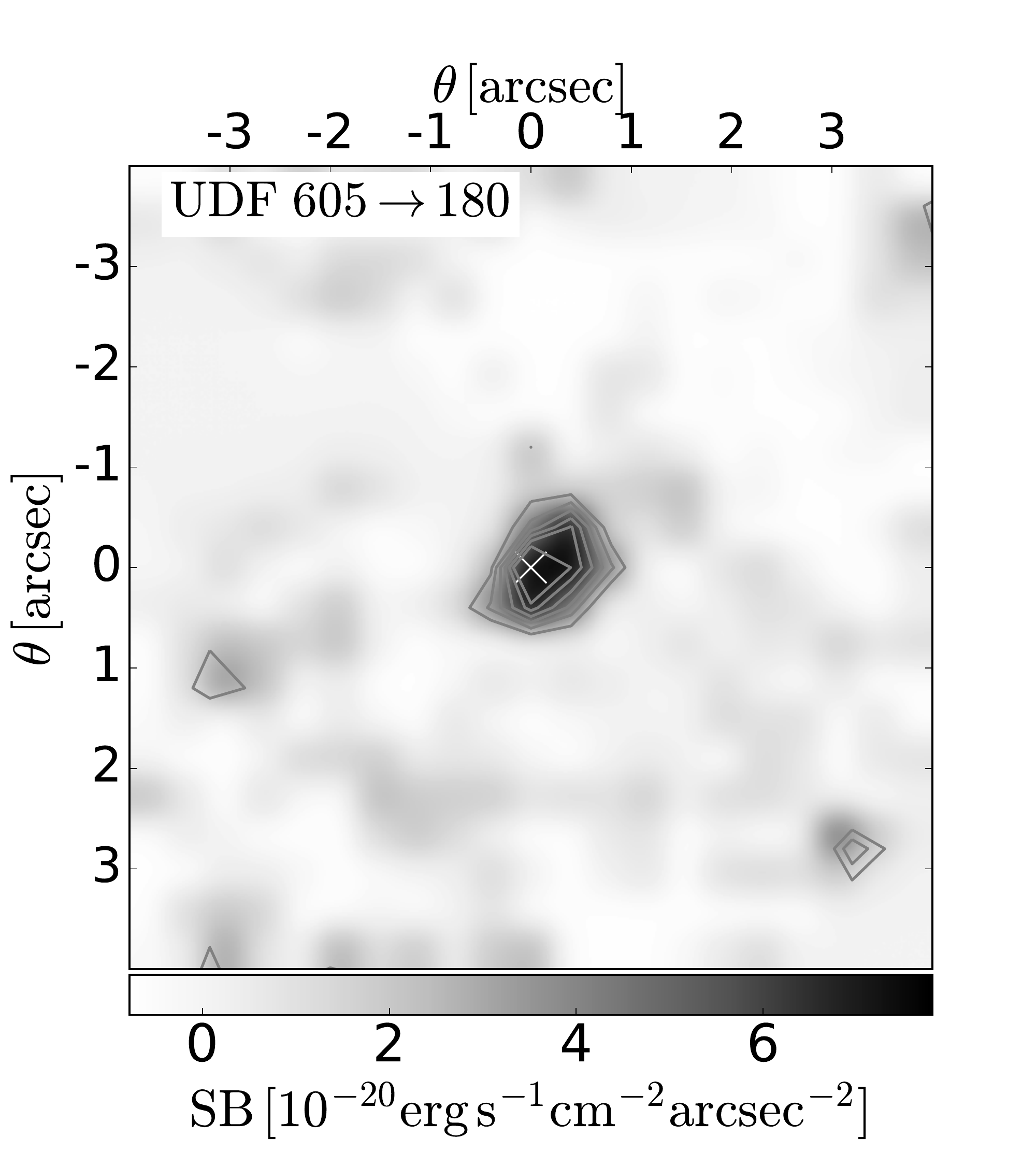}
\includegraphics[trim={2.4cm .4cm 1.98cm 3cm},clip, width=1.61in]{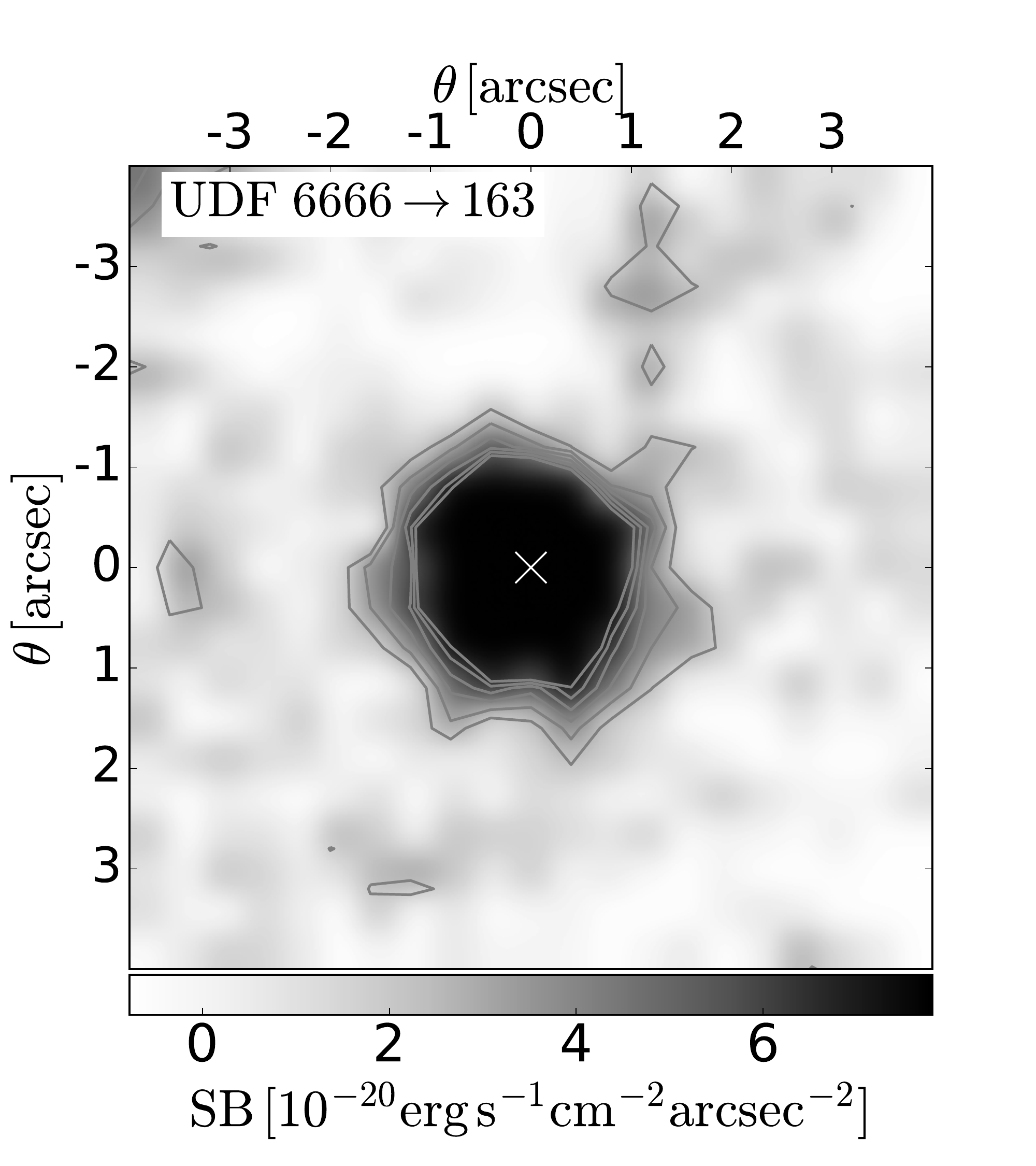}
\includegraphics[trim={2.4cm .4cm 1.98cm 3cm},clip, width=1.61in]{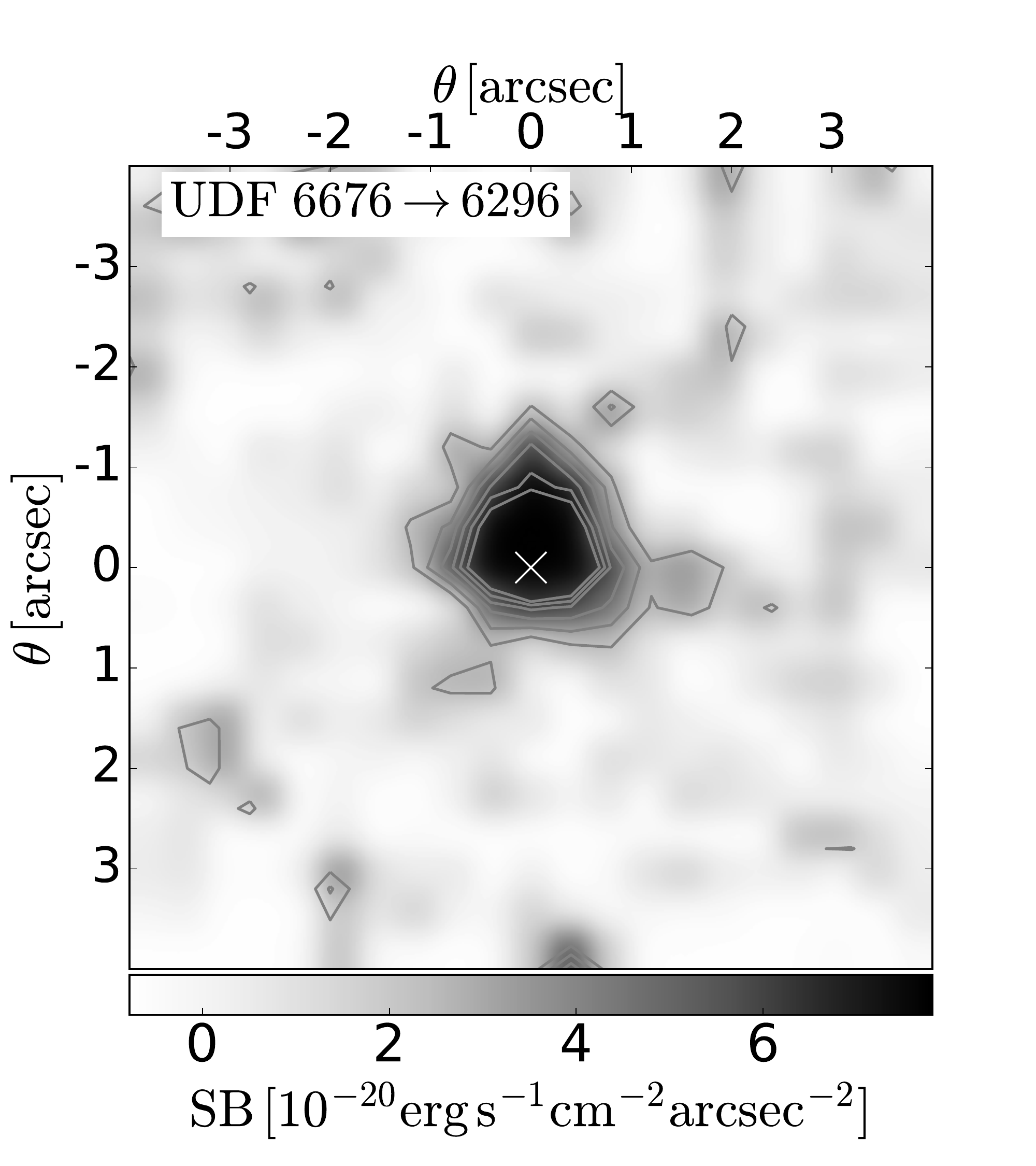}

   \caption{Examples of pseudo NB images with a wavelength-width of $6.25\,\mathrm{\AA}$ obtained from the oriented and resampled subcubes. The label indicates the field and ids of the LAE, at the center of the image, and its neighbour, located along the positive side of the x-axis (outside the image). The cross at the origin of coordinates represents the position of the peak of the Ly$\alpha$ emission.  
   The images are smoothed with a Gaussian filter of $\sigma=1$ pixel ($0.4''$) to improve visualization. Contour levels range from 2 to 6 times the noise levels of the smoothed image.
   }
   \label{LAEs}
\end{center}
\end{figure*}

With this approach, LAEs close to the border will not have available data on some part of the subcube outside of our region of interest (i.e. the transformed voxels on the positive part of the x-axis). 
In this case, we fill the missing values with NaNs. 

The new $x^*$ and $y^*$ coordinates are resampled with a bin size twice as large as the original one to avoid empty voxels, whereas the wavelength coordinate $z^*$ is preserved. 
The spatial sampling of the transformed subcubes is therefore $0.4''\times0.4''$.
The voxel value is assigned to the nearest new voxel to conserve the flux with respect to the initial cube. 

As a consequence of the resampling method plus the rotations involved in the coordinate transformation, the number of the original voxels contributing to each of the transformed voxels will not be completely uniform, and therefore we expect that the noise will not be uniform across
our subcubes. However, this should have a minimal effect in the propagated noise because subcube orientations are largely independent of each other. 

In Figure \ref{LAEs}, we show a few examples of oriented individual subcubes where the neighbour (which is outside the image) is always located along the positive side of the x-axis at distances larger than $16''$. 

Finally, we stack all our subcubes applying an averaged-sigma-clipping algorithm with a single iteration discarding values above and below $\pm3\sigma$, where $\sigma$ is calculated for each voxel.

\section{Results}

\begin{figure*}
\begin{center}
\includegraphics[trim={.2cm 1.5cm 2.01cm 3cm},clip, width=3.696in]{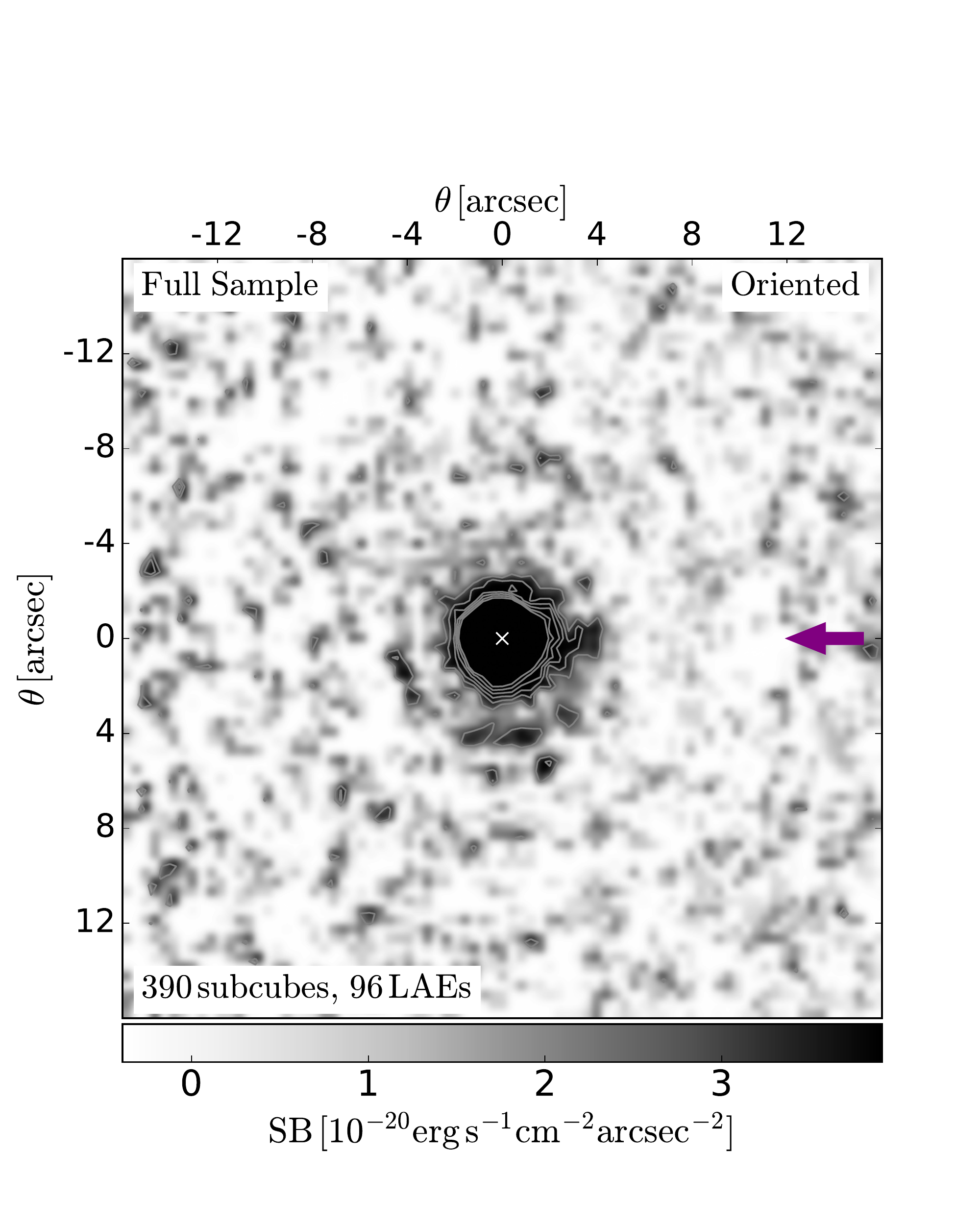}
\includegraphics[trim={2.5cm 1.5cm 2.0cm 3cm},clip, width=3.23in]{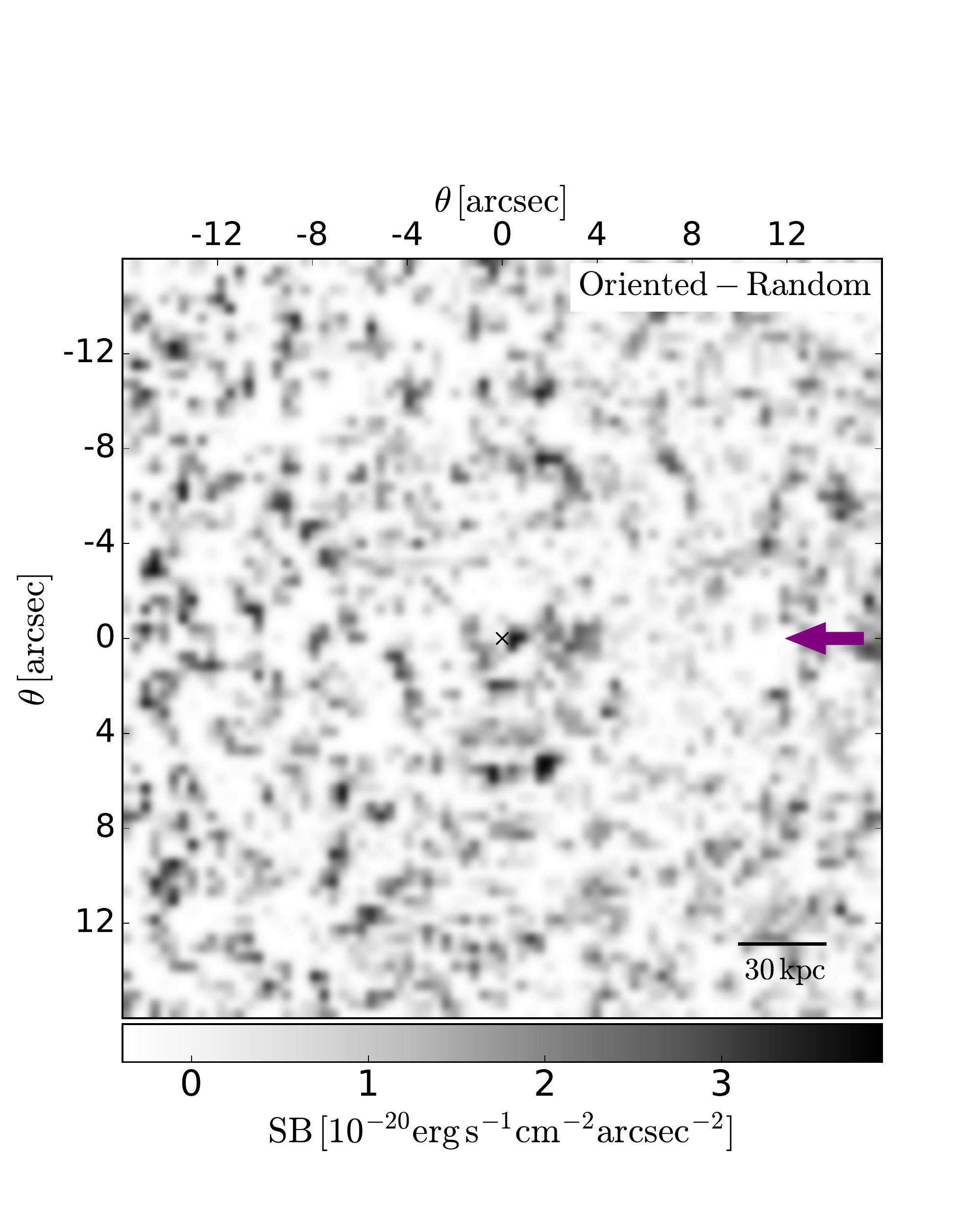}
   \caption{Left panel: pseudo NB image of the oriented stack using the full sample of subcubes. As in Figure \ref{LAEs}, the wavelength width of the images corresponds to  $6.25\,\mathrm{\AA}$. The subcubes have been oriented in such a way that the galaxy 
   neighbours are always positioned on the positive part of the x-axis at distances larger than $16''$. The purple arrow indicates the expected position of filaments connecting neighbouring galaxies. 
  Notice that the noise is not uniform and that it is higher on the negative part of the x-axis because of edge effects (since neighbours are always inside the cube, the edges will be on the negative x-axis side).
  Despite this effect, a clear asymmetry towards galaxy neighbours is present in the light distribution around the central galaxy. 
  Right panel: pseudo NB image after subtracting a combination of randomly oriented stacks (``super-random" stack, see text for details) from the oriented one presented in the left panel. 
  As expected, systematic effects (e.g., the ring-like structures present in the oriented stack) are significantly reduced in this image. However, the asymmetry in the central emission 
  towards the neighbouring galaxies remains. 
  }
   \label{stack_full}
\end{center}
\end{figure*}
In left panel of Figure \ref{stack_full}, we present the pseudo NB image of the oriented stack-cube using our full sample of subcubes. This image has been obtained by collapsing the stack-cube along 5 layers in the $z$-direction, i.e. $6.25\,\mathrm{\AA}$, centred on the peak of the galaxy Ly$\alpha$ emission.

The wavelength width of the pseudo NB has been chosen to maximise the expected Signal to Noise Ratio (SNR) taking into account the possible width of the intergalactic Ly$\alpha$ emission \citep[e.g.,][]{Cantalupo2005} and wavelength shifts with respect to the LAE peak.
We have experimented different NB wavelength widths and found that using 5 layers gives the best results both in terms of noise and detectability of Ly$\alpha$ emission as we will show in this section.

Clearly, there are no indications of significant emission at distances larger than $4''$ from the center at the predicted position, i.e. the expected location of emitting filaments with respect to the central, LAE emission (indicated by the purple arrow). 
A closer look at the central part of the stack shows the presence of ring-like emission features and slight asymmetric emission distribution in the direction of the neighbouring galaxies
(up to a scale of about $4''$).
The most prominent of the ring-like features is at a distance of $4''$ from the center. 
In order to understand if these features are due to systematics in our stacking procedure we produced a set of 200 new stacks using the same sample of 390 subcubes obtained with \emph{random} orientations. We combined these 200 randomly-oriented stacks into a single ``super-random" stack
in order to boost the systematic effects with respect to Poisson noise. 

In the right panel of Figure \ref{stack_full}, we show the resulting pseudo NB image after subtracting this ``super-random" stack from the oriented one. We notice that
the ring-like features present on the oriented stack are mostly suppressed suggesting a non-physical nature of this emission. Because a single LAE can be repeated in the stack several times at different orientations,
any non-circularly-symmetric emission can indeed appear as a ring-like feature in the final cube (notice that a single asymmetric object repeated at an infinite number of random orientations will create perfect rings). 
However, we notice that the asymmetry in the emission towards the neighbouring galaxies in the light distribution remains. 

\begin{figure*}
\centering Full Sample
\begin{center}
\includegraphics[trim={2cm 0cm 3.2cm 0cm},clip,width=3.4in]{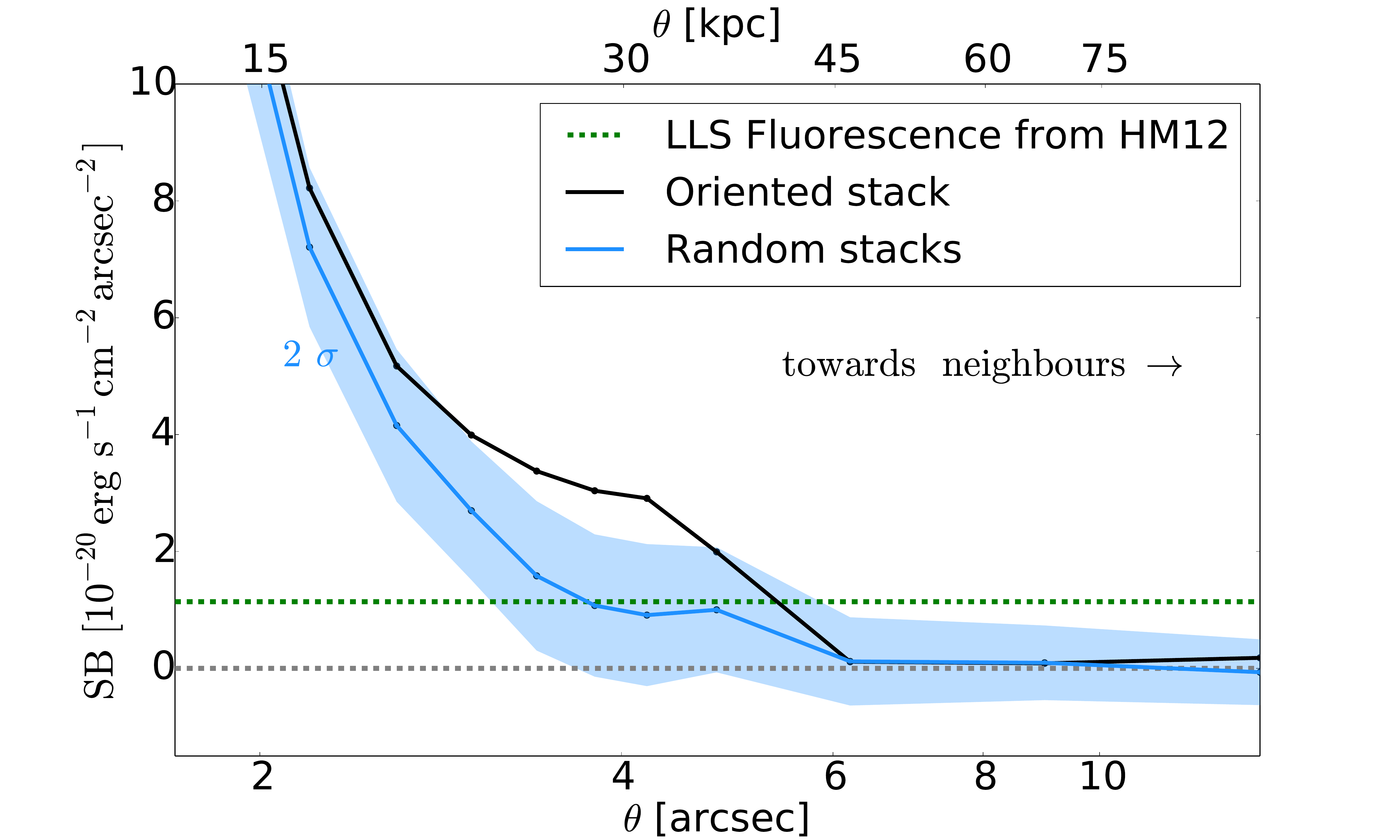}
\includegraphics[trim={2cm 0cm 3.2cm 0cm},clip,width=3.4in]{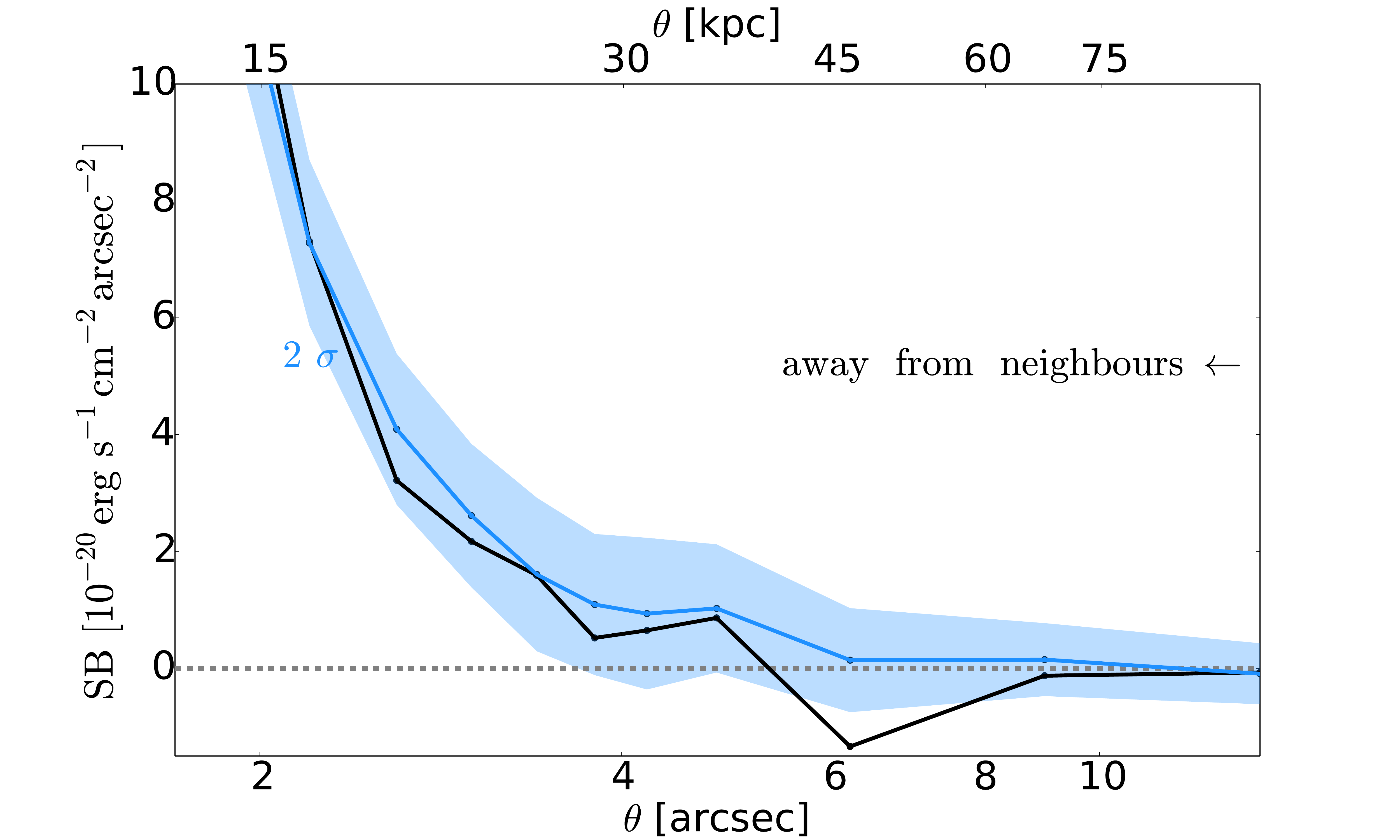}
   \caption{Left panel: Surface brightness profile of the oriented (black line) and ``super-random" (blue line) full-sample stacks obtained using apertures with increasing sizes (from $0.4''\times 2''$ at the center to $2''\times 2''$ at $\theta$ larger than $10''$).
   The oriented SB profile  is calculated starting from the center towards the neighbouring LAEs. The light blue shadowed areas represent the two sigma deviations from the average of the random stacks. The green dotted line indicates the expected
   Ly$\alpha$ fluorescent SB from LLSs illuminated by the Haardt \& Madau (2012) UVB.  A clear excess of emission in the oriented stack is present up to scales of about $4''$ (30 projected kpc) from the central galaxies compared to the randomly-oriented stack.
   No significant emission is detected at scales larger than $5''$ up to 2$\sigma$ levels that are well below the expected UVB fluorescence values. 
   Right panel: SB profile in the direction opposite to the neighbours. Similarly to other directions (except the one towards neighbours), there is no significant emission excess at any scales with respect to the randomly-oriented stack.
   }
   \label{SB_full}
\end{center}
\end{figure*}

In order to assess the significance of this asymmetry we examine the surface brightness (SB) profile integrated over a spatial aperture of vertical height of $2''$ and increasing horizontal widths (from $0.4''$ to $2''$) for both the oriented and ``super-random" stacks. 
In the left panel of Figure \ref{SB_full}, we show as a black line the SB profile obtained for the oriented stack along the positive side of the x-axis (right direction) and as a blue line the same profile obtained for the ``super-random". The shaded area represents
the 2$\sigma$ standard deviation of the average of the ``random" stacks. 
Notice that this value is well below the expected fluorescence from UVB (green line in Figure \ref{SB_full}) and therefore gives us constraints on either the value or the UVB or the presence of LLSs in our subcubes.
The integrated 2$\sigma$ limit considering a region of 1 arcsec$^{2}$ area between $6''$ and $12''$ corresponds to $\UVB$\SBcgs, i.e. a factor of about 18
deeper than the individual cubes in the same spatial aperture and wavelength width (see e.g., Bacon et al. 2017, submitted). 
Notice that this is consistent with the expected decrease for non-correlated noise given the amount of subcubes in our stack (i.e., a factor of 19.7). 
Under the extreme and unlikely hypothesis that all our galaxies are connected to each other by LLS filaments the limit given above is 
about a factor of three below the expected fluorescent Ly$\alpha$ SB from the HM12 UVB ($\UVBHM$\SBcgs\ at $z=3.5$, see Section 1).
We will discuss the implications of this result in Section 5.

Focusing again at the closest region around galaxies, we notice that the oriented stack shows excess emission between 3 and 4$\sigma$ with respect to the ``random" orientation up to a scale of about $4''$ (corresponding to about 30 projected kpc at $z\sim3.5$.
The comparison of the SB profile in the opposite direction (right panel of Figure \ref{SB_full}) shows no excess with respect to random orientations, reinforcing the hypothesis of a physical origin for the oriented CGM emission (see Section 5 for discussion).

\begin{table}
\centering
\caption{Median values for the main properties of the full sample of subcubes included in our stack. Because some LAEs are repeated multiple times in our sample of subcubes, our median values are not the same as the ones from the selected sample of LAEs, but instead biased towards LAEs with more neighbours.}
\label{table}
 \begin{tabular}{||c|c||}
  \hline    
  \hline    
    Property    & Median Value \\ 
  \hline    
    Projected distance to the neighbour & $32''$\\ 
    Comoving distance to the neighbour & $8\,\mathrm{Mpc}$\\ 
    Redshift & $3.5$\\ 
    Number of neighbours & $8$ \\ 
    Luminosity & $9.1\times10^{41}\,\mathrm{erg\,s^{-1}}$\\ 
    \hline
    \hline
 \end{tabular}

\end{table}

\begin{figure}
\begin{center}
\includegraphics[trim={2.5cm 4.2cm 2cm 15cm},clip, width=3.3in]{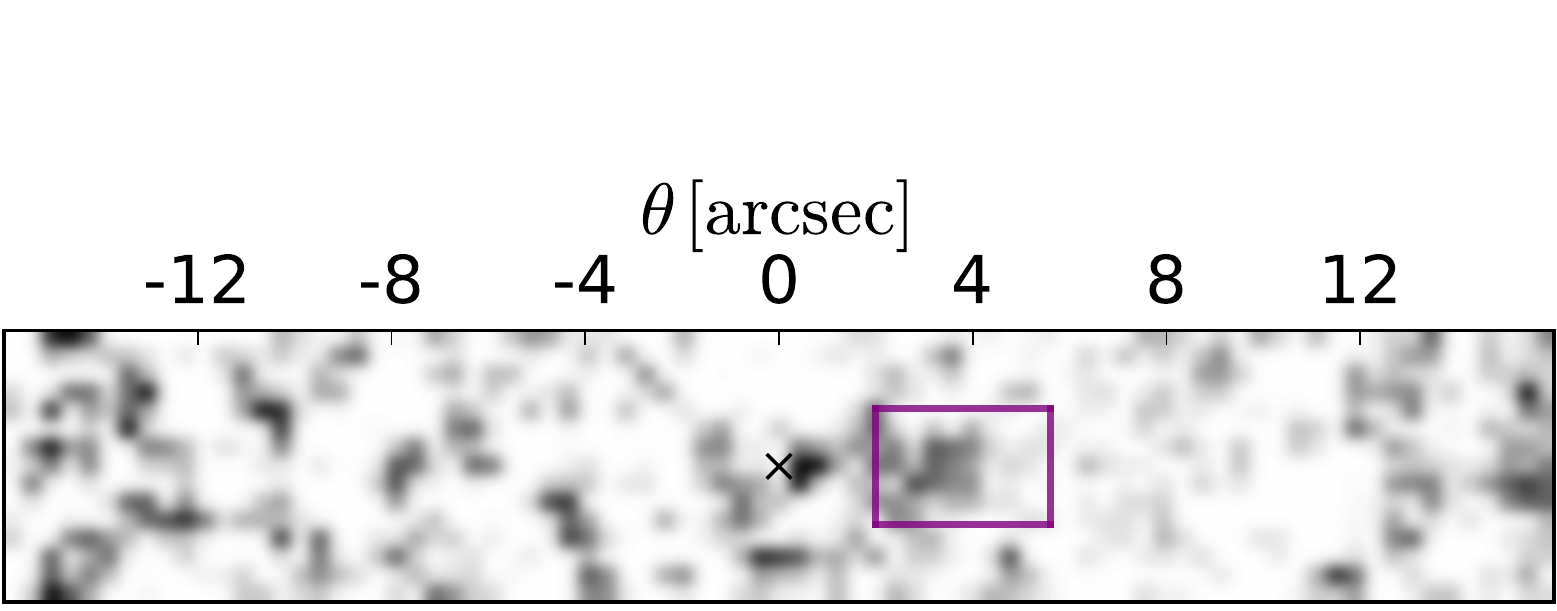}
\includegraphics[trim={0.2cm 0cm .5cm 0cm},clip, width=3.3in]{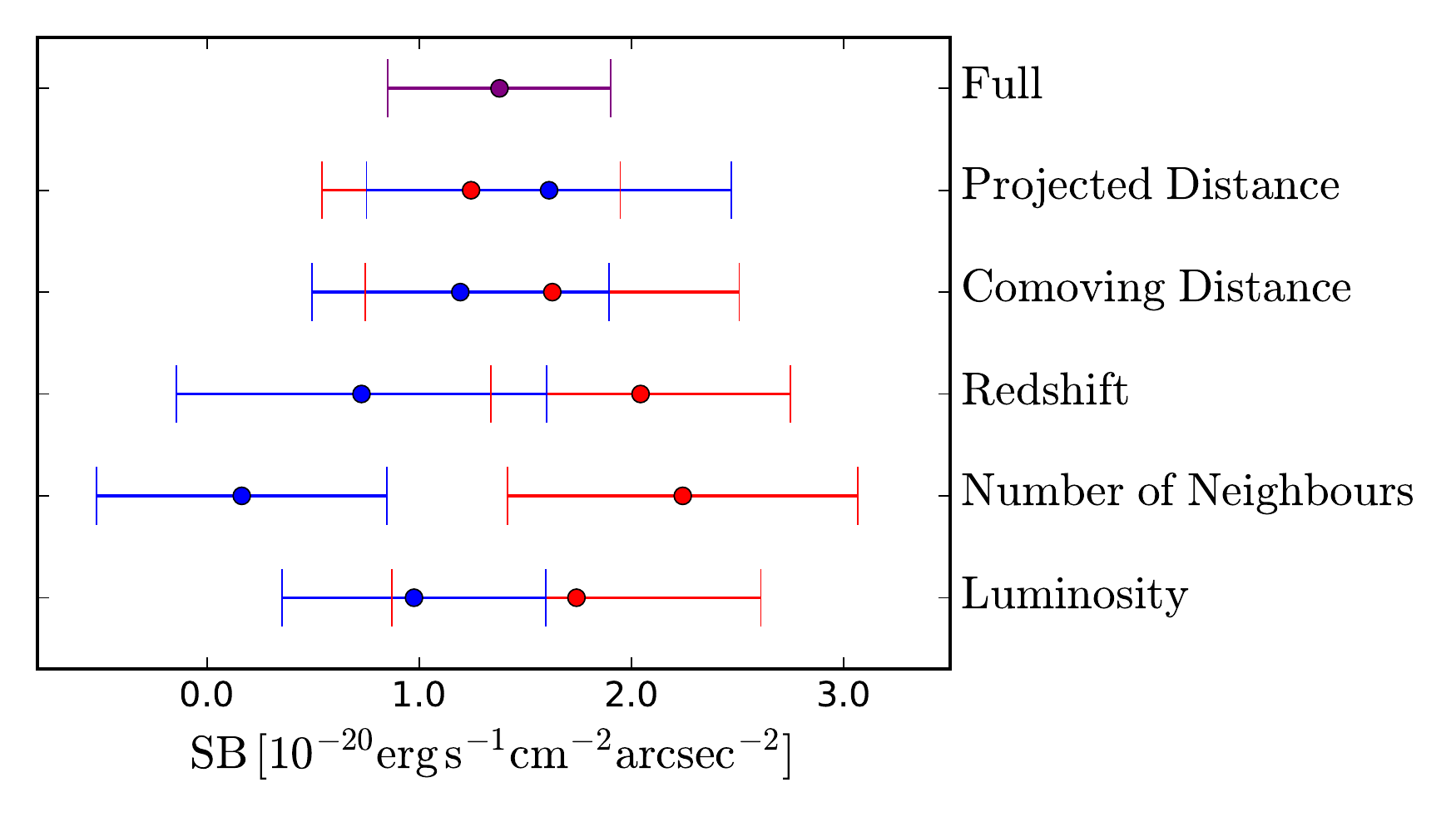}
   \caption{Top panel: image extracted from Figure 4 indicating the region of excess emission (purple square) around galaxies in the oriented stack with respect to the ``super-random" one.
   Bottom panel: SB values within the region indicated above for each subsample of subcubes splitted by the median values of galaxies properties (see text for details) and for the full sample. 
   Among all the properties examined, splitting the sample by the number of neighbours shows the largest variation and, in particular, the subsample with a number of neighbours larger than 8
presents the brightest and most significant signal in the region of interest.  The SB and the 1 sigma error bars are obtained by combining all randomly oriented stacks with a bootstraping of the oriented stack (Figure \ref{stack_full}, left panel), 
therefore it represents the excess of emission without the average CGM contribution. The red and blue dots represent subsamples with parameter values higher and lower than the median, respectively. 
The median values are: projected distance to the neighbour $32''$, comoving distance to the neighbour $8\,\mathrm{Mpc}$, redshift $3.5$, number of neighbours $8$ and luminosity $9.1\times10^{41}\,\mathrm{erg\,s^{-1}}$.
   }
   \label{halfs}
\end{center}
\end{figure}

\begin{figure*}
\begin{center}
\includegraphics[trim={.2cm 4.2cm 2.01cm 3cm},clip, width=3.696in]{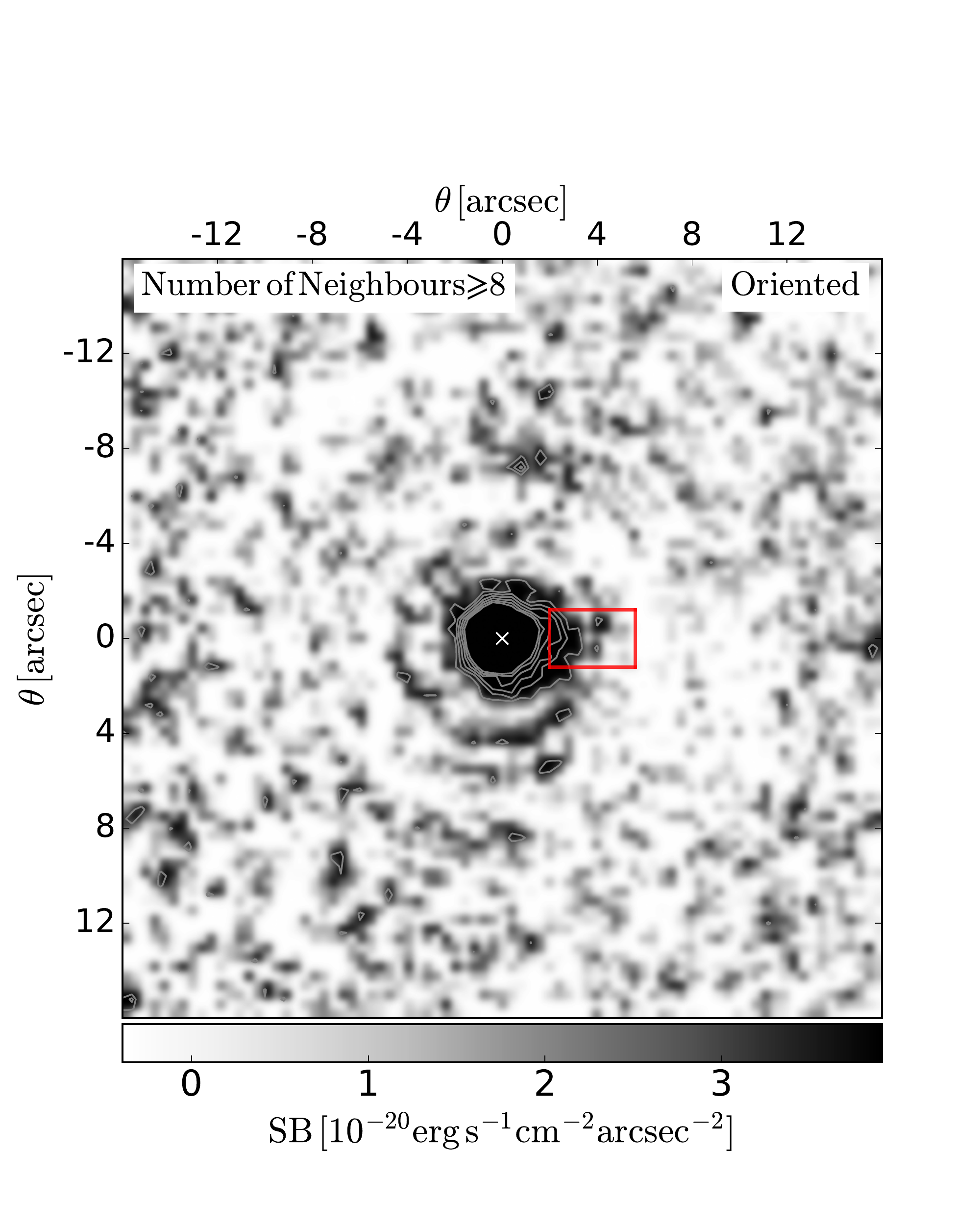}
\includegraphics[trim={2.5cm 4.2cm 2cm 3cm},clip, width=3.23in]{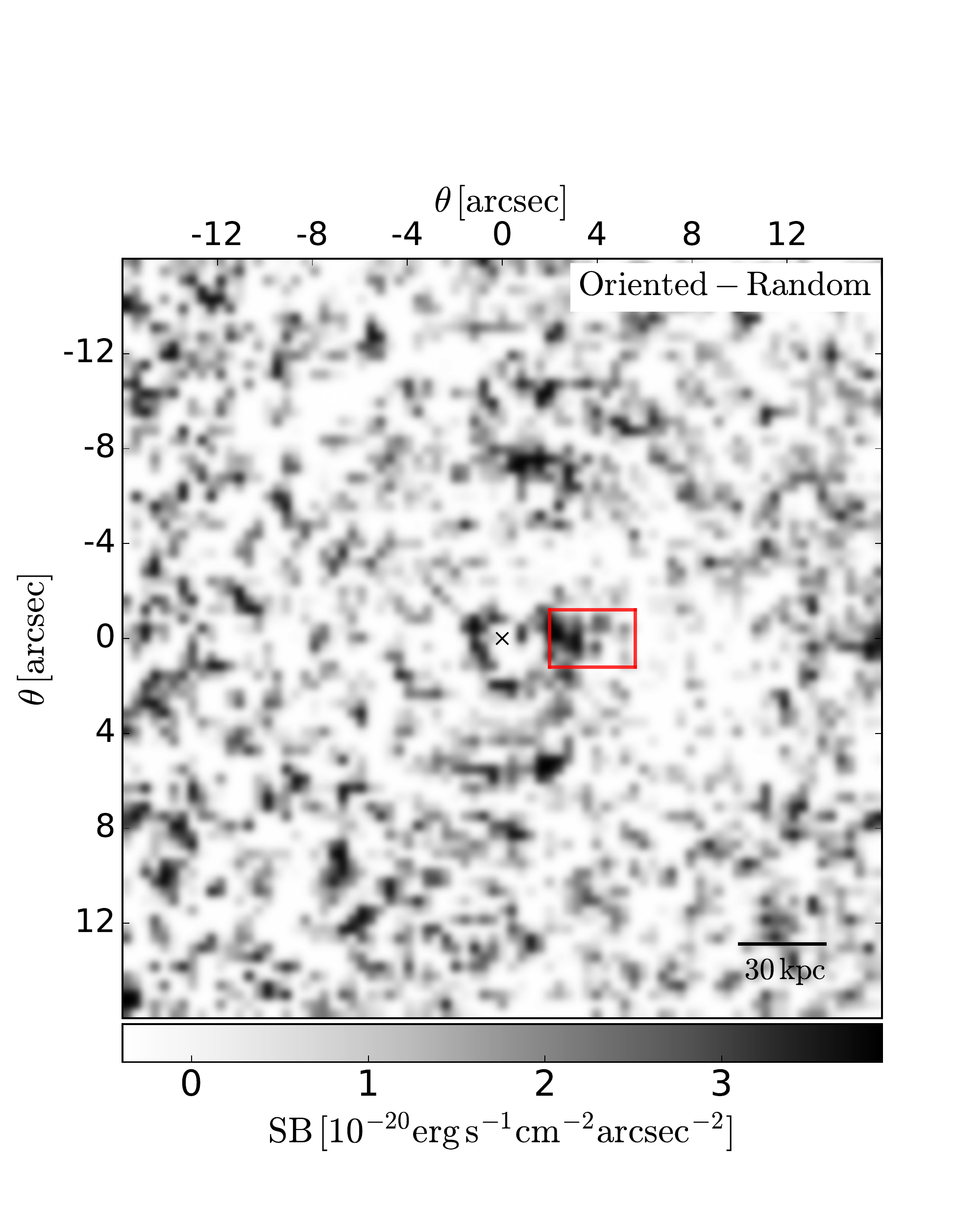}
\includegraphics[trim={.2cm 1.5cm 2.01cm 5.3cm},clip, width=3.696in]{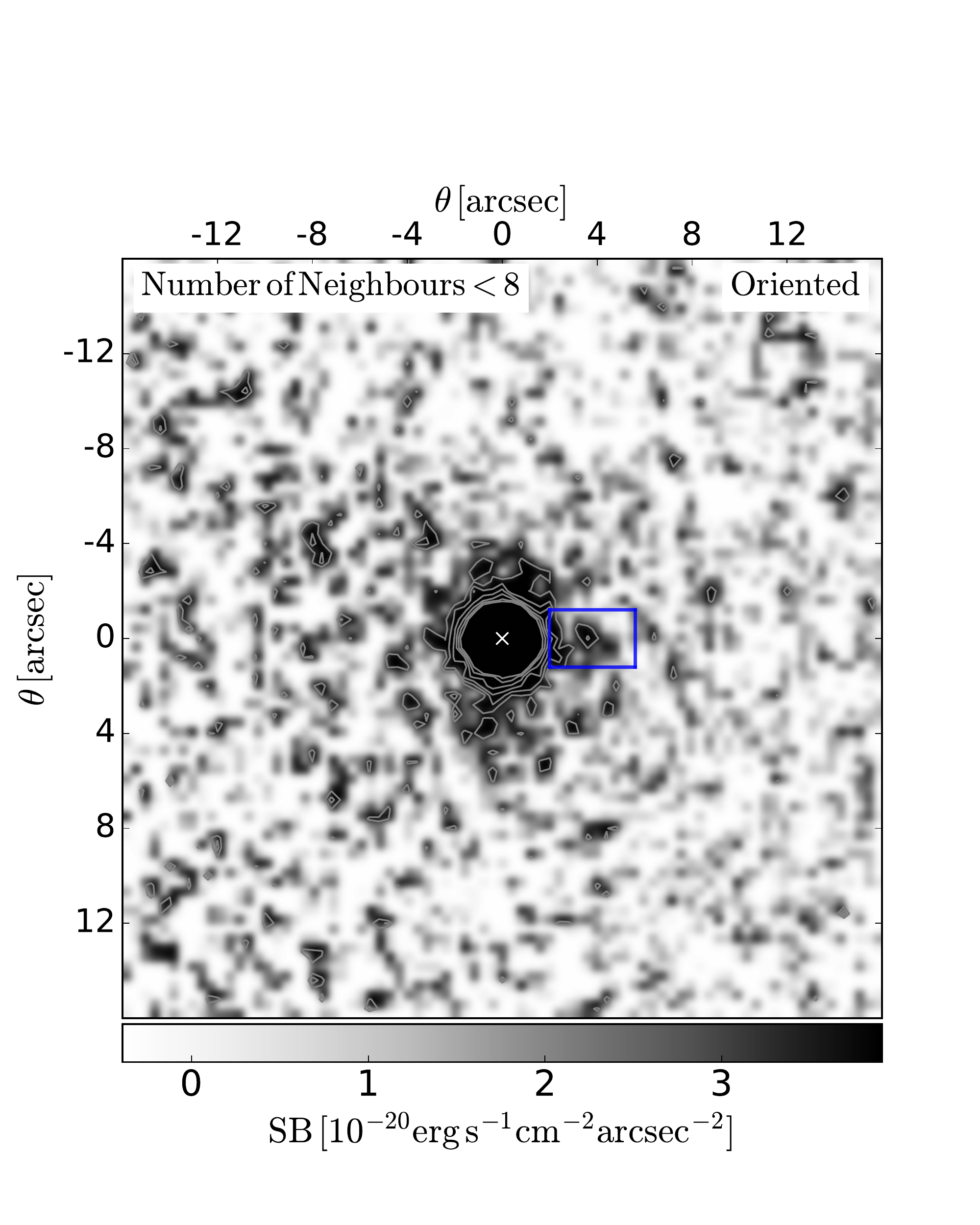}
\includegraphics[trim={2.5cm 1.5cm 2cm 5.3cm},clip, width=3.23in]{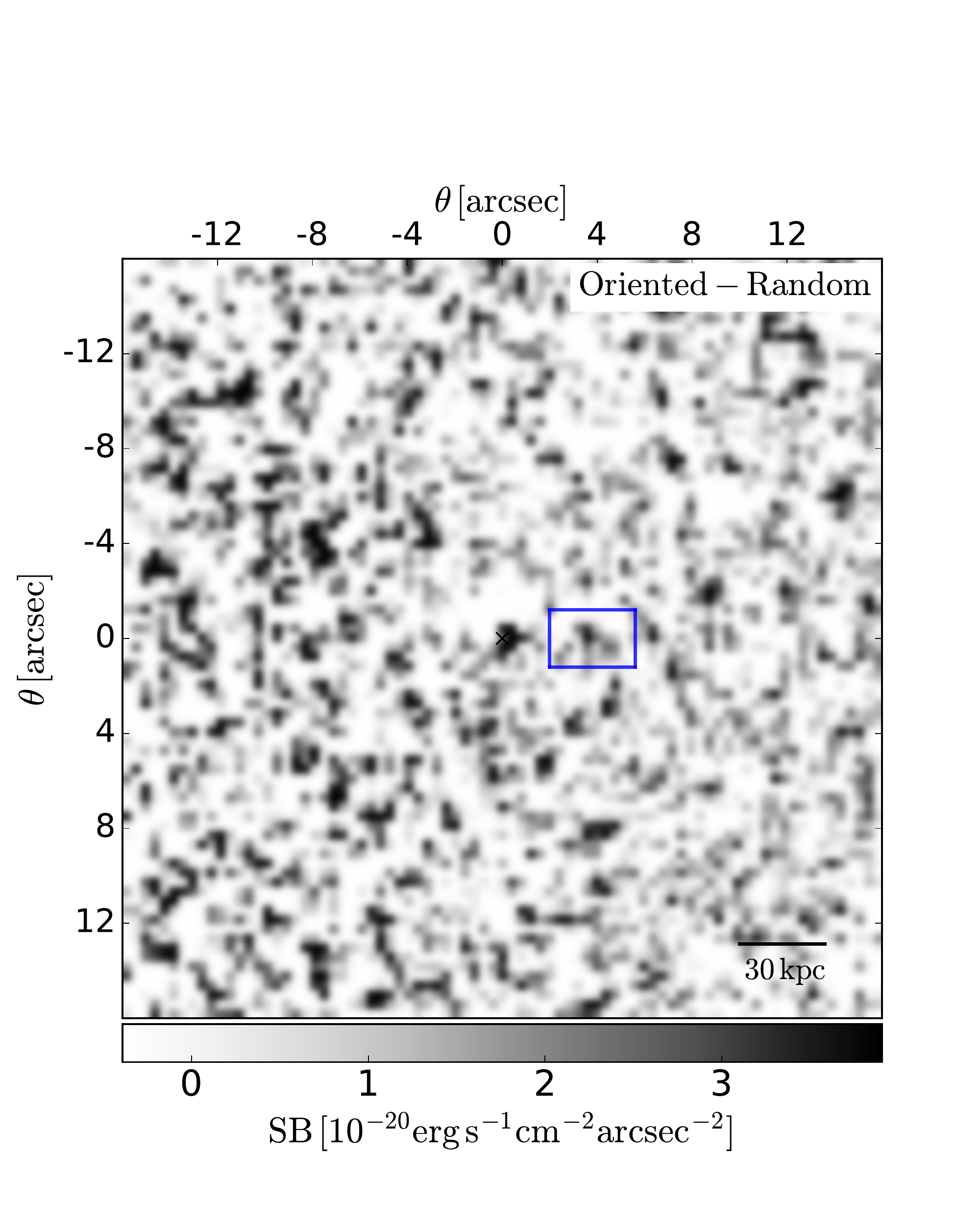} 
   \caption{As in Figure \ref{stack_full} but for the subsamples of subcubes with number of neighbours larger (top-panels) and smaller (bottom-panels) than the median value (8). The red and blue squares indicate the region of excess emission examined in Figure \ref{halfs}.}
  \label{stacks}
\end{center}
\end{figure*}

Motivated by this result, we perform a new set of stacks splitting our sample into halves.
The subcubes for each of the half-samples have been selected by looking at the median of the
following observational properties of the LAEs (see Table 1): 
i) line of sight comoving and projected (ii) distance to the neighbours, iii) redshift, iv) luminosity and, v) number of neighbours per galaxy (within a distance range $0.5\,\mathrm{cMpc}<d<20\,\mathrm{cMpc}$). 
In particular, we group together all the subcubes with values below and above the medians.
In Figure \ref{halfs}, we show the SB in the rectangular region indicated in the top panel (the region with the strongest asymmetric emission) for the full sample and for each of the subsamples. 
Among all the properties examined, splitting the sample by the number of neighbours shows the largest variation and, in particular, the subsample with a number of neighbours larger than 8
presents the brightest and most significant signal in the region of interest. In the top-left panel of Figure \ref{stacks}, we present the pseudo NB image obtained from this subsample
compared to the other half of the sample (bottom-left panel) and the corresponding ``oriented-random" pseudo NB image. The results obtained with the other subsamples are presented
in the appendix. 

\begin{figure*}
  \centering Sample with 8 neighbours or more
\begin{center}
\includegraphics[trim={2cm 0cm 3.2cm 0cm},clip,width=3.4in]{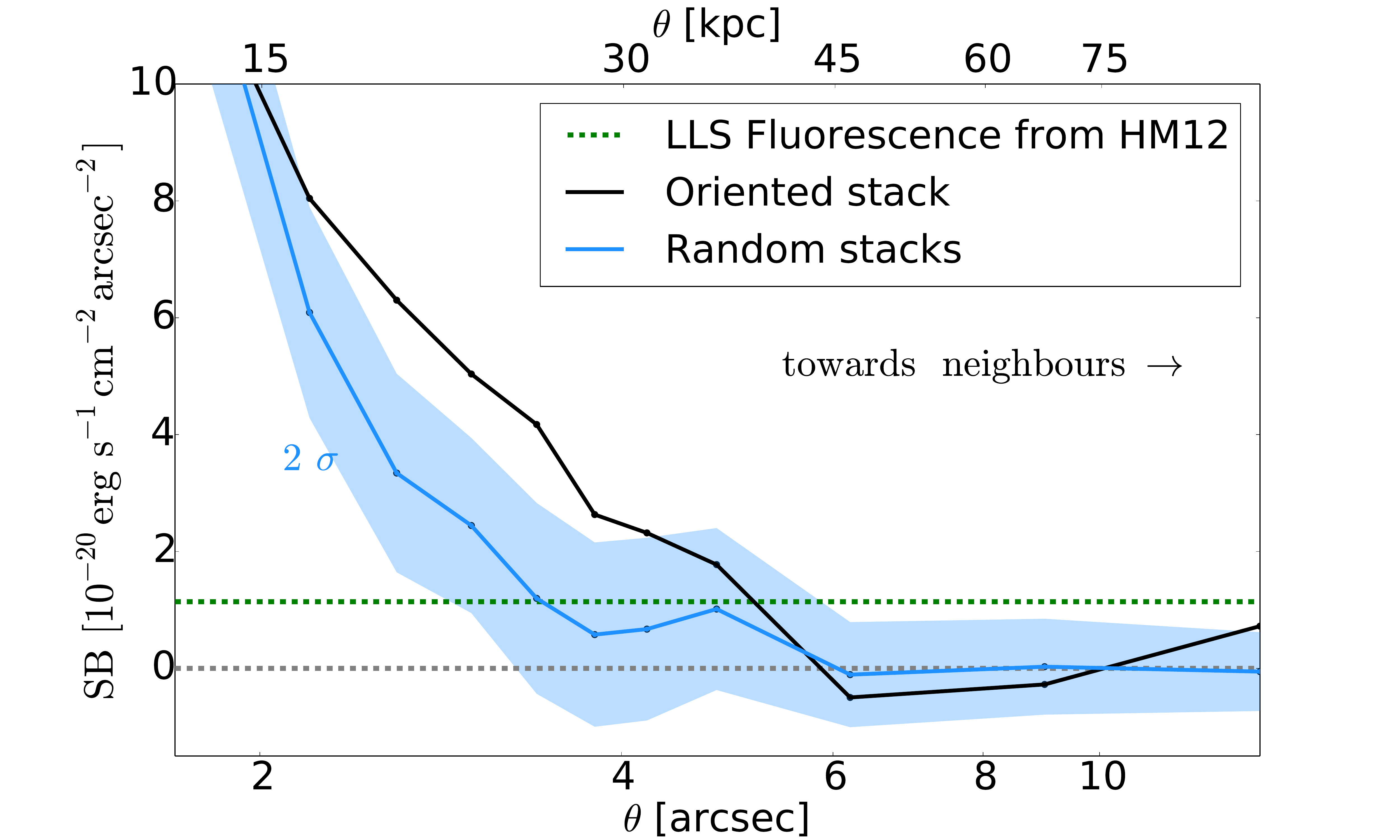}
\includegraphics[trim={2cm 0cm 3.2cm 0cm},clip,width=3.4in]{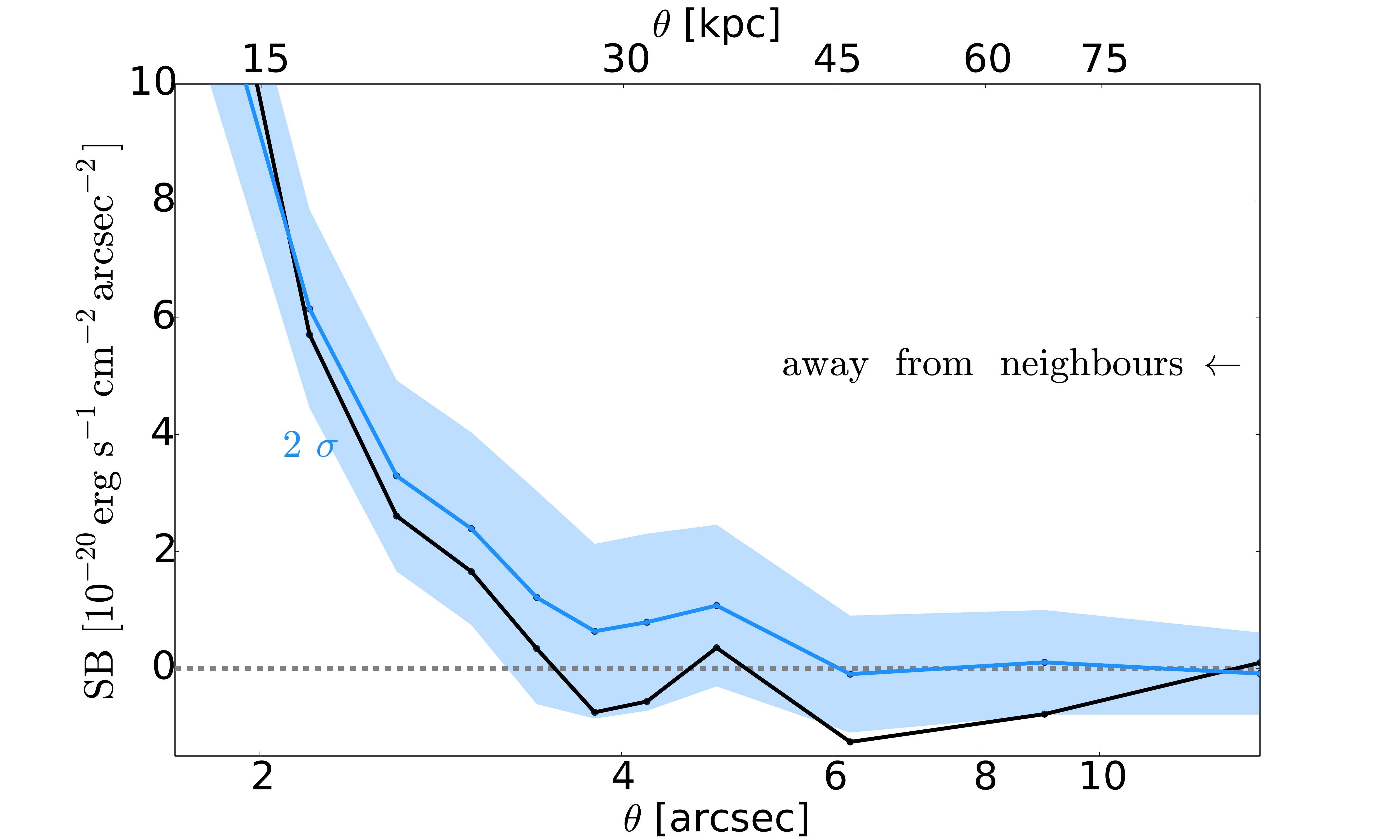}
   \caption{
   Same as Figure \ref{SB_full} but for the subsample of subcubes with 8 or more neighbours. As with the full sample, we clearly see an excess of emission in the oriented stack compared to random orientations, but now extended between 2′′ and 4′′ at more than 3$\sigma$ level at each spatial position.}
   \label{SBnmin}
\end{center}
\end{figure*}

In left panel of Figure \ref{SBnmin}, we show the SB profile for the subsample of galaxies with number of neighbours larger than 8. As in Figure \ref{SB_full}, this profile has been
obtained by integrating over a spatial aperture of vertical height of $2''$ and increasing horizontal widths (from $0.4''$ to $2''$). Compared to the ``super-random'' stack SB profile derived from the
same subsample of galaxies, we clearly see an excess of emission between $2''$ and $4''$ at more than 3$\sigma$ level at each spatial position.

\begin{figure}
\begin{center}
\includegraphics[trim={3cm 0cm 4cm 0cm},clip,width=3.3in]{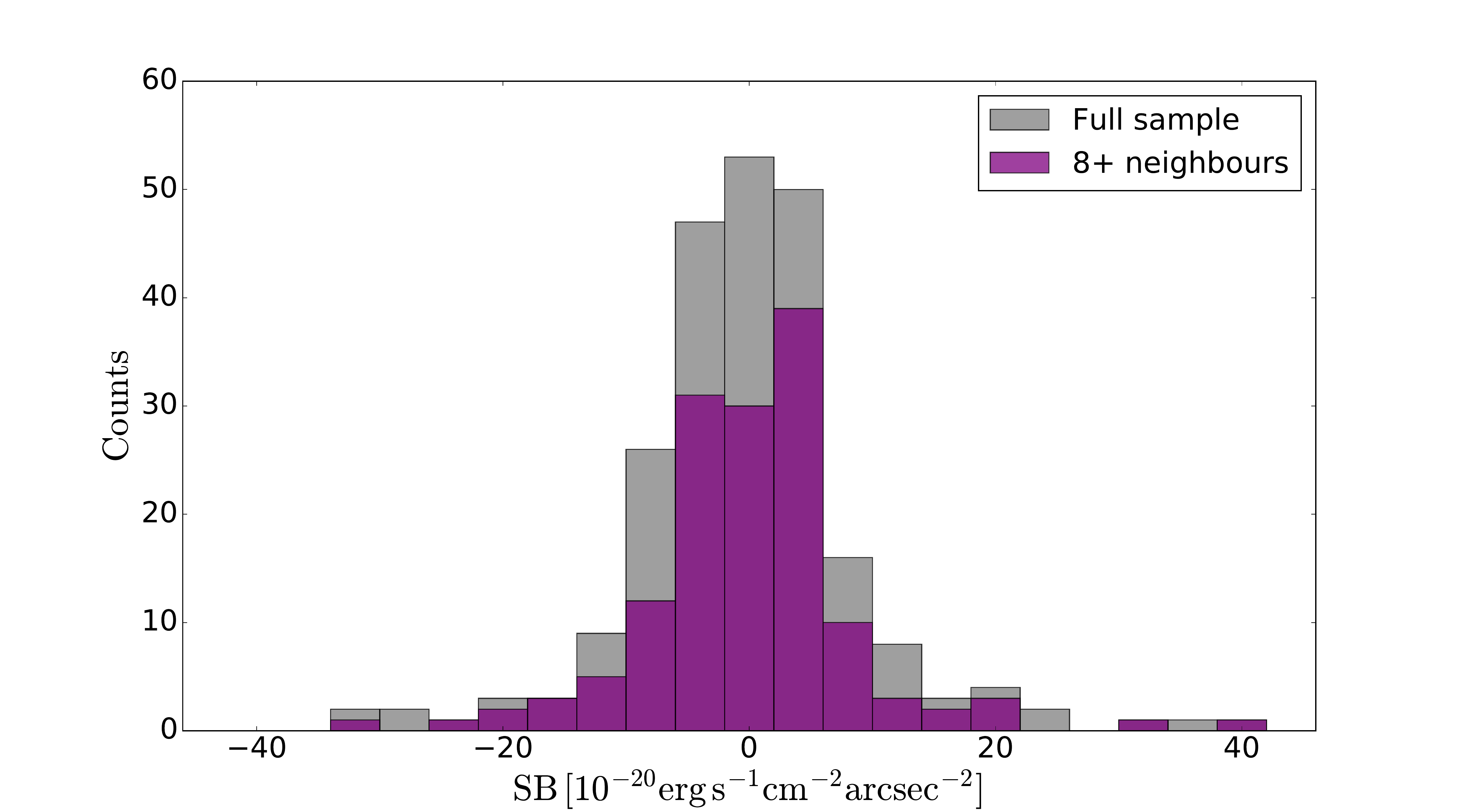}
   \caption{Distribution of SB in the aperture defined in Figure \ref{halfs} for each individual subcube in the full sample (gray) and the sample with 8 or more neighbours (purple).}
   \label{subcubes}
\end{center}
\end{figure}

In order to assess if this ``oriented'' excess of emission comes just from a small set of subcubes, we show in Figure \ref{subcubes} the individual SB values in the region of interest
for each of the oriented subcubes of both the full sample and the subsample with 8 or more neighbours.
The distribution of SB values is approximately Gaussian for both samples with very few outliers that do not contribute significantly to the
overall emission. A closer look for the sample with 8 or more neighbours shows a slight asymmetry on the positive part of the distribution 
at small SB values ($\sim4\times10^{-20}$\SBcgs) consistent with the measured emission in the stack.

\section{Discussion}

The main results of our stacking analysis presented in the previous section are: 
i) a lack of detectable extended emission on IGM scales, 
ii) the presence of a significant, statistical excess of CGM Ly$\alpha$ emission in the direction of neighbouring galaxies up to distances of about $4''$, i.e. about 30 projected kpc.
In this section we discuss the implications of our results in the context of intergalactic structures around galaxies and the possible origin of the CGM emission excess.

\subsection{LLSs and cosmic UVB constraints}

As discussed in Section 1, our stacking analysis should maximise the detectability of intergalactic filaments illuminated by the cosmic UVB 
in the hypothesis that a significant fraction of the galaxies in our sample are connected to each other by filaments with column densities 
similar or higher than LLSs. 
Note that because of the particular geometry of the observed volume in our survey (limited to $\sim450\times450\,\mathrm{kpc}^2$ in the plane of the sky) 
the majority of our galaxies may be connected to other sources that are \emph{outside} our field of view. Therefore, those galaxies and corresponding
subcube orientations may be missing in our stacking analysis.

Indicating with $f_{\mathrm{conn}}$ the fraction of possible galaxy-neighbour orientations with LLS filaments, our result should provide an
upper limit on the product of $f_{\mathrm{conn}}$ and the cosmic UVB photoionisation rate ($\Gamma_\mathrm{HI}$). 
In the extreme and unlikely case that $f_{\mathrm{conn}}=1$, then we would obtain a 2$\sigma$ upper limit of $\Gamma_\mathrm{HI}=0.2\times10^{-12}$ s$^{-1}$. 
This value is about a factor of three below the HM12 and four times below the latest empirical estimates based on the comparison of the Ly$\alpha$ forest mean flux
with cosmological simulations \citep[e.g.,][]{Becker2013}{\begin{footnote}{
Possibly by coincidence our 2$\sigma$ upper limit for $f_{\mathrm{conn}}=1$ is close to the HM12 estimates based on a model that includes only quasars. Taken a face value, this would
imply a Lyman-continuum escape fraction from galaxies at $z\sim3.5$ close to zero and therefore such a model would have serious problems 
for Hydrogen reionization if this result would be extended to higher redshifts.} 
\end{footnote}.

How can we constrain the possible value of $f_{\mathrm{conn}}$ and the spatial distribution of LLS around our galaxies?
From an observational perspective, we can compare the estimates of the 
incidence of LLSs per line of sight and redshift $dn/dz\sim1.5$ at $z\sim3.5$ \citep{Prochaska2010}
with the number of galaxy-neighbour orientations in our sample.}

Let us first hypothesise that each one of our galaxies is surrounded by a circularly symmetric distribution of gas with the column density of LLSs
and that no other regions in our datacubes are covered by LLSs, i.e. that $f_{\mathrm{conn}}=0$.
In order to reproduce the observed $dn/dz$ we would therefore require that each of the 96 galaxies in our sample, plus the 11 galaxies 
with a high-confidence flag that were discarded by our neighbouring-distance selection criteria, should be surrounded by a LLS with a radius of about $6''$. 
If this were the case, then we should have been able to detect fluorescent emission from the UVB up to this
radius at every possible angle around our galaxies. Figures \ref{SB_full} and \ref{SBnmin} do not show evidence for an excess
extending up to $6''$ with values compatible with the UVB and therefore LLSs cannot all be confined into circular regions around our galaxies. 
We notice, however, that there is an excess up to about $5''$ in the stack made with randomly oriented subcubes. In the conservative
hypothesis that this is due to fluorescence from the HM12 UVB instead of being produced by processes related to the central galaxies,
we would obtain a $dn/dz$ that is slightly larger than half of the observed value. Therefore, unless un-detected faint galaxies substantially
contribute to $dn/dz$, we think that it is likely that $f_{\mathrm{conn}}$ is not equal to zero.

On the other end, if we assume that the HM12 UVB is correct, we can use our result to provide a
2$\sigma$ upper limit on $f_{\mathrm{conn}}\approx0.3$.
We notice that restricting the sample to the half of the subcubes at smaller neighbouring distances (up to 8 cMpc, see Appendix) 
or splitting the sample in two halves based on other galaxy properties (see Appendix)
does not give any detectable intergalactic emission and therefore we cannot obtain a better constraint on $f_{\mathrm{conn}}$.
However, the increased noise of these subsamples do not allow 
a detailed analysis as in the case of the full sample. We will repeat this split-sample analysis in the future with the new, much larger 
sample of LAE emitters that will be detected in the UDF mosaic region (see e.g., Bacon et al. 2017 submitted, Leclercq et al. in prep)
and we will include an analysis of cosmological simulations to guide our stacking analysis and to better constrain 
the value of $f_{\mathrm{conn}}$.

\subsection{Origin of the oriented CGM emission excess}

The analysis of the SB profile of the oriented stack for the full sample revealed a significant excess of emission towards 
galaxy neighbours with respect to the ``random" stacks (see Fig. \ref{SB_full}) up to distances of about $4''$ from the galaxies. 
This excess is more pronounced when the stack is performed on the subsample
of subcubes that are surrounded by the largest number of neighbours (see Fig. \ref{SBnmin}). 
What is the origin of this ``statistical excess" of oriented CGM emission? 

We first consider the possibility that this excess is due to the Ly$\alpha$ emission from aligned and undetected satellite galaxies with Ly$\alpha$
fluxes below the detection limit. 
Using the results of \citet{Wisotzki2016} and Leclercq et al. in prep., we know that the circularly-averaged
UV emission from galaxies can be described by an exponential profile with a typical scale length of $r_{\mathrm{UV}}\sim0.3$ kpc.
If this extended UV profile contains the contribution of undetected satellites then we expect that their Ly$\alpha$ emission
should be at least a few orders of magnitude below the observed value in our stack at a distance of  $4''$, i.e. 30 projected kpc,
from the central galaxy in the direction of the neighbours. This applies also in the extreme case in which we 
place all the possibly undetected satellite galaxies in the region of excess emission (see Fig. 6). 
In this calculation, we have assumed that the Equivalent Width (EW) of the undetectable satellite galaxies is
similar to the measured EW of our galaxies. In order to obtain the observed Ly$\alpha$ emission
in the region of excess emission, the EW of the satellite galaxies should have been much larger than what
normal stellar population could produce (see \citeauthor{Cantalupo2012} \citeyear{Cantalupo2012} for discussion)
and therefore we exclude a satellite-galaxy origin for this excess emission.

As an alternative possibility, let us consider the hypothesis that the CGM Ly$\alpha$ emission is produced by fluorescence due to
the ionizing photons from the central galaxies. Given the average Ly$\alpha$ luminosities of our sample ($\sim10^{42}$ erg s$^{-1}$)
we expect average star formation rates of about 0.6 M$_{\odot}$ yr$^{-1}$ (using the standard SFR to H$\alpha$ conversion factors
and assuming Case B recombination line ratios between H$\alpha$ and Ly$\alpha$)
and therefore intrinsic ionisation rates of about 10$^{53.5}$ photon s$^{-1}$ (from Starburst99 assuming continuous SFR and
an age larger than 10$^{7}$ yr, Leitherer et al. 1999). To explain the observed Ly$\alpha$ SB at 30 projected kpc
in the oriented stack ($\approx3\times10^{-20}$\SBcgs) with galaxy-fluorescence emission for self-shielded gas, we estimate that a Lyman-continuum
escape fraction from the galaxy's Interstellar Medium of $f^{\mathrm{ISM}}_{\mathrm{esc}}\sim2\times10^{-2}$ would be sufficient.
We notice that this escape fraction is an upper limit of the measurable escape fraction because it does not include absorption
by the CGM. Although there are no direct measurements of Lyman-continuum photons escaping from high-z galaxies, 
a value of $f^{\mathrm{ISM}}_{\mathrm{esc}}\sim2\times10^{-2}$ is totally consistent with current upper limits \citep[see e.g.,][]{Siana2015}
and with the required value for the reionization of hydrogen (see e.g., HM12).
Larger values of $f^{\mathrm{ISM}}_{\mathrm{esc}}$ can produce highly ionised gas. In this case, 
the expected fluorescent SB will scale with the gas density squared. The excess of emission towards the galaxy neighbours
could then be simply explained by an increased gas density along this direction. In particular, the increased SB 
in the oriented stack by a factor of about 3 with respect to the ``random" stack would imply statistically higher
densities by a factor of about 1.7 towards the galaxy neighbours. 

If the Ly$\alpha$ emission is due to scattering in a neutral medium instead of fluorescence, our result would again imply
that CGM densities towards galaxy neighbours should be statistically larger than in any other direction. 
We note however, that the lack of a correlation between Ly$\alpha$/UV luminosities and 
halo exponential scale lengths does not clearly favour a scattering scenario
(see e.g., \citeauthor{Wisotzki2016} \citeyear{Wisotzki2016} and Leclercq et al. in prep.).

In both cases, an increased density on scales of 30 projected kpc around galaxies in the direction of much more distant
neighbours (on average, 10 comoving Mpc) seems a surprising result for which a detailed comparison with simulations
will be needed. As discussed in Section 4, we do not detect any correlation between 
the strength of this oriented CGM excess emission and any properties of the galaxies, including neighbour-distances,
with the exception of the environment, as measured using the number of neighbours. 
A possible origin of this trend may be due to the larger dark matter haloes of the more clustered LAEs 
that, therefore, could have larger and denser filaments in their circumgalactic environments. 
If the CGM excess is connected to the distribution of gas on IGM scales, e.g. cosmological filaments, 
then the derived densities above will be degenerate with the value of $f_{\mathrm{conn}}$. 
In particular, we expect that the implied densities due to fluorescence or Ly$\alpha$ scattering
in the direction of the neighbours will scale as $f_{\mathrm{conn}}^{-1}$. 
Another dilution effect of the expected signal in our stacking analysis, on both CGM and 
IGM scales, could be due to the possibility that filaments are bended. 
Also in this case, a detailed comparison with simulations will be needed to asses the
importance of these effects for the implications of our results.

\section{Summary}

Cosmological simulations suggest that the gas distribution between galaxies is filamentary
and that the filaments are oriented preferentially towards neighbouring galaxies \citep[e.g.,][]{Gheller2015}, a property that can be intimately linked to the initial conditions of the cosmic density field \citep[e.g.,][]{Bond1996}. 
Illuminated by the cosmic UVB, these filaments are expected to emit fluorescent Ly$\alpha$ radiation with SB levels that 
are, unfortunately, one or two orders of magnitude below current observational limits for individual detections.

We presented and developed the idea of an ``oriented stacking" approach using Ly$\alpha$ emitting galaxies (LAEs) 
away from quasars at redshift $3<z<4$ detected in deep MUSE cubes.
We stacked three-dimensional regions (subcubes) around LAEs in the HDFS and UDF-10 MUSE fields (\citeauthor{Bacon2015} \citeyear{Bacon2015}, Bacon et al. 2017 submitted)
with orientations determined by the position of LAE neighbours within a line of sight comoving distance of $0.5<d<20$ cMpc (assuming pure Hubble flow).
If neighbouring galaxies are connected by filaments and these filaments are Lyman-Limit Systems (LLSs), then our oriented-stacking
method should boost the signal-to-noise ratio of UVB-induced Ly$\alpha$ fluorescence by about the square root of the number of stacking elements. 

By stacking 390 individual, ``re-oriented" subcubes we achieved a $3\sigma$ sensitivity level of $\mathrm{SB}\approx0.78\times10^{-20}$\SBcgs\  
in an aperture of $0.4\,\mathrm{arcsec^2}$ for a pseudo NB of width $6.25\,\mathrm{\AA}$, three times below the expected 
fluorescent signal from the values of the cosmic UVB at $z\sim3.5$ estimated by HM12 in the extreme hypothesis that all our galaxies are connected to each other by LLS filaments.
No detectable emission is found on intergalactic scales (i.e. at distances larger than 40 and up 120 projected kpc from galaxies) at significant levels,
implying that at least two thirds of our subcubes should not contain oriented LLSs for a HM12 cosmic UVB.
This result is independent of all galaxy properties that we have investigated in this study
(projected and comoving distances from neighbours, redshifts, numbers of neighbours and luminosities). 

However, significant emission is detected in the circum-galactic medium (CGM) of galaxies (up to about 30 projected kpc) 
 at SB levels of $\approx3\times10^{-20}$\SBcgs\ in the direction of galaxy neighbours but not in other directions. 
 The signal is stronger (4$\sigma$ level) at radii up to $4''$ when the sample is splitted considering only the galaxies with 
 a number of neighbours equal or larger than 8, while it seems independent of any of the other galaxy properties 
mentioned above.
We investigated the possible origin of this excess emission and we found that  ``preferentially oriented" Ly$\alpha$ emission 
from un-detected satellite galaxies is at least two orders of magnitude below the observed value. 
We estimated that a very modest escape fraction of Lyman-continuum photons from the ISM of the central galaxies 
(i.e. $f^{\mathrm{ISM}}_{\mathrm{esc}}\sim2\times10^{-2}$) should be sufficient to produce enough Ly$\alpha$ emission by photoionising at least part of the CGM up to 30 kpc.
In this case,  the excess of CGM emission towards the galaxy neighbours
can simply be explained by an increased gas density along this direction by a factor of about two, on average.
The dependence of this excess on the galaxy environments may suggest a connection with the 
host halo of the LAEs in terms of filament sizes and densities.

The methods and the idea developed in this first study will be extended in several directions in 
future works with the goal of understanding the origin and nature of the oriented CGM emission
excess and to provide better constraints on the presence and properties of intergalactic filaments. 
In particular, we plan to increase the observational sample of LAEs for our stacking analysis
with the new catalogues and data in the MUSE UDF-mosaic region (Bacon et al. 2017 submitted)
and other MUSE cubes with similar exposure times.  
This new data will provide a one order of magnitude increase in the number of galaxies
and spatial coverage, albeit at a lower sensitivity level (10 hours exposure time per field, 
versus the 30 hours per field used in this study).
At the same time, we plan to use high-resolution cosmological simulations to guide future 
stacking analyses by estimating the probability that galaxies with given properties are 
connected by LLSs. A positive detection would provide constraints on the morphological 
and physical properties of the cosmic web away from quasars and, at the same time, 
a direct measurement of the amplitude of the cosmic UVB at high redshift.

\section*{Acknowledgments}

This work has been supported by the Swiss National Science Foundation. In particular,
SC gratefully acknowledges support from Swiss National Science Foundation grant PP00P2\_163824.
LW acknowledges funding by the Competitive Fund of the Leibniz Association through grant SAW-2015-AIP-2. ERC Grant agreement 278594-GasAroundGalaxies.
\bibliographystyle{mnras}
\bibliography{refs.bib}

\appendix

\section{Half Stacks}

In this section, we present the pseudo NB images for the subsample of subcubes splitted by the following galaxy properties (See Table 1): i) luminosity (Figure A1),
 ii) redsfhit (Figure A2), iii) comoving line of sight distance (Figure A3), and iv) projected distance (Figure A4).
\begin{figure*}
\begin{center}
\includegraphics[trim={.2cm 4.2cm 2.01cm 3cm},clip, width=3.696in]{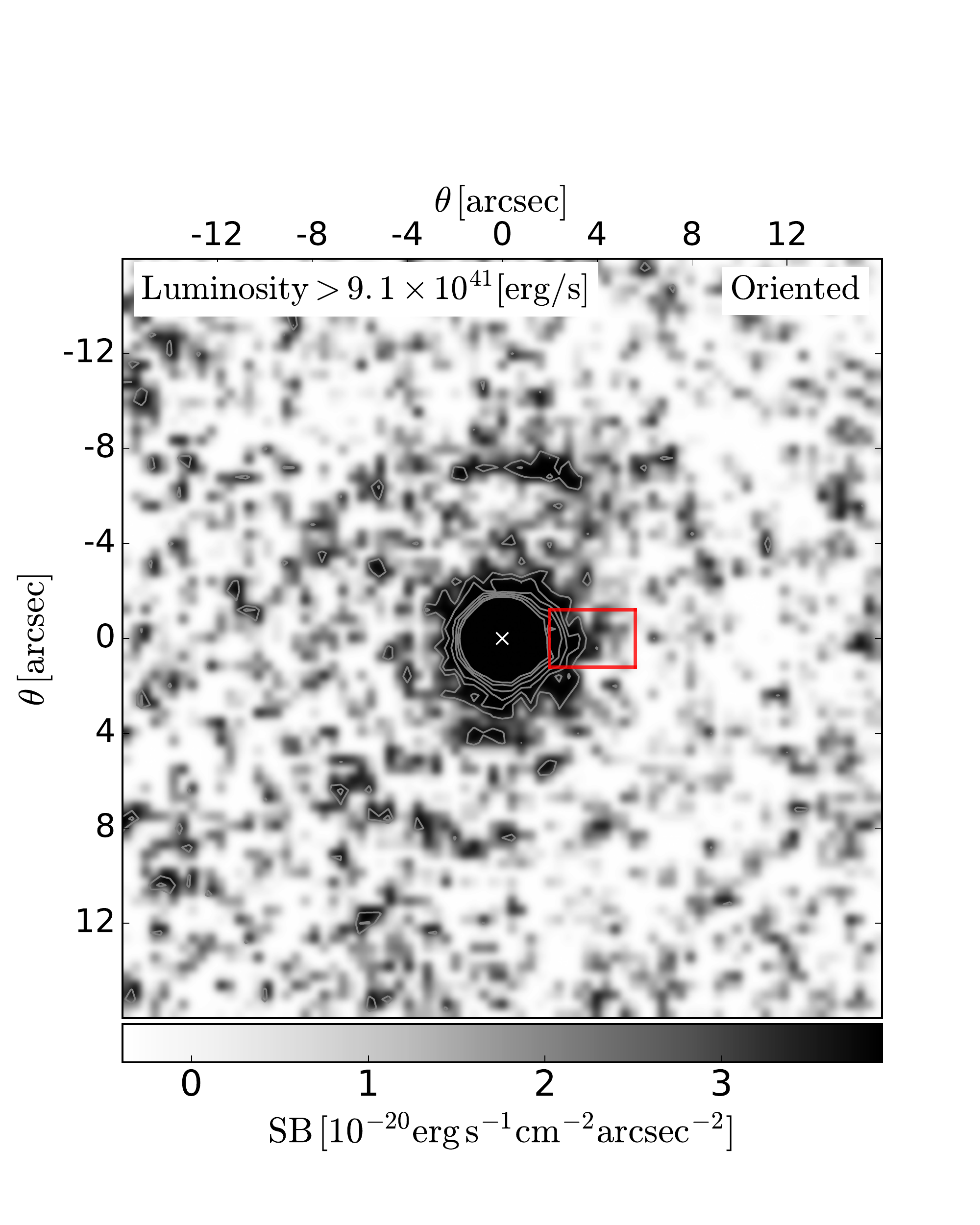}
\includegraphics[trim={2.5cm 4.2cm 2cm 3cm},clip, width=3.23in]{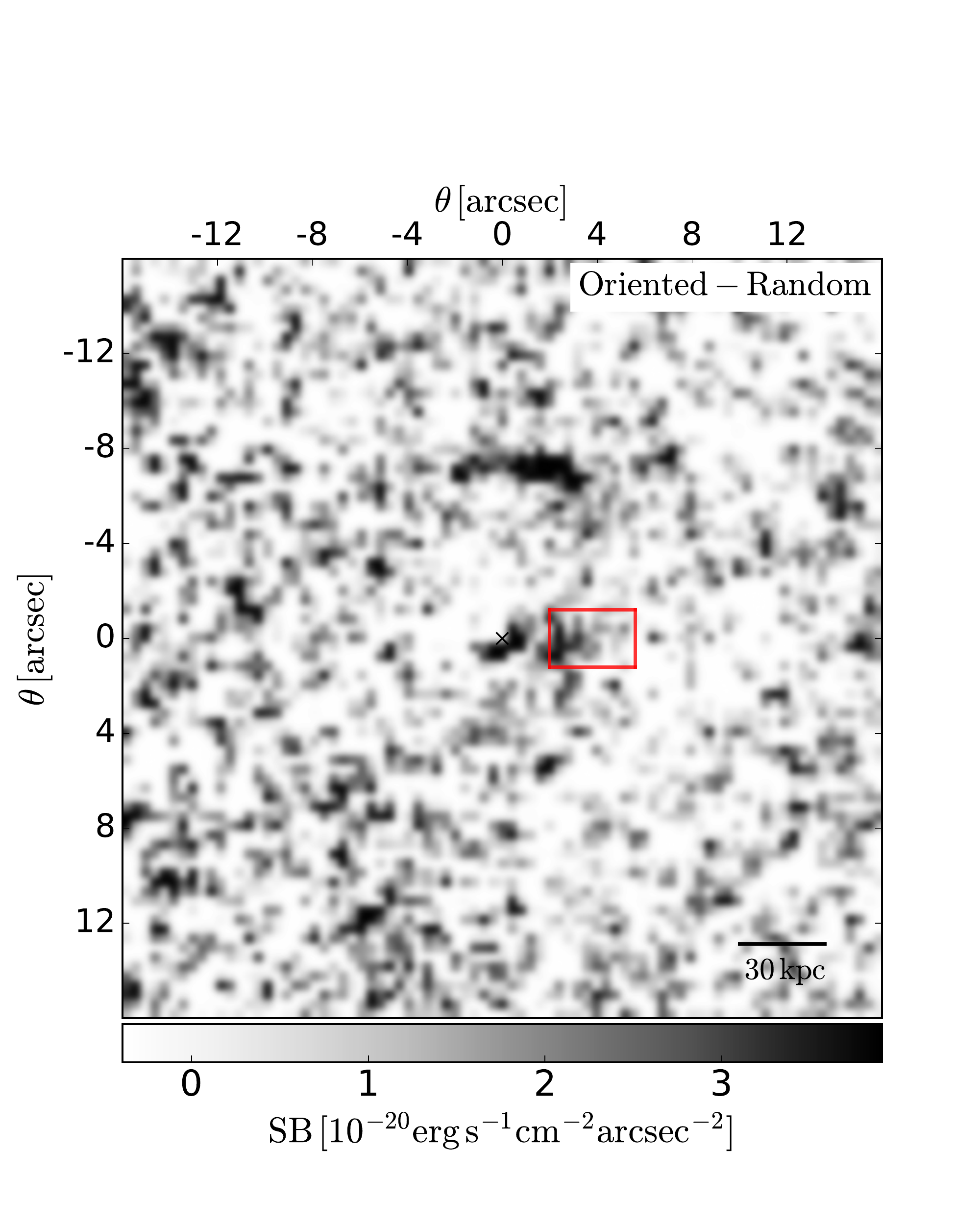}
\includegraphics[trim={.2cm 1.5cm 2.01cm 5.3cm},clip, width=3.696in]{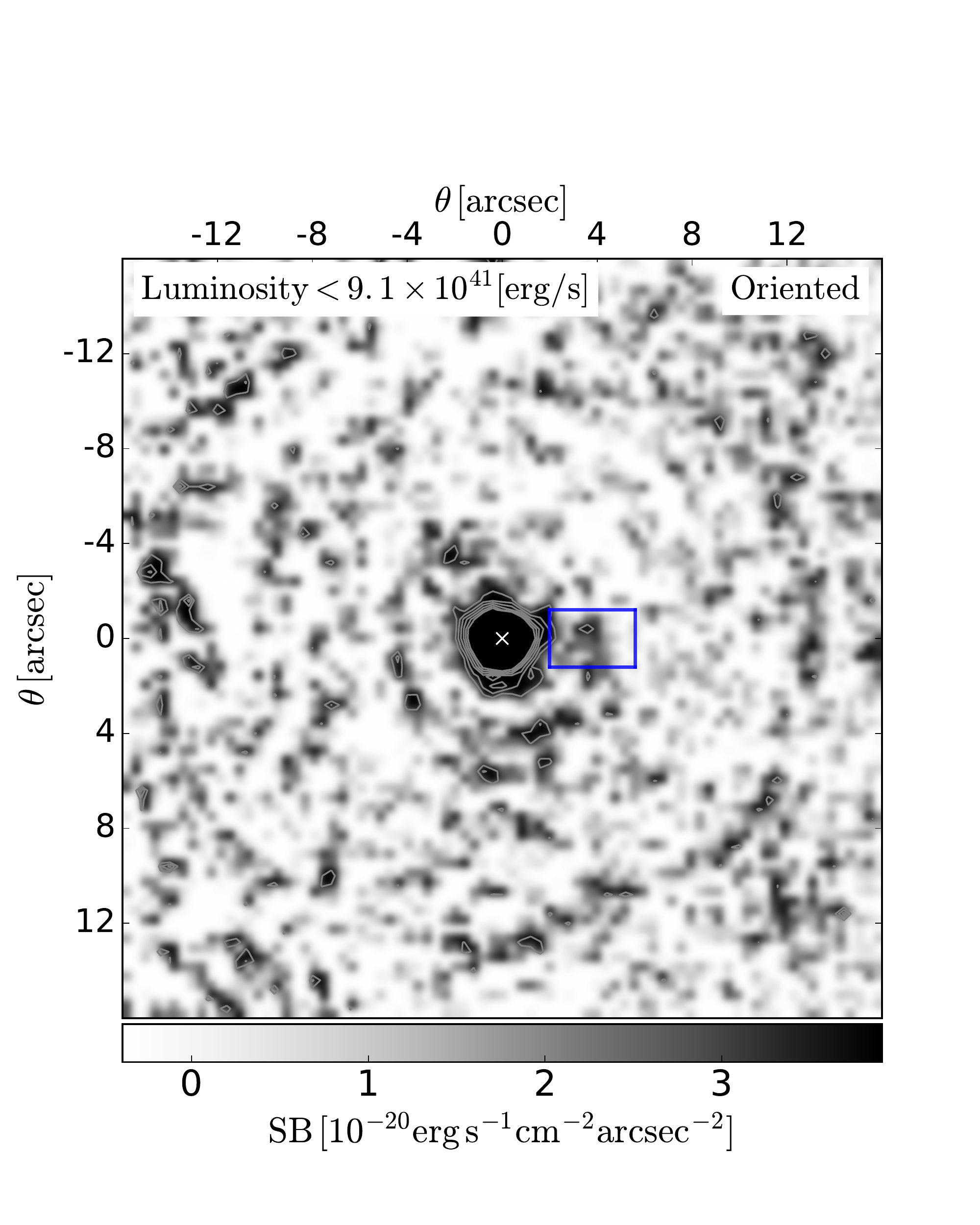}
\includegraphics[trim={2.5cm 1.5cm 2cm 5.3cm},clip, width=3.23in]{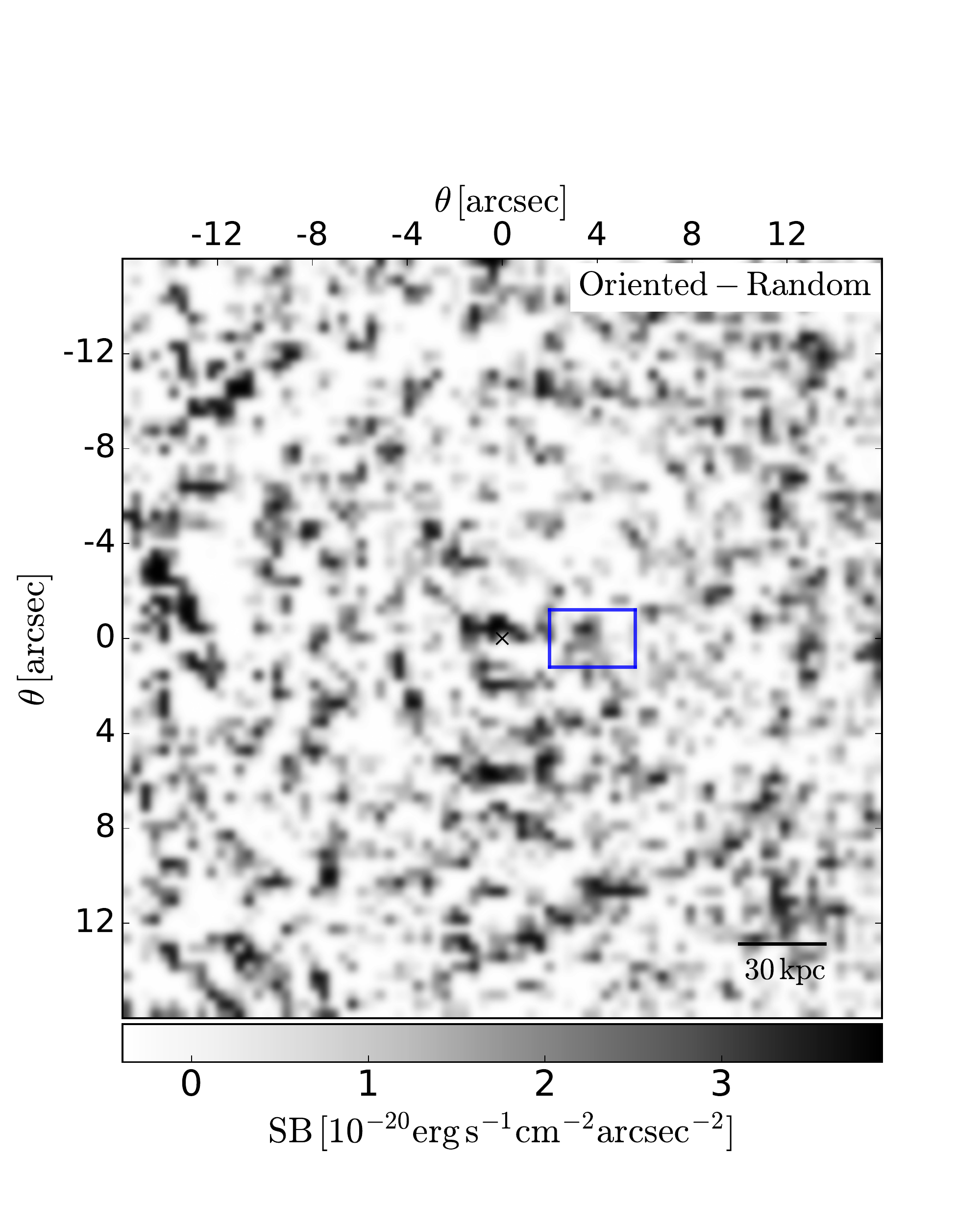}
   \caption{Same as Figure \ref{stacks} for a median luminosity of $9.1\times10^{41}\,\mathrm{erg\,s^{-1}}$.}
  \label{stacks1}
\end{center}
\end{figure*}

\begin{figure*}
\begin{center}
\includegraphics[trim={.2cm 4.2cm 2.01cm 3cm},clip, width=3.696in]{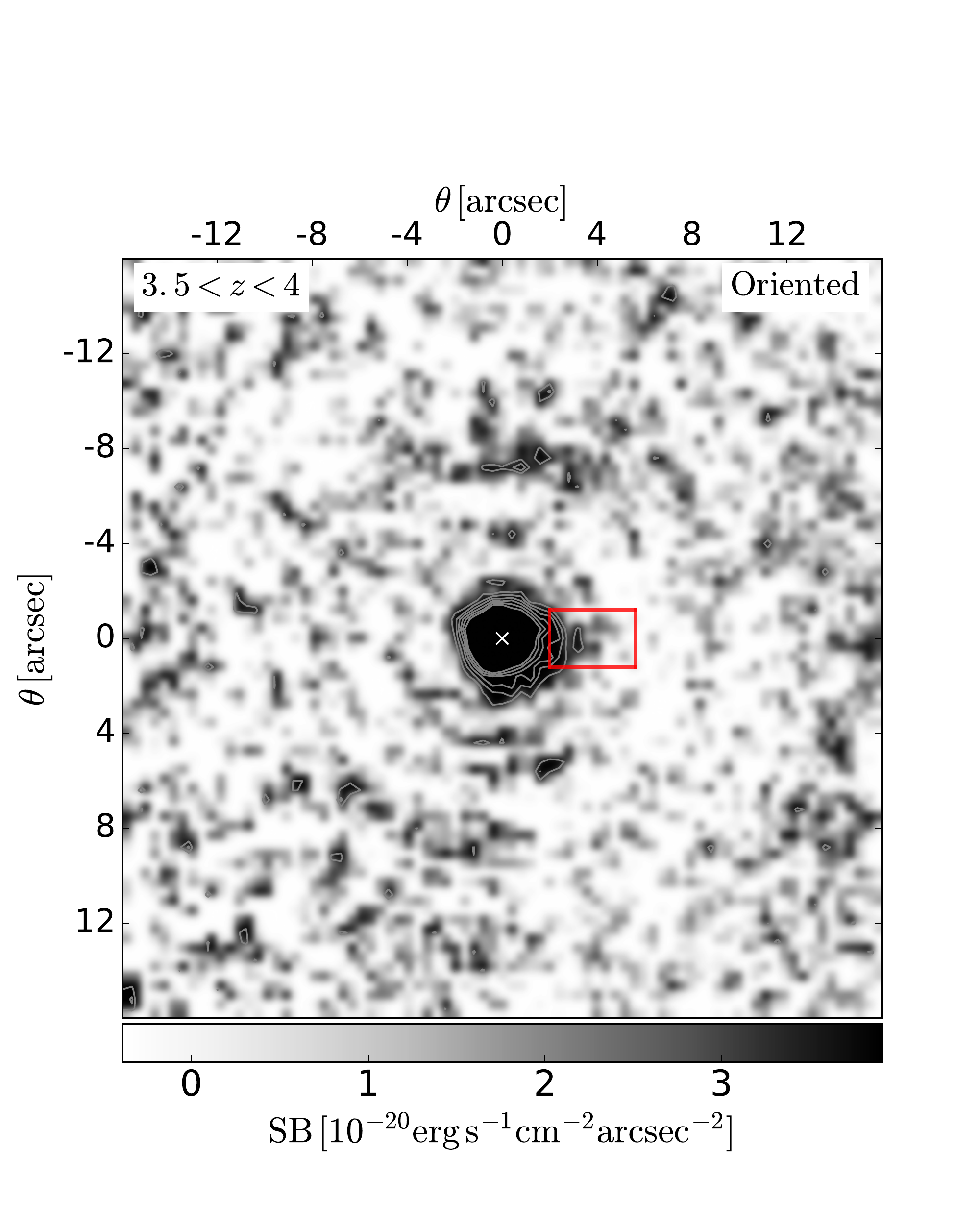}
\includegraphics[trim={2.5cm 4.2cm 2cm 3cm},clip, width=3.23in]{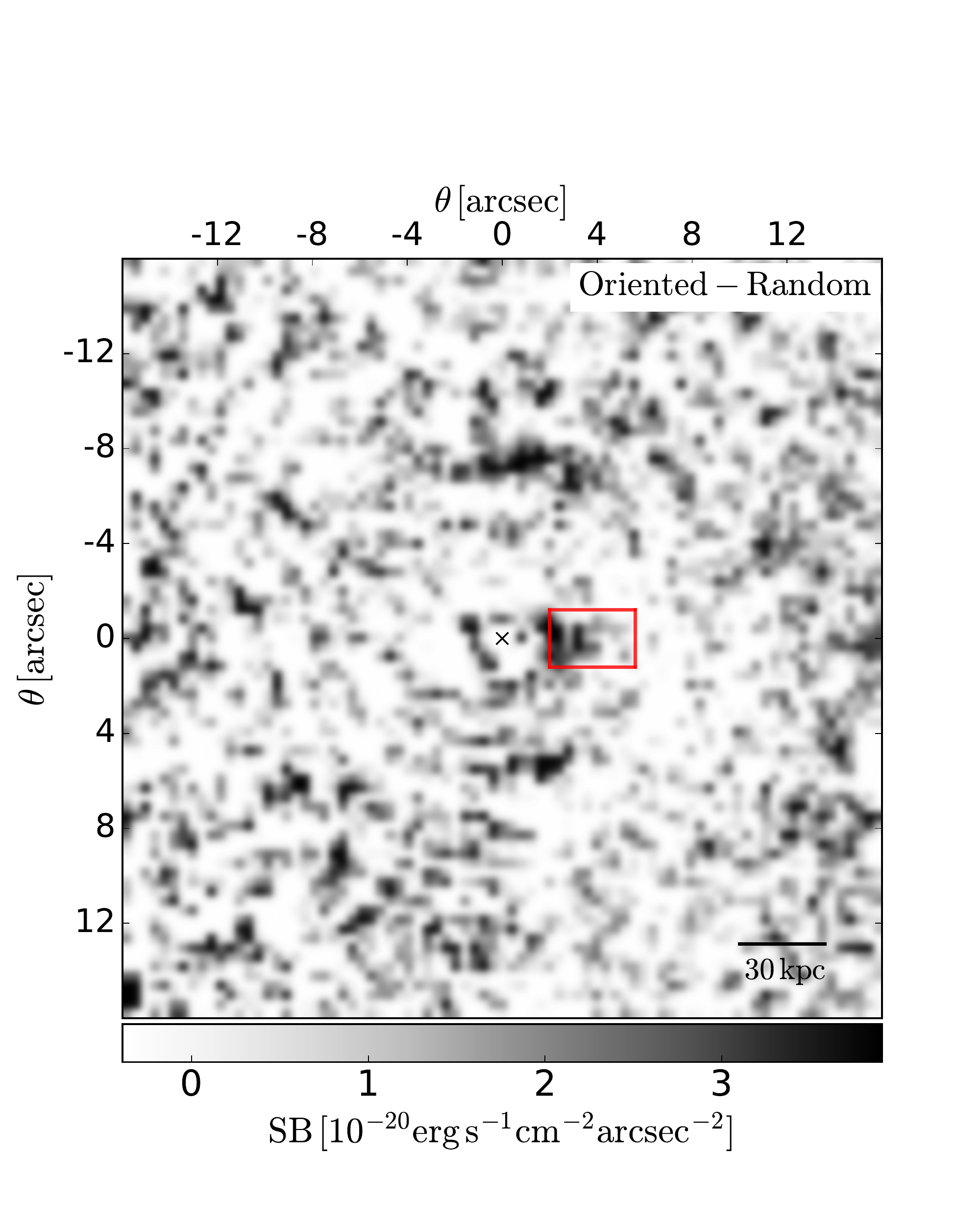}
\includegraphics[trim={.2cm 1.5cm 2.01cm 5.3cm},clip, width=3.696in]{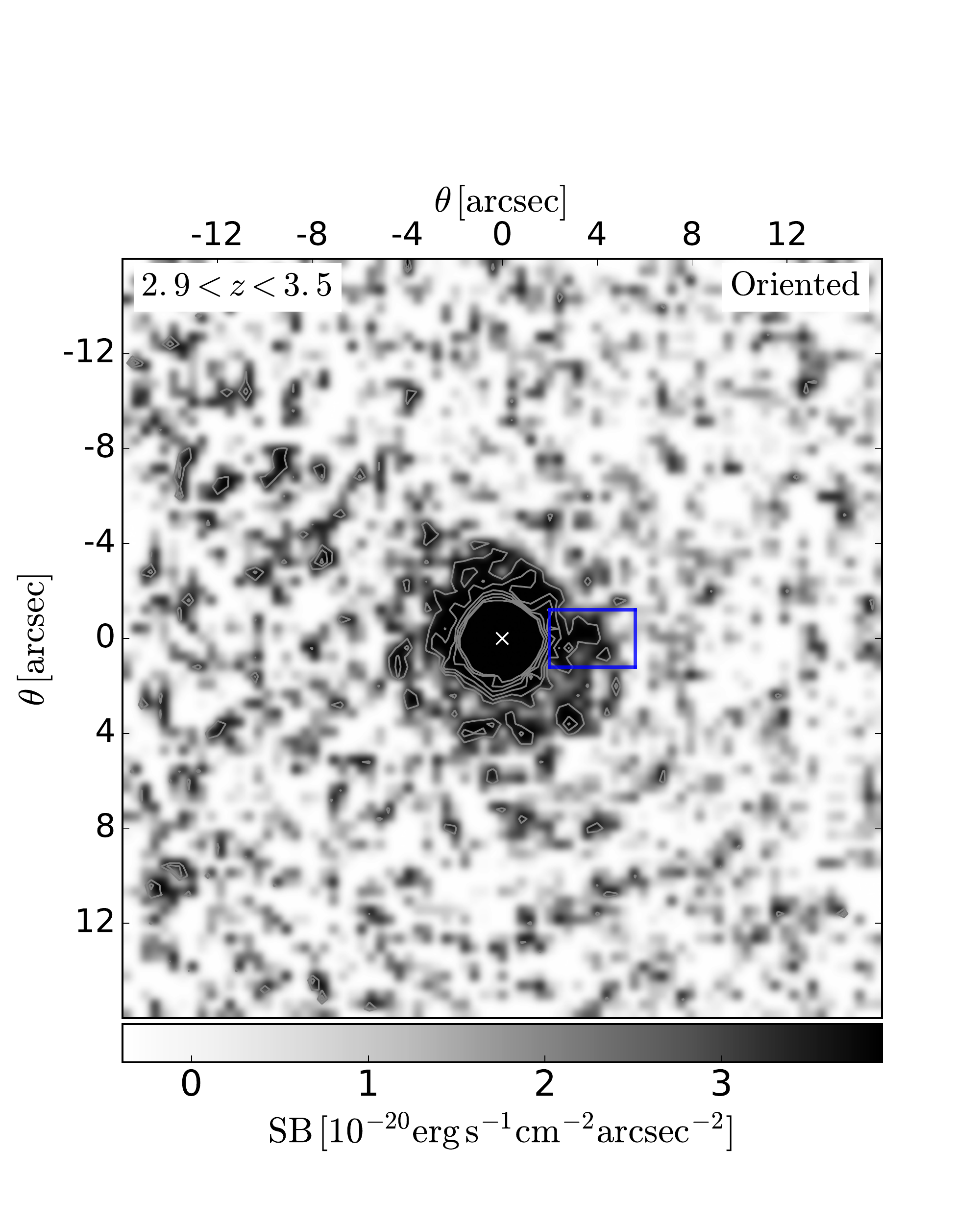}
\includegraphics[trim={2.5cm 1.5cm 2cm 5.3cm},clip, width=3.23in]{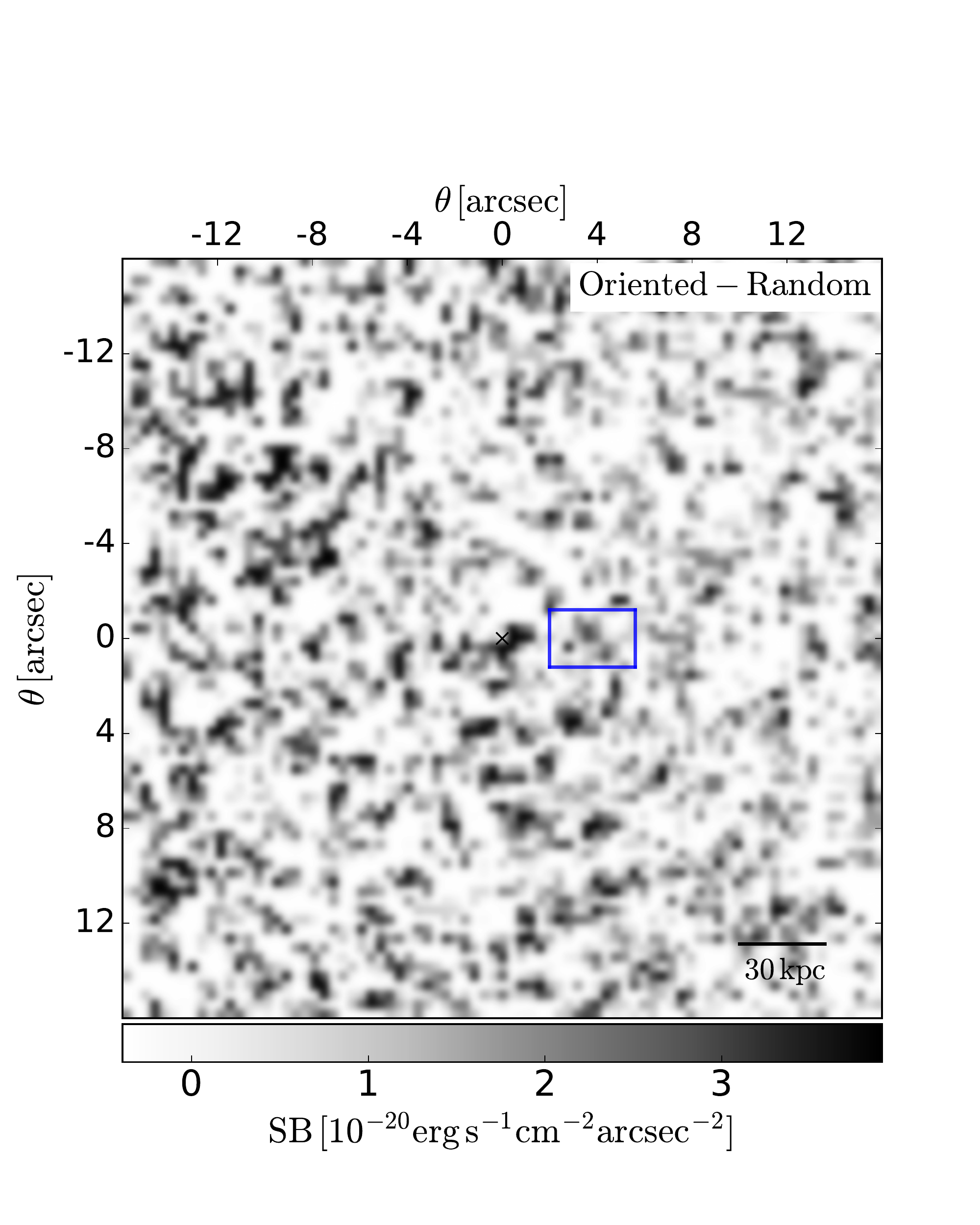}
   \caption{Same as Figure \ref{stacks} for a median redshift of $z=3.5$.}
  \label{stacks2}
\end{center}
\end{figure*}

\begin{figure*}
\begin{center}
\includegraphics[trim={.2cm 4.2cm 2.01cm 3cm},clip, width=3.696in]{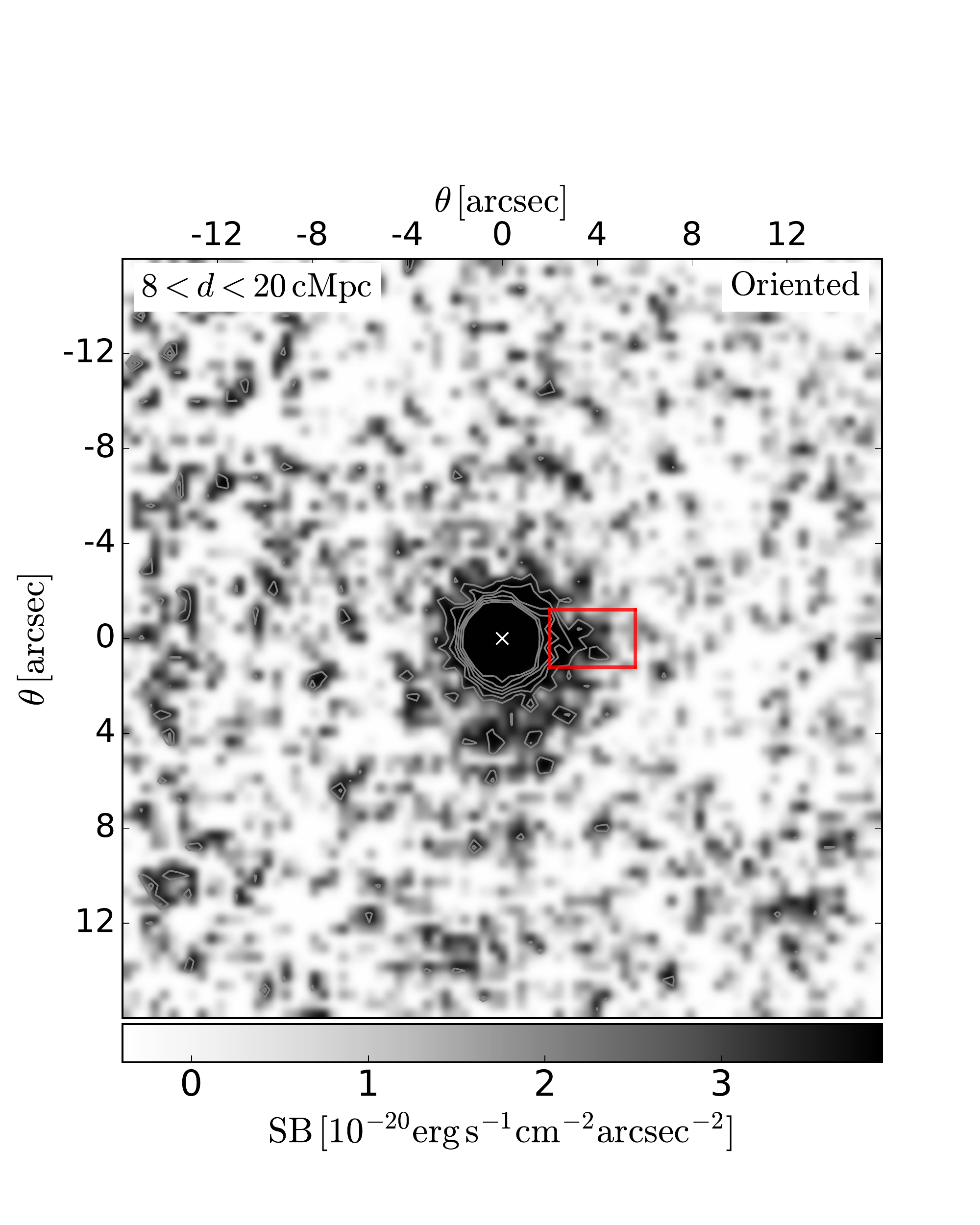}
\includegraphics[trim={2.5cm 4.2cm 2cm 3cm},clip, width=3.23in]{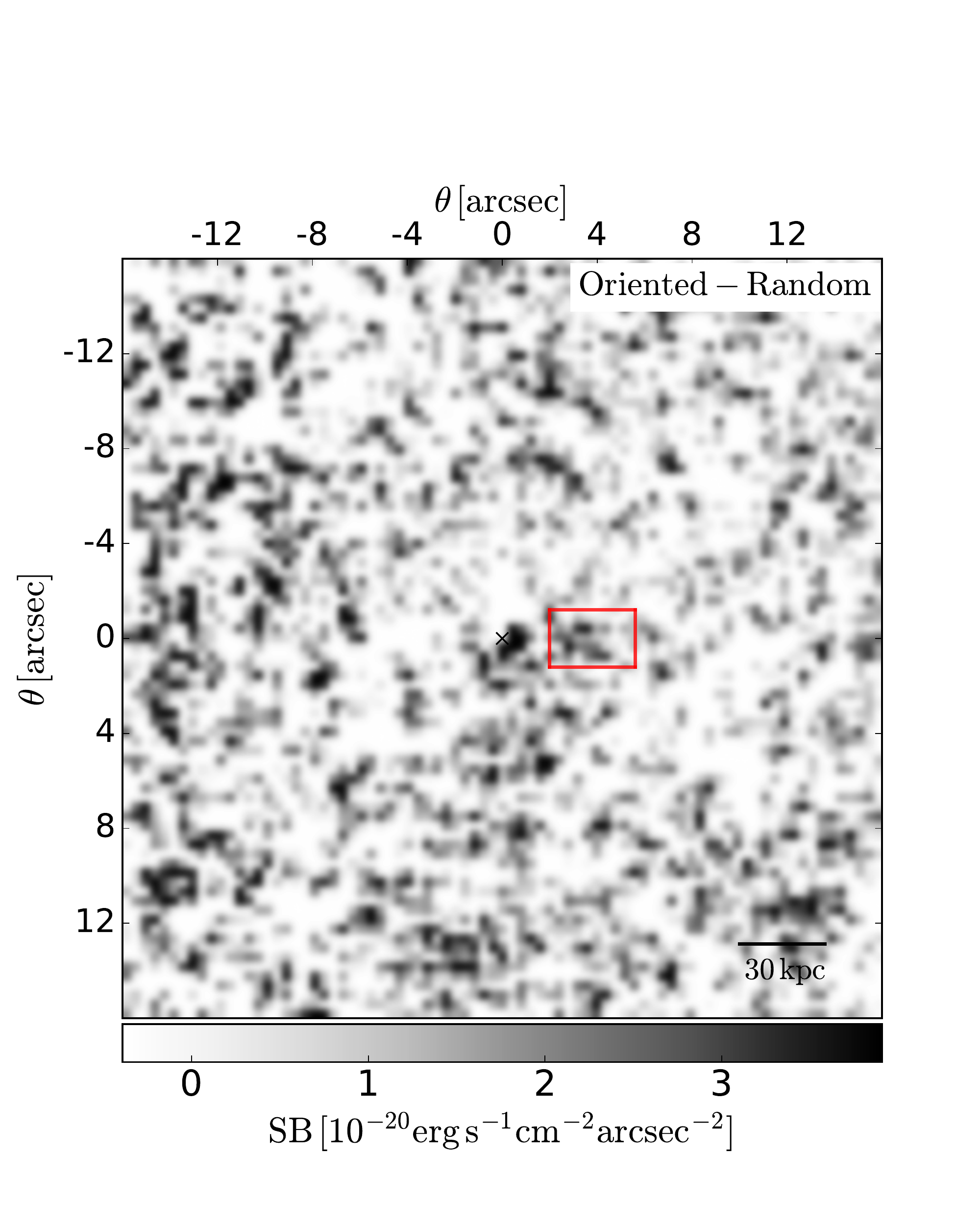}
\includegraphics[trim={.2cm 1.5cm 2.01cm 5.3cm},clip, width=3.696in]{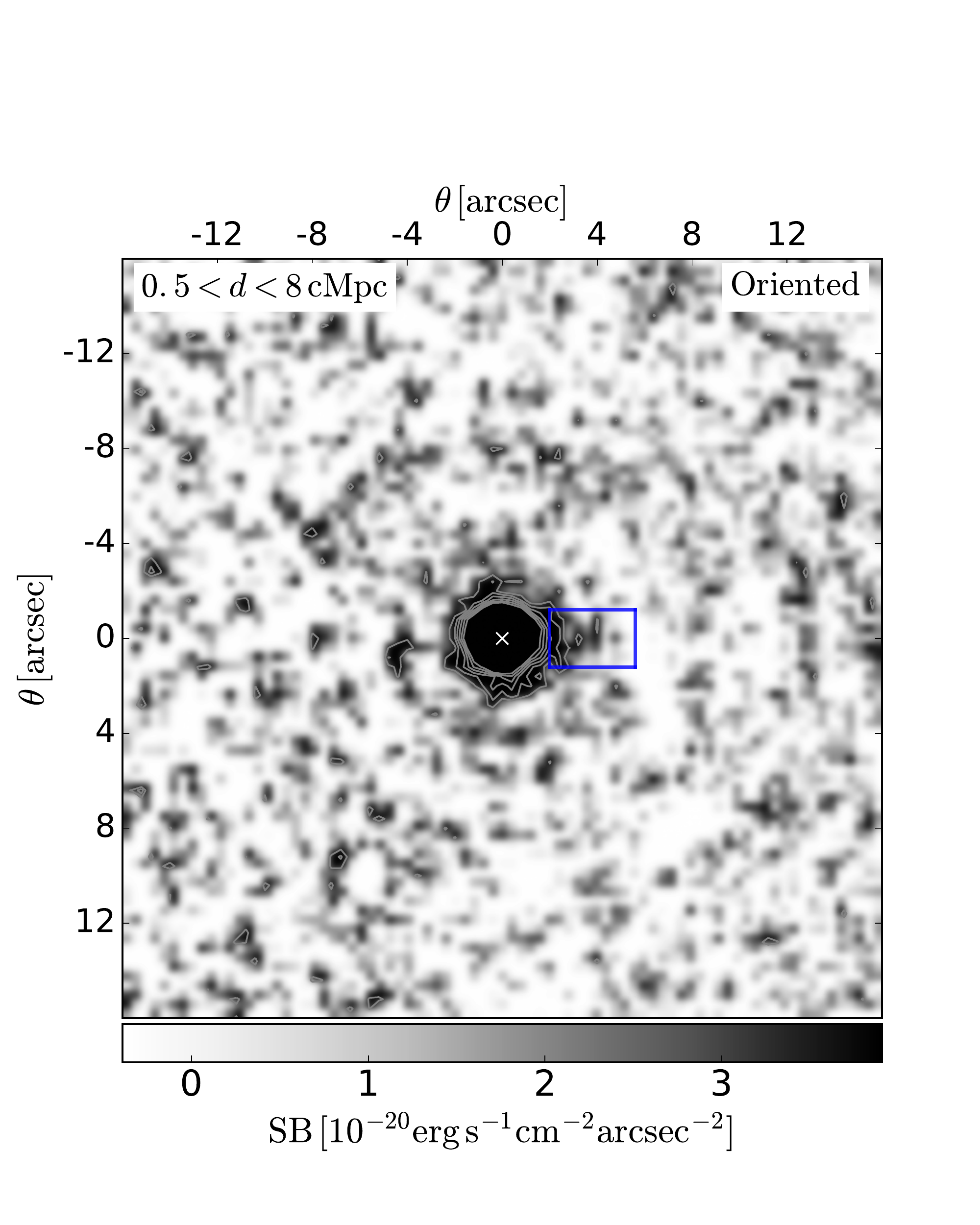}
\includegraphics[trim={2.5cm 1.5cm 2cm 5.3cm},clip, width=3.23in]{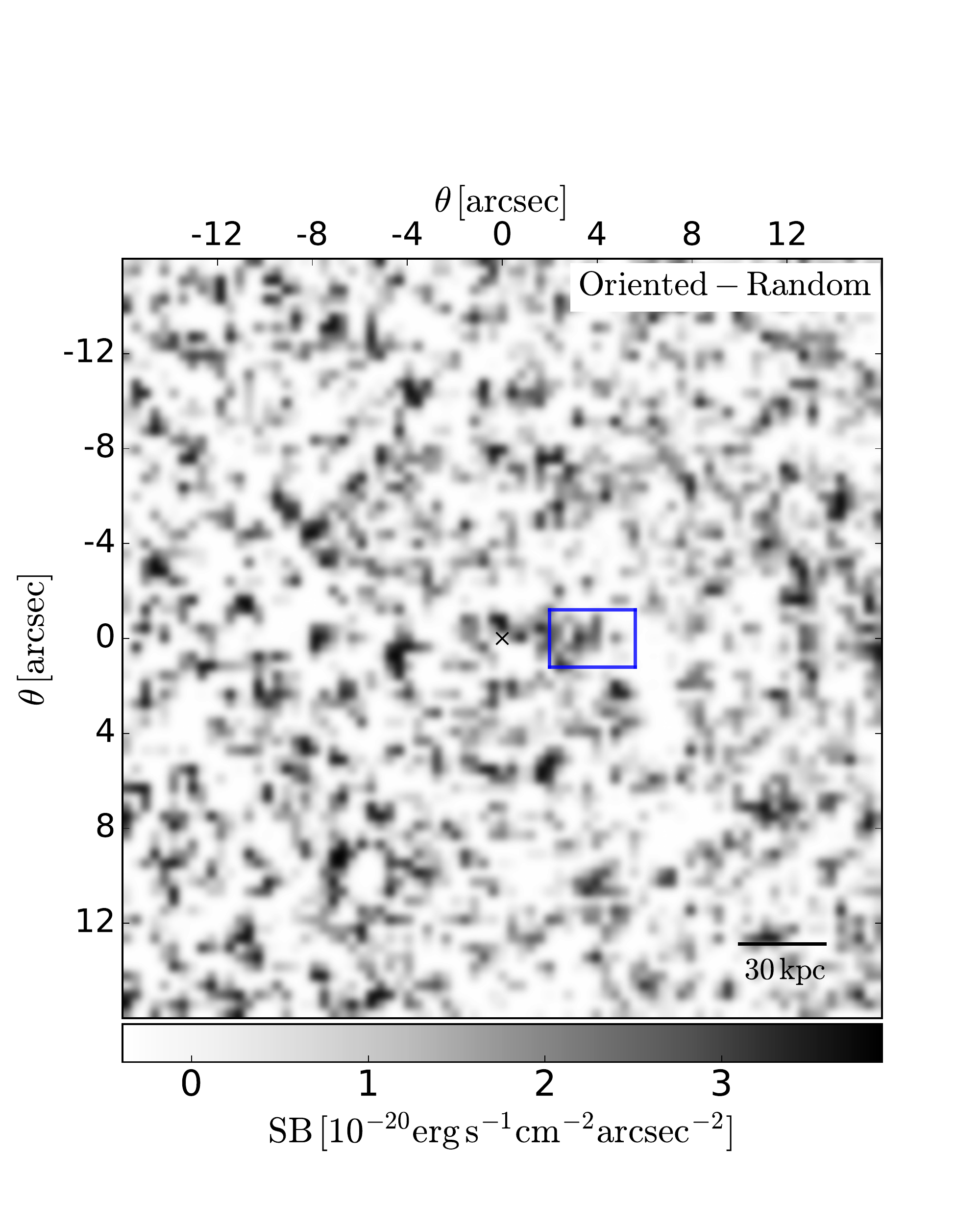}
    \caption{Same as Figure \ref{stacks} for a median comoving distance to the neighbour of $8\,\mathrm{cMpc}$.}
  \label{stacks3}
\end{center}
\end{figure*}

\begin{figure*}
\begin{center}
\includegraphics[trim={.2cm 4.2cm 2.01cm 3cm},clip, width=3.696in]{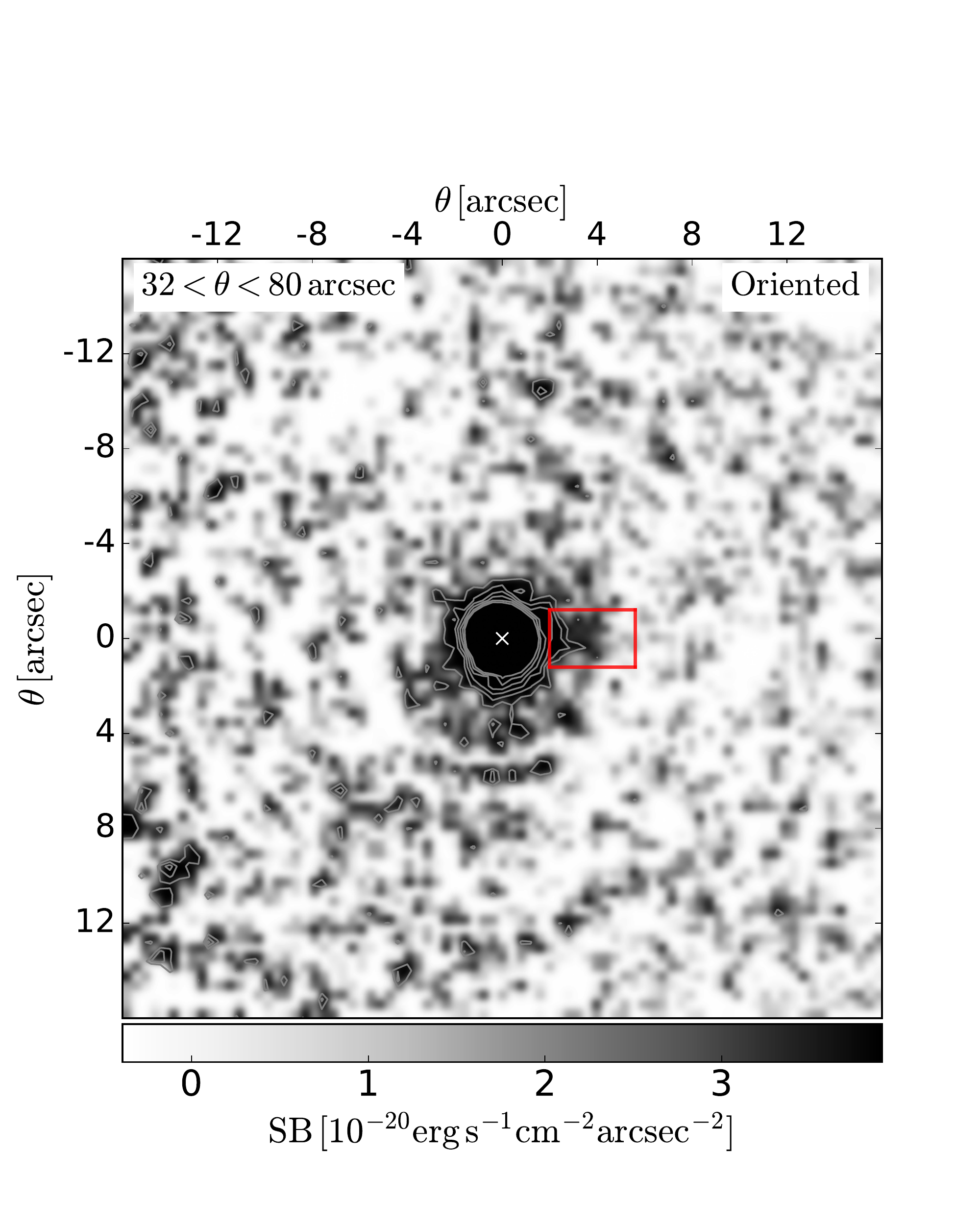}
\includegraphics[trim={2.5cm 4.2cm 2cm 3cm},clip, width=3.23in]{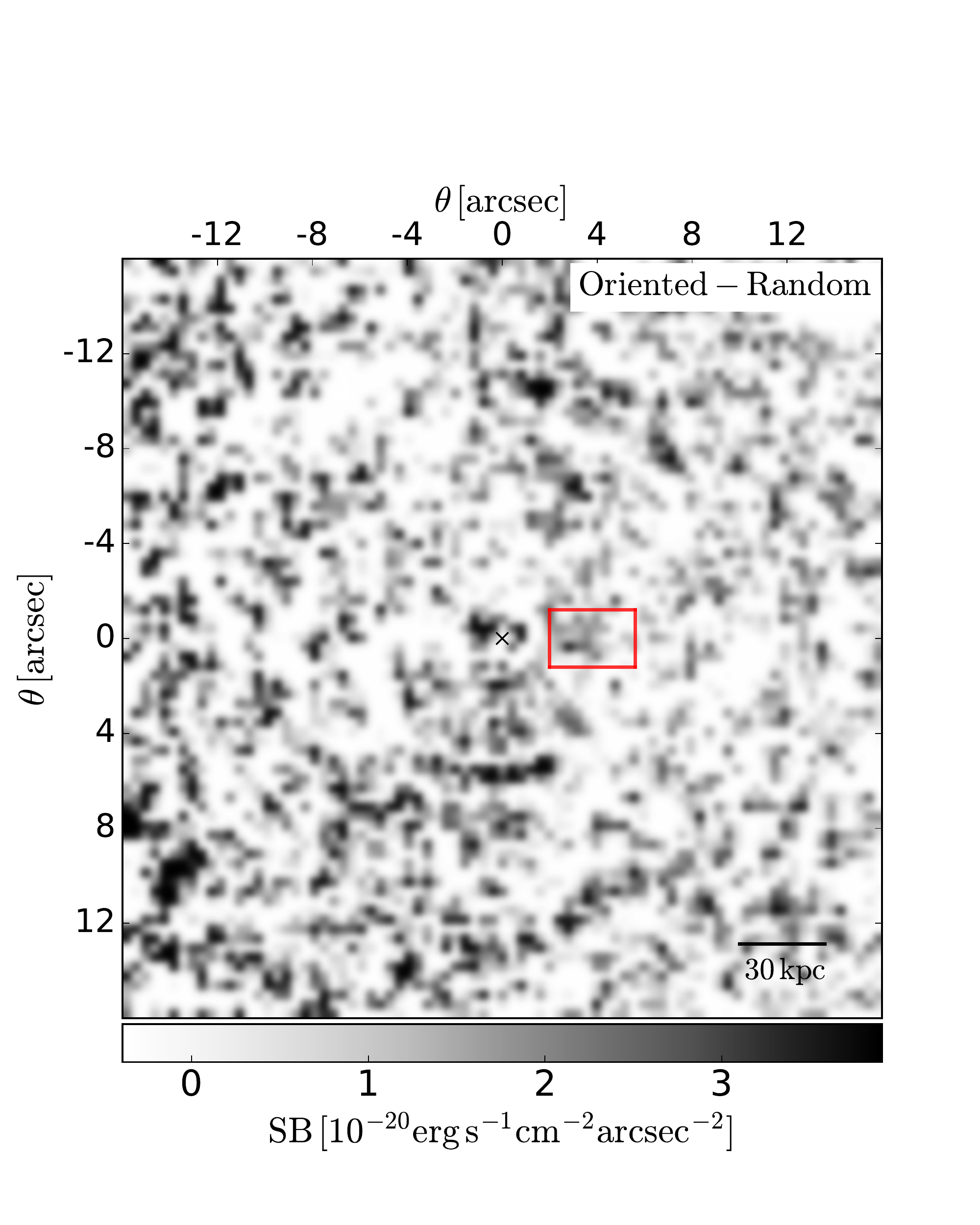}
\includegraphics[trim={.2cm 1.5cm 2.01cm 5.3cm},clip, width=3.696in]{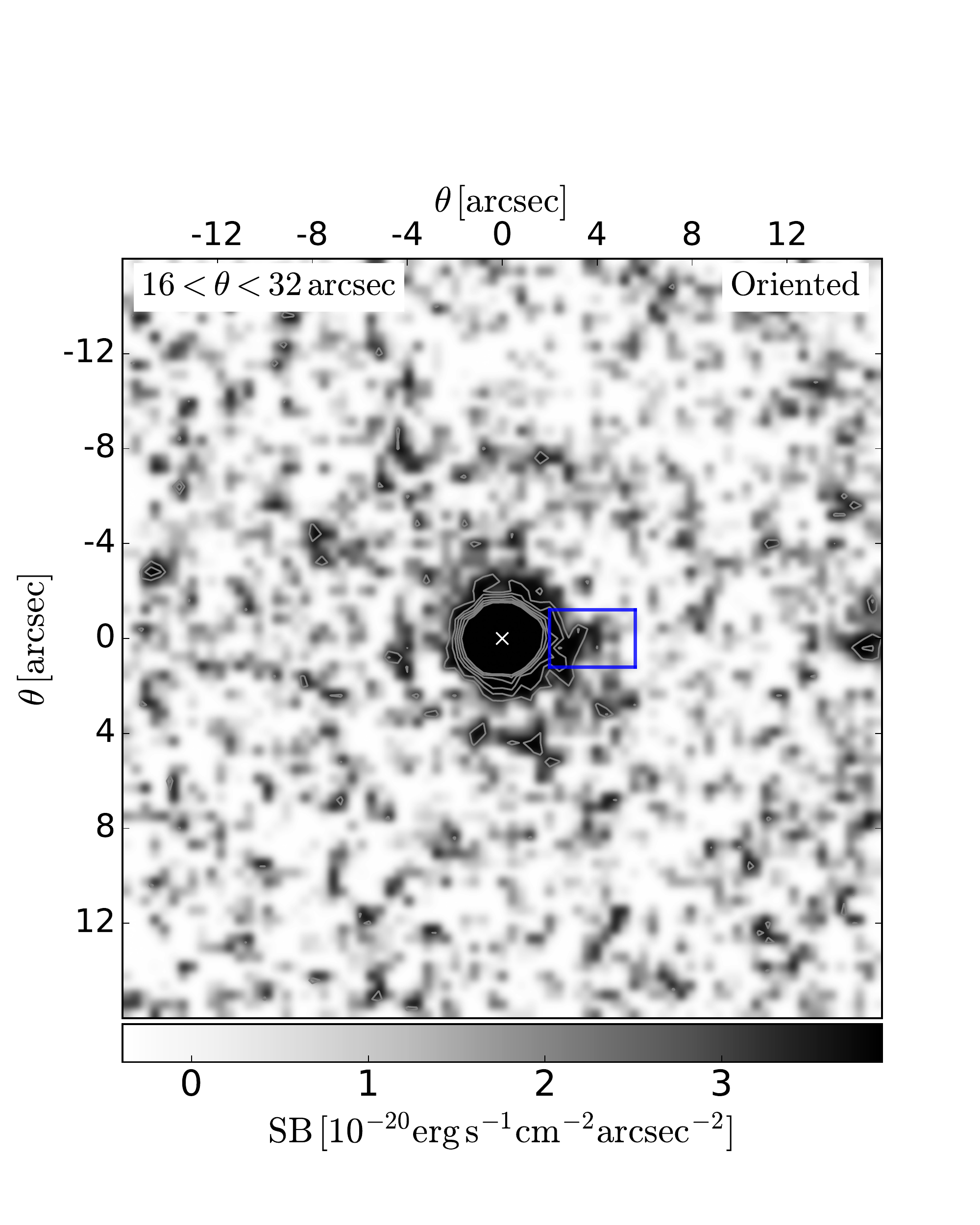}
\includegraphics[trim={2.5cm 1.5cm 2cm 5.3cm},clip, width=3.23in]{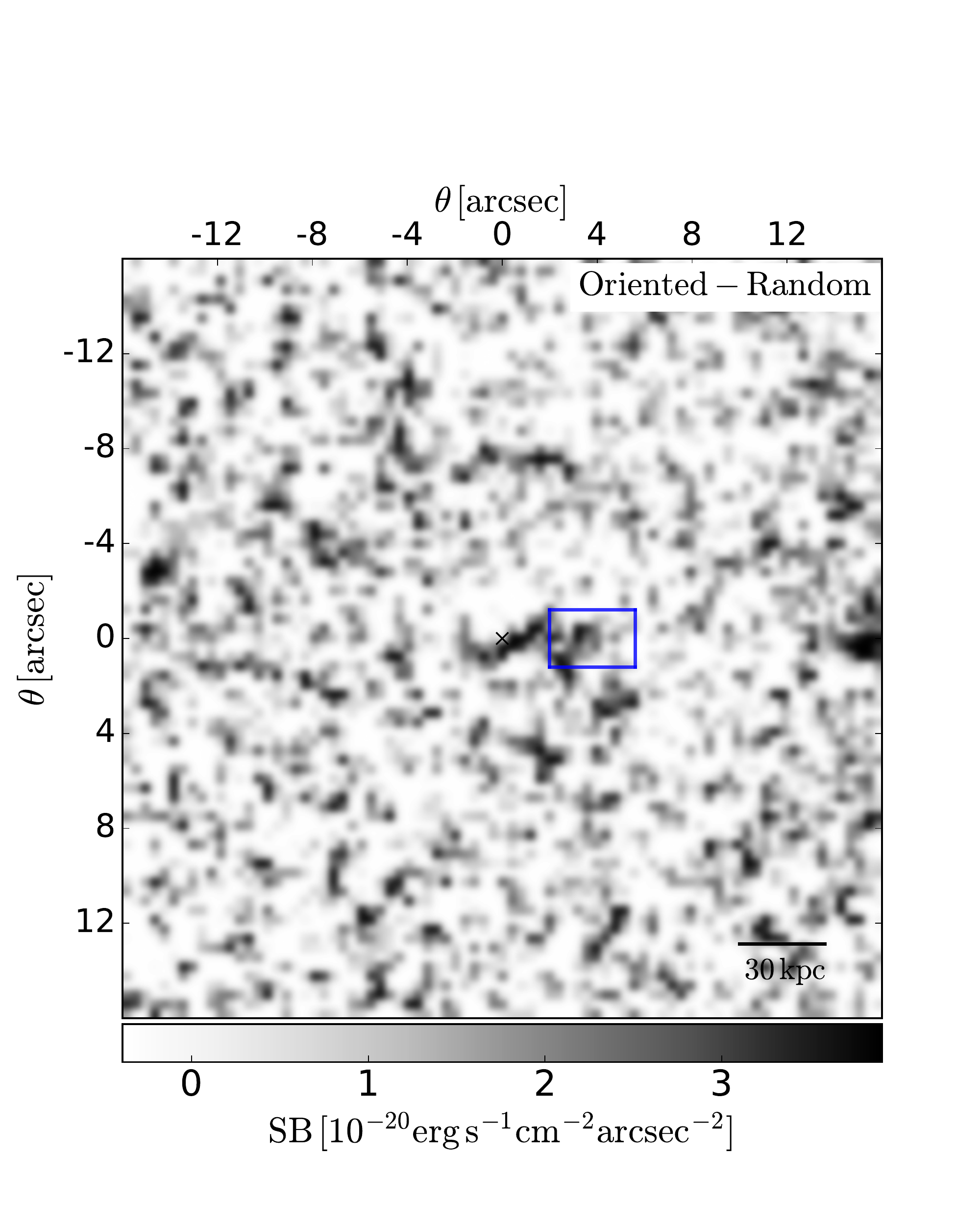}
   \caption{Same as Figure \ref{stacks} for a median projected distance to the neighbour of $32''$.}
  \label{stacks4}
\end{center}
\end{figure*}

\label{lastpage}

\end{document}